\documentclass[epj]{webofc}
\usepackage[varg]{txfonts}

\wocname{EPJ Web of Conferences}

\woctitle{ICNFP 2017}

\begin{document}

\begin{titlepage}

\def\thefootnote{\fnsymbol{footnote}}       % symbols for footnotes

\begin{center}
\mbox{ }

\end{center}
\begin{center}
\vskip 1.0cm
{\Huge\bf
SUSY (ATLAS)
}
\vskip 1cm
{\LARGE\bf 
Andr\'e~Sopczak on behalf of the ATLAS Collaboration
\bigskip
}

\Large 
Institute of Experimental and Applied Physics, \\
Czech Technical University in Prague, Czech Republic

\vskip 1.0cm
\centerline{\Large \bf Abstract}
\end{center}

\vskip 0.6cm
\hspace*{-0.5cm}
\begin{picture}(0.001,0.001)(0,0)
\put(,0){
\begin{minipage}{\textwidth}
\Large
\renewcommand{\baselinestretch} {1.2}
During the LHC Run-II data-taking period, several searches for 
supersymmetric particles were performed by the ATLAS collaboration.
The results from these searches are concisely reviewed.
Model-independent and model-dependent limits on new particle production are set,
and interpretations in supersymmetric models are given.
\renewcommand{\baselinestretch} {1.}

\normalsize
\vspace{3.5cm}
\begin{center}
{\sl \large
Presented at the 6th International Conference on New Frontiers in Physics (ICNFP2017), 
Crete, Greece 
\vspace{-6cm}
}
\end{center}
\end{minipage}
}
\end{picture}
\vfill

\end{titlepage}

\newpage
\pagestyle{headings} 
%\markboth{left head}{right head}
\setcounter{page}{1}

\selectlanguage{english}
\title{SUSY (ATLAS)}

\author{Andr\'e Sopczak on behalf of the ATLAS Collaboration\inst{1}\fnsep\thanks{\email{andre.sopczak@cern.ch}}
}

\institute{Institute of Experimental and Applied Physics, Czech Technical University in Prague
}

\abstract{%
During the LHC Run-II data-taking period, several searches for 
supersymmetric particles were performed by the ATLAS collaboration.
The results from these searches are concisely reviewed.
Model-independent and model-dependent limits on new particle production are set,
and interpretations in supersymmetric models are given.
}
\maketitle
\section{Introduction}
\label{intro}
\begin{figure}[h]

\centering
\includegraphics[width=0.44\textwidth]{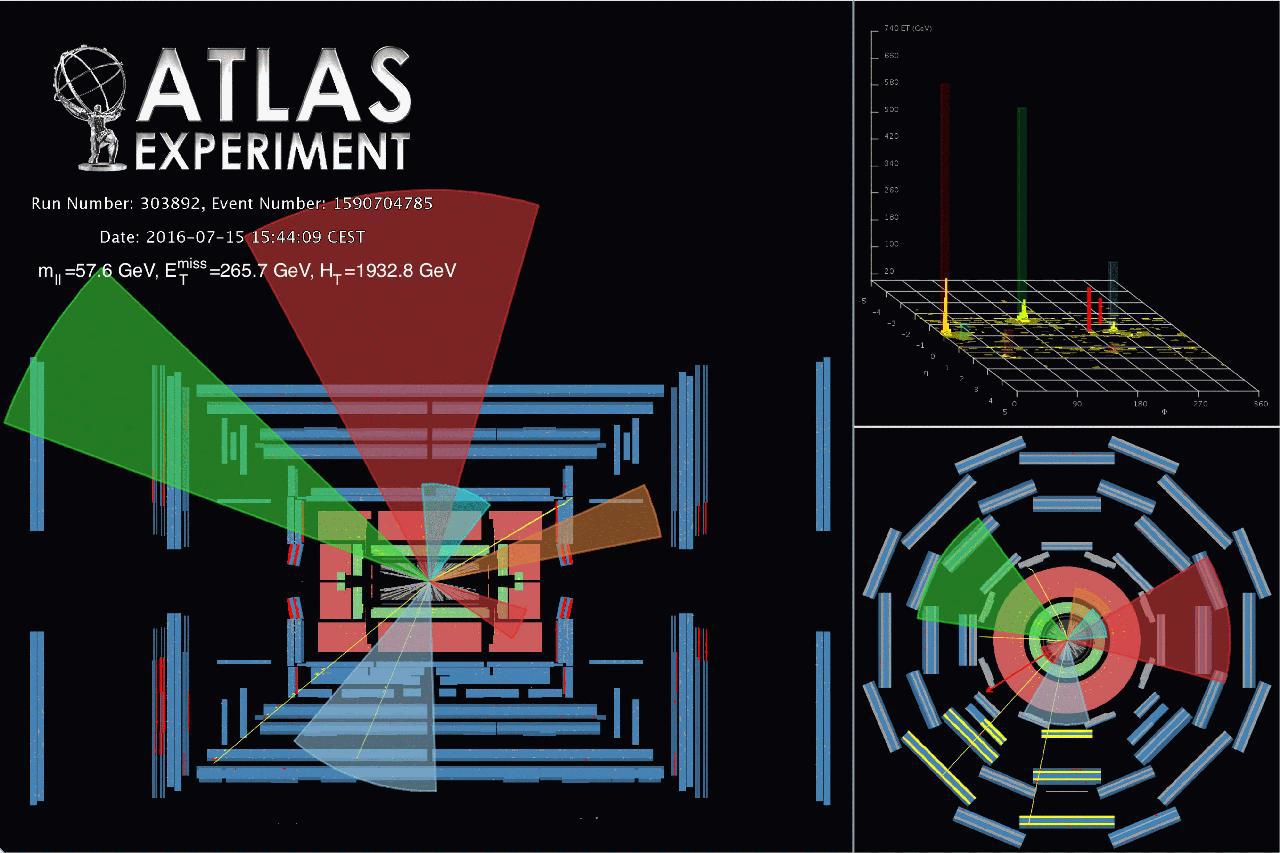}
\includegraphics[width=0.5\textwidth]{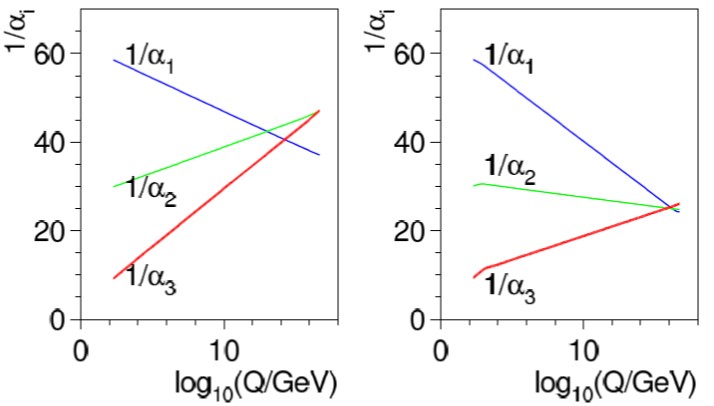}
\includegraphics[width=0.38\textwidth]{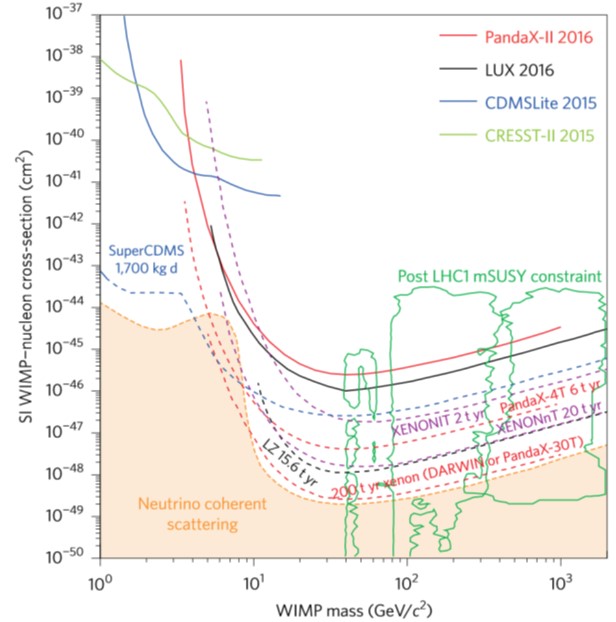} 
\includegraphics[width=0.54\textwidth]{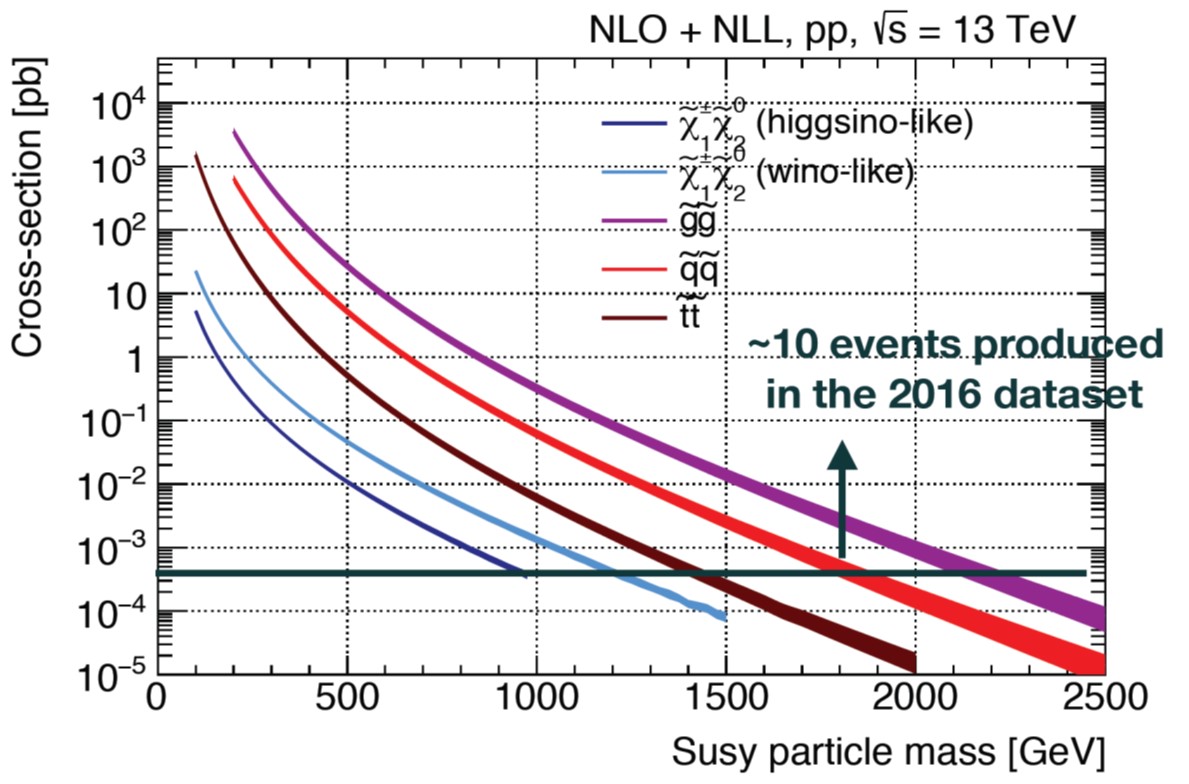}\\
\includegraphics[width=0.6\textwidth]{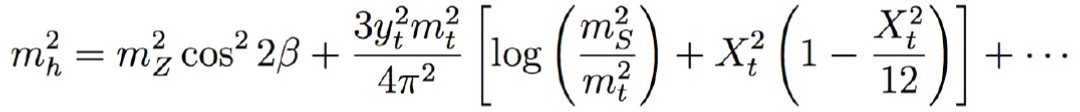}
\caption{ATLAS event display of a SUSY candidate with two muons, 
         jets and large missing energy~\cite{Aaboud:2016ejt}.
Illustration of gauge coupling unification~\cite{deBoer:2003xm}, 
dark matter candidates~\cite{Liu:2017drf} and
naturalness (light Higgs boson mass)~\cite{naturalness}.
Typical production cross-sections for SUSY reactions~\cite{Borschensky:2014cia}.
}
\label{fig:candidate}
\vspace*{-5mm}
\end{figure}

Several searches for supersymmetric (SUSY) particles were performed
with data taken by the ATLAS experiment~\cite{Aad:2008zzm}.
For illustration, an example of a SUSY candidate event display is shown 
in Fig.~\ref{fig:candidate}~\cite{Aaboud:2016ejt}. 
The event display shows two muons, jets and large missing energy.

The search for SUSY particles is highly motivated in three aspects 
1) gauge coupling unification~\cite{deBoer:2003xm}, 
2) dark matter candidates~\cite{Liu:2017drf} and 
3) naturalness (light Higgs boson mass)~\cite{naturalness}. 
These aspects are illustrated in Fig.~\ref{fig:candidate}.
An overview of all ATLAS SUSY searches is given in Ref.~\cite{atlassusy}.
The major expected production cross-sections for SUSY reactions are illustrated 
in Fig.~\ref{fig:candidate}~\cite{Borschensky:2014cia}.

This article is structured in the following sections.
\ref{sec:inclusive}: inclusive production (0, 1, 2 leptons),
\ref{sec:stop}: direct stop production,
\ref{sec:bottom}: direct sbottom production,
\ref{sec:small}: small mass differences,
\ref{sec:ew}: electroweak direct production,
\ref{sec:photpnic}: photonic signatures,
\ref{sec:long}: long-lived particles,
\ref{sec:parity}: R-parity violation,
\ref{sec:displaced}: displaced vertices,
\ref{sec:pmssm}: pMSSM,
\ref{sec:prospects}: prospects
\ref{sec:conclusions}: conclusions and outlook.

\section{Inclusive production (0, 1, 2 leptons)}
\label{sec:inclusive}
\begin{figure}[h]
\centering
\includegraphics[width=0.24\textwidth]{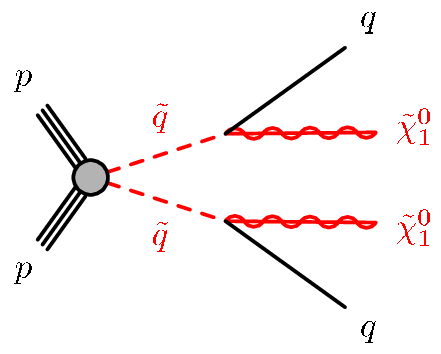}
\includegraphics[width=0.24\textwidth]{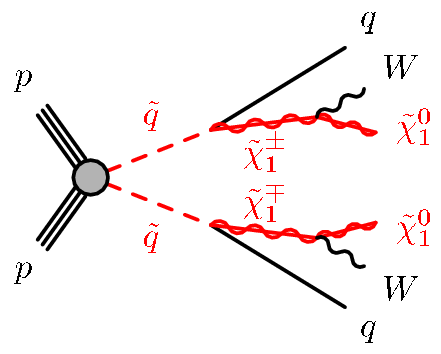}
\includegraphics[width=0.24\textwidth]{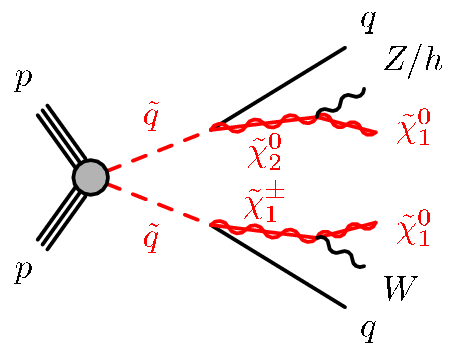}
\includegraphics[width=0.24\textwidth]{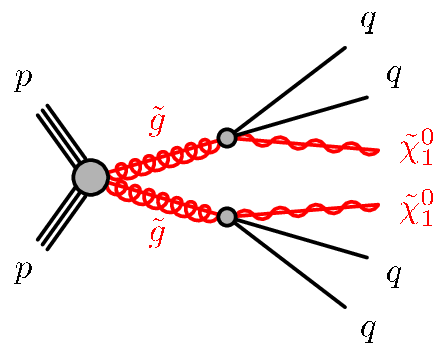}
\includegraphics[width=0.24\textwidth]{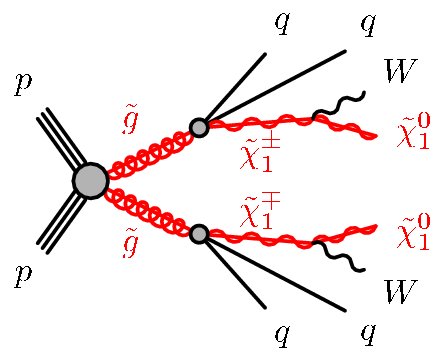}
\includegraphics[width=0.24\textwidth]{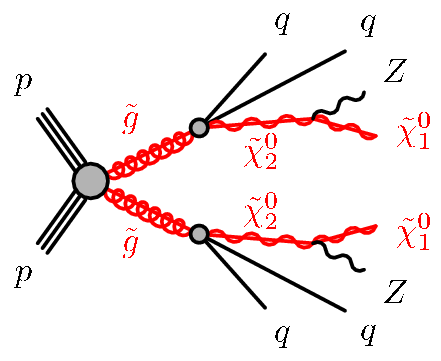}
\includegraphics[width=0.24\textwidth]{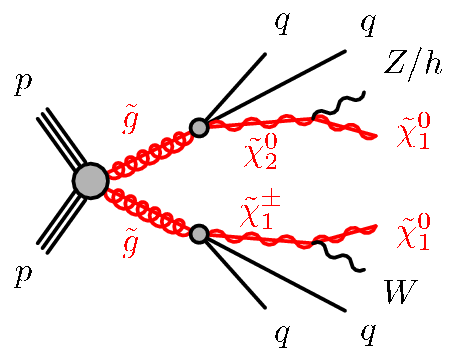}\\
\includegraphics[width=0.24\textwidth]{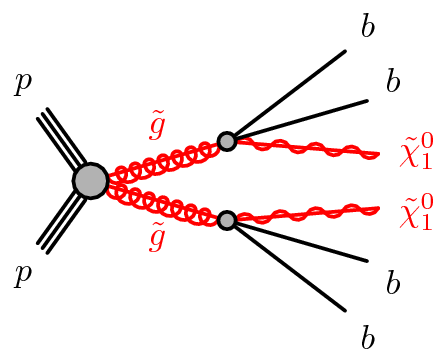}
\includegraphics[width=0.24\textwidth]{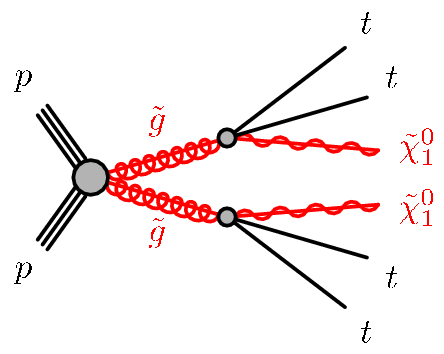}
\includegraphics[width=0.24\textwidth]{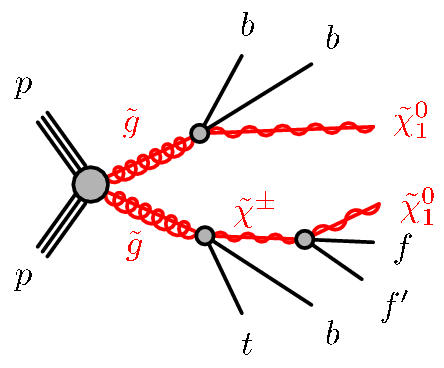}
\includegraphics[width=0.24\textwidth]{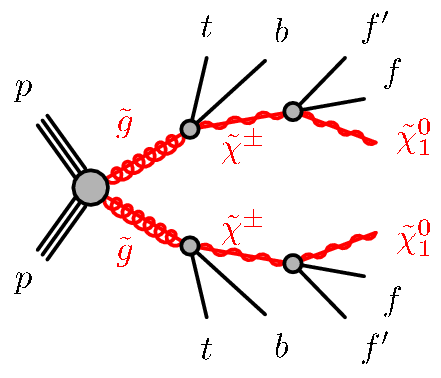}
\includegraphics[width=0.24\textwidth]{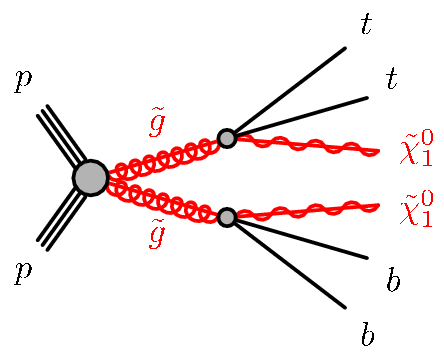}
\includegraphics[width=0.24\textwidth]{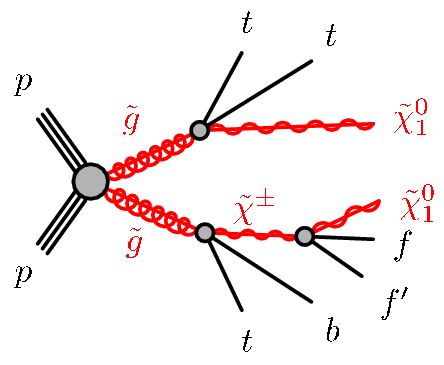}\\
\includegraphics[width=0.27\textwidth]{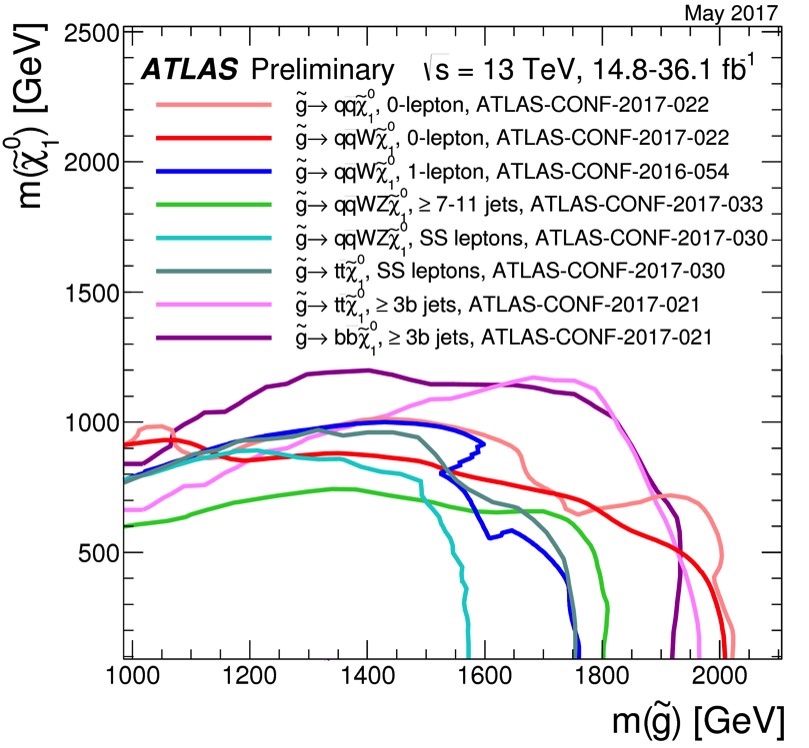}
\includegraphics[width=0.38\textwidth]{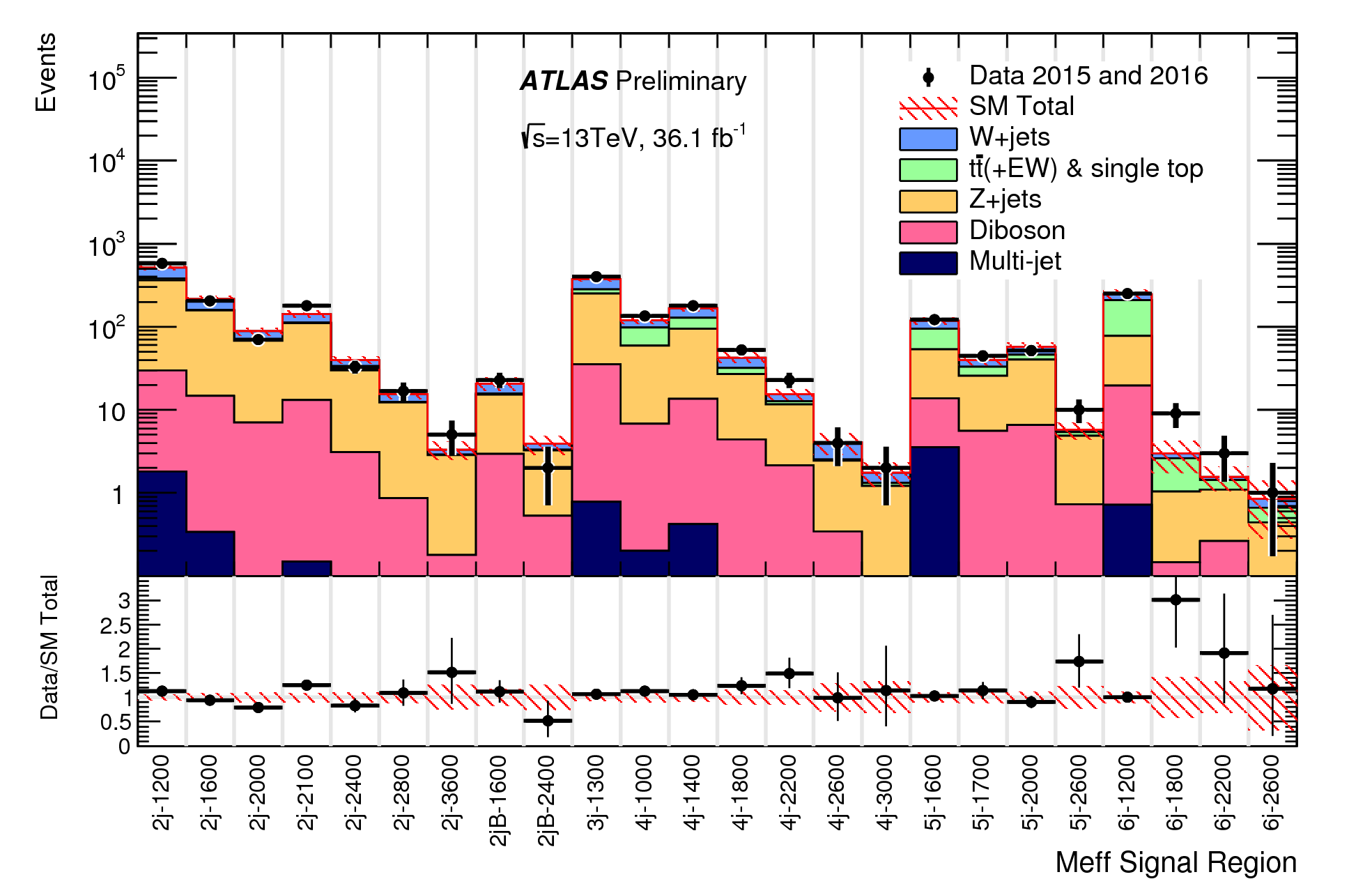}
\includegraphics[width=0.32\textwidth]{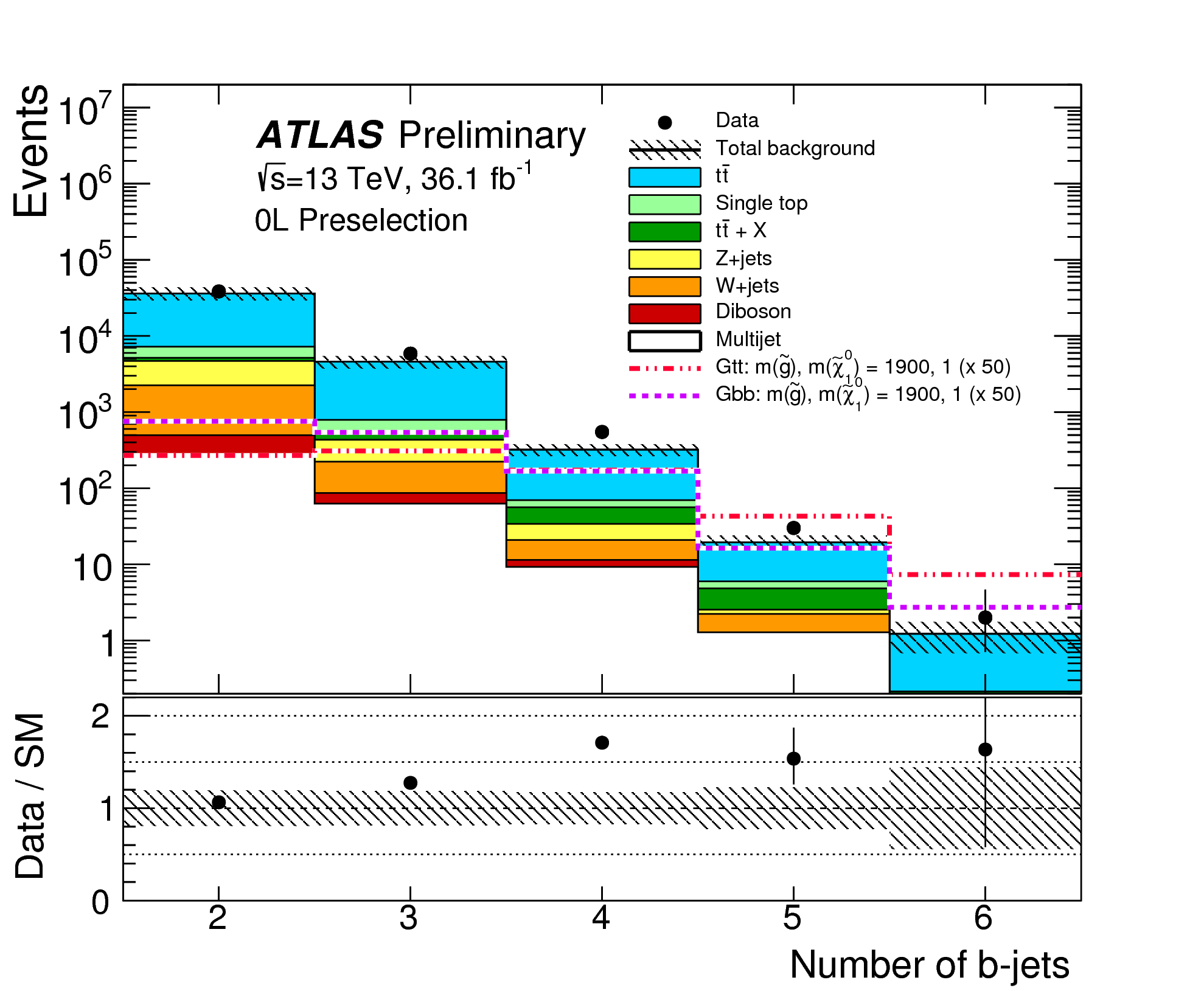}
\includegraphics[width=0.38\textwidth]{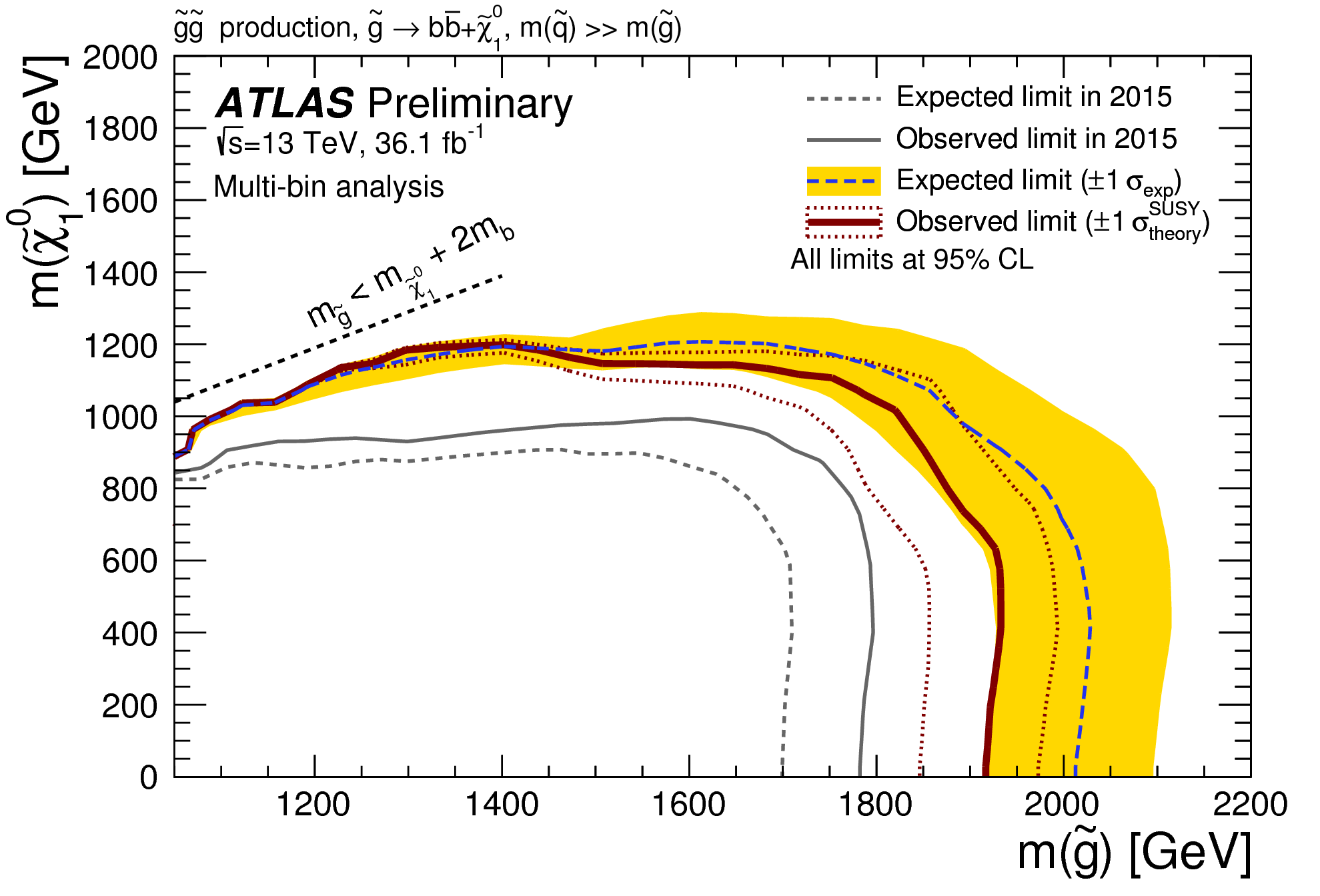}
\includegraphics[width=0.38\textwidth]{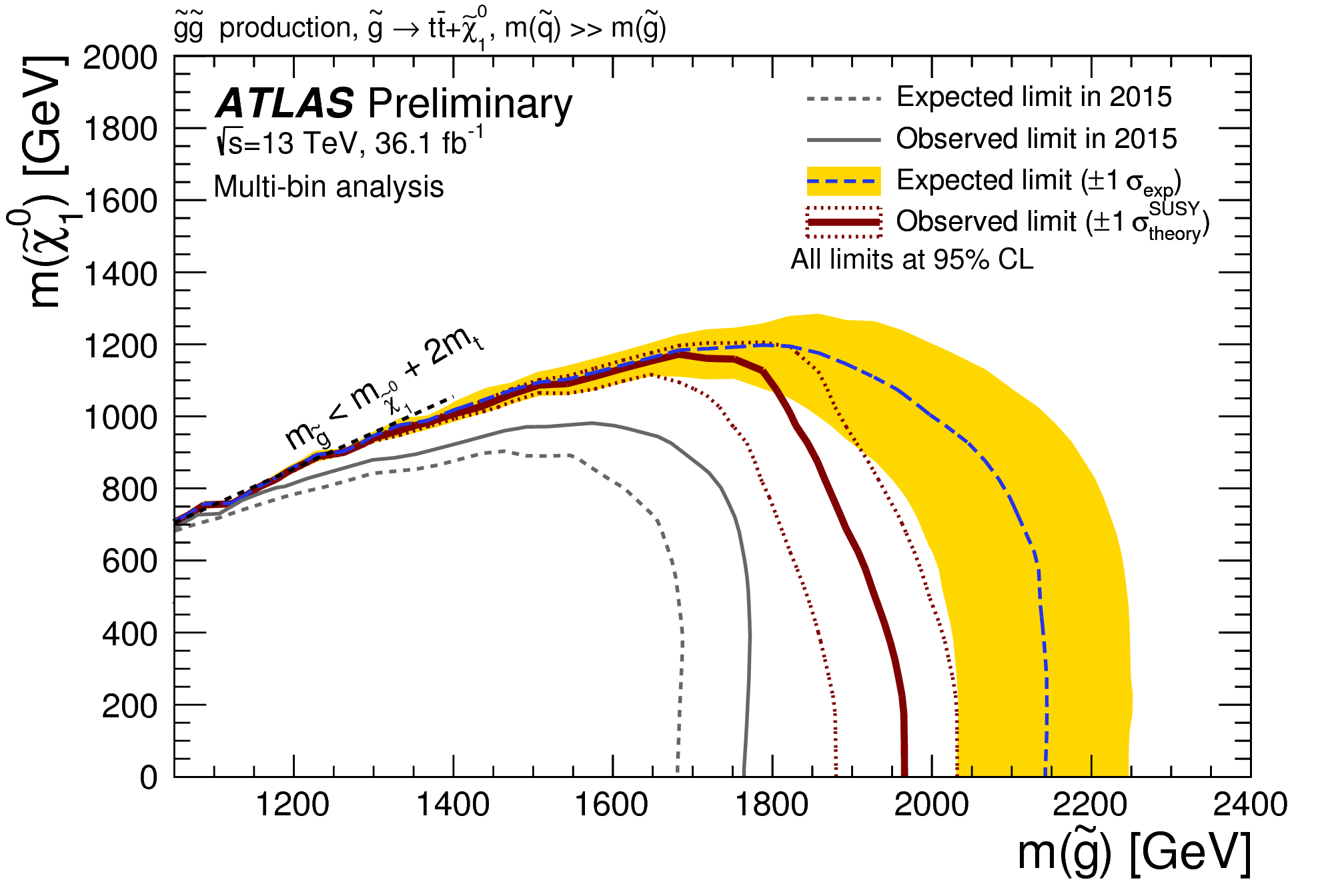}
\caption{Inclusive production decay topologies,
        overview of inclusive searches,
        comparison of the observed and expected event yields 
        as a function of signal region,
        number of selected b-jets and exclusion contours. 
         }
\label{fig:inclusive}
\end{figure}

Inclusive 2-6 jets searches for the third generation gluino mediation
were performed. 
Feynman diagrams, an overview and results are summarized in Fig.~\ref{fig:inclusive}~\cite{ATLAS:2017cjl,ATLAS:2017vjw}.

\section{Direct stop production}
\label{sec:stop}

\begin{figure}[h]
\centering
\includegraphics[width=0.24\textwidth]{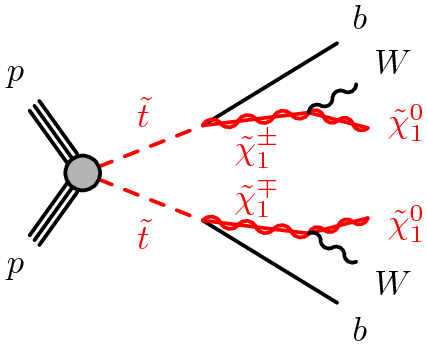}
\includegraphics[width=0.24\textwidth]{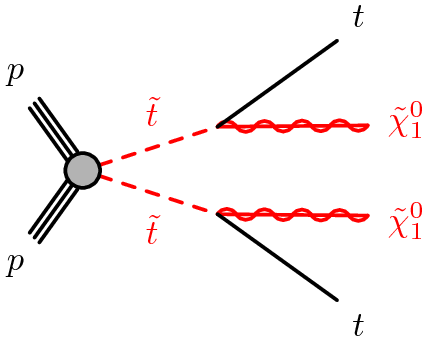}
\includegraphics[width=0.24\textwidth]{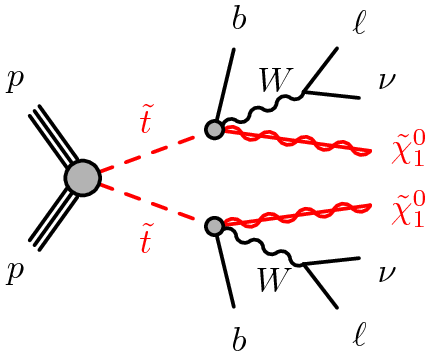}
\includegraphics[width=0.24\textwidth]{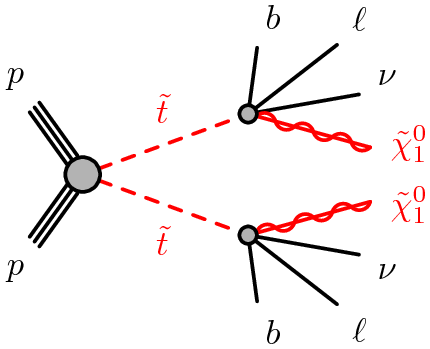}
\includegraphics[width=0.32\textwidth]{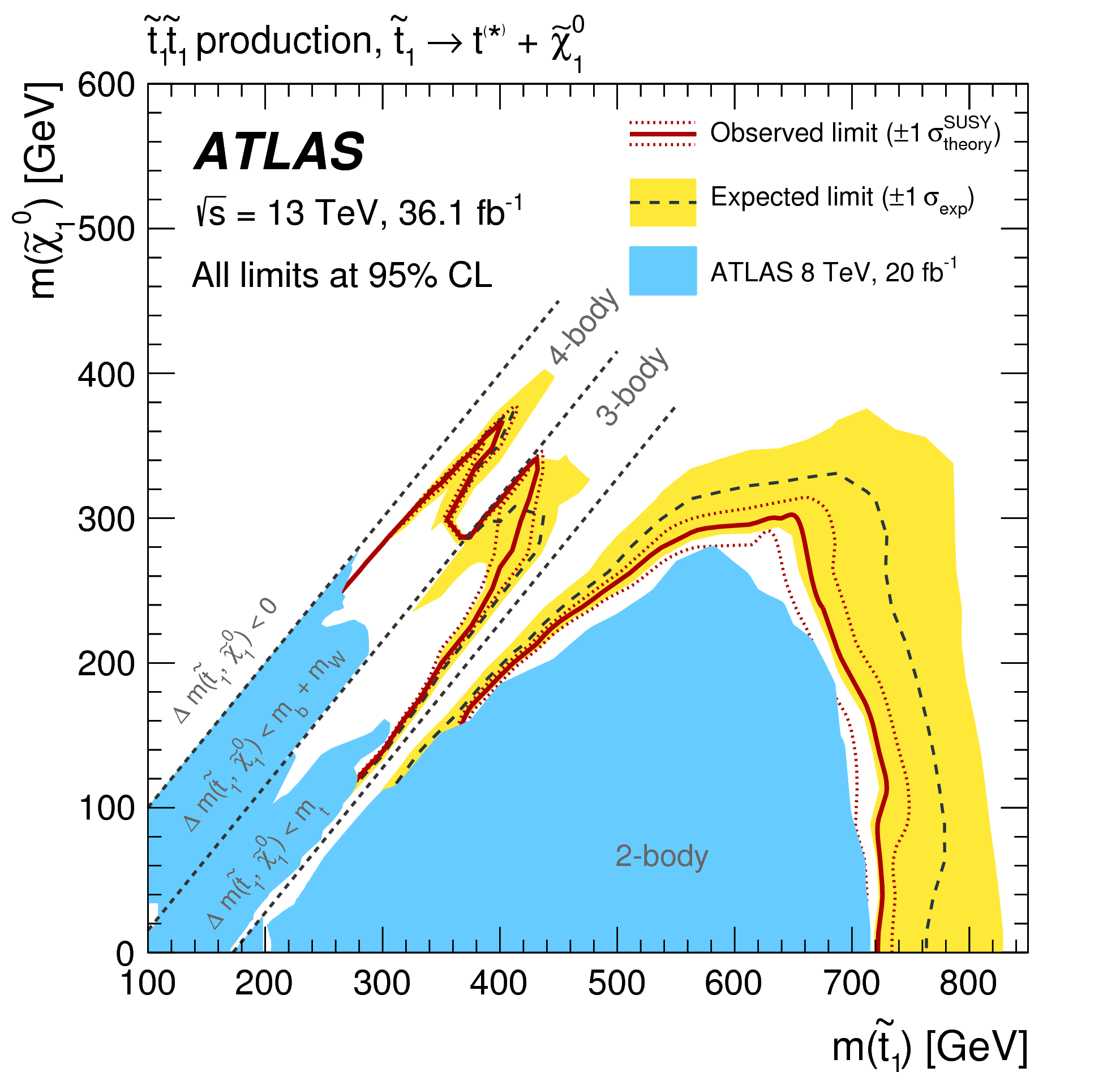}
\includegraphics[width=0.32\textwidth]{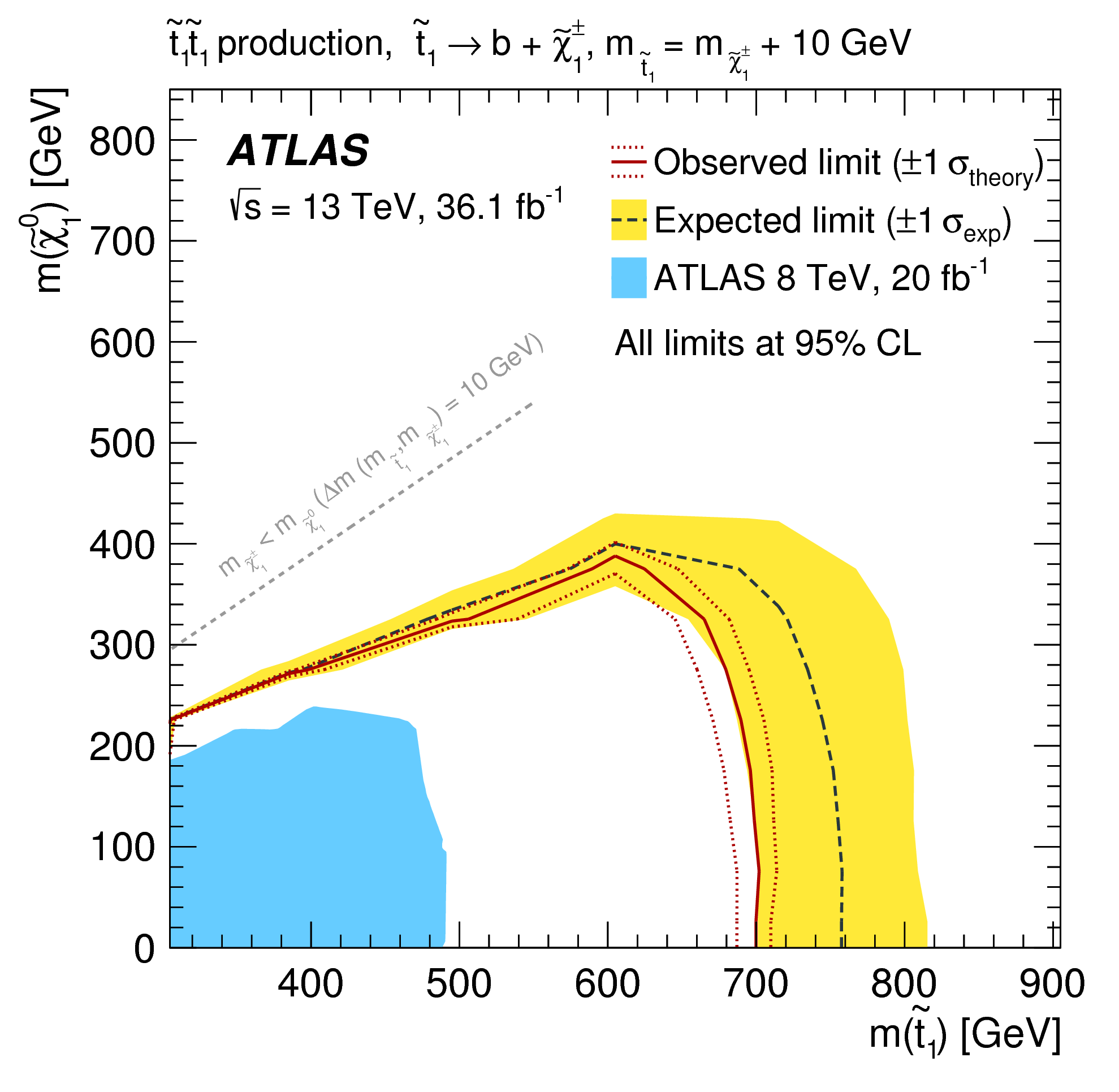}
\includegraphics[width=0.32\textwidth]{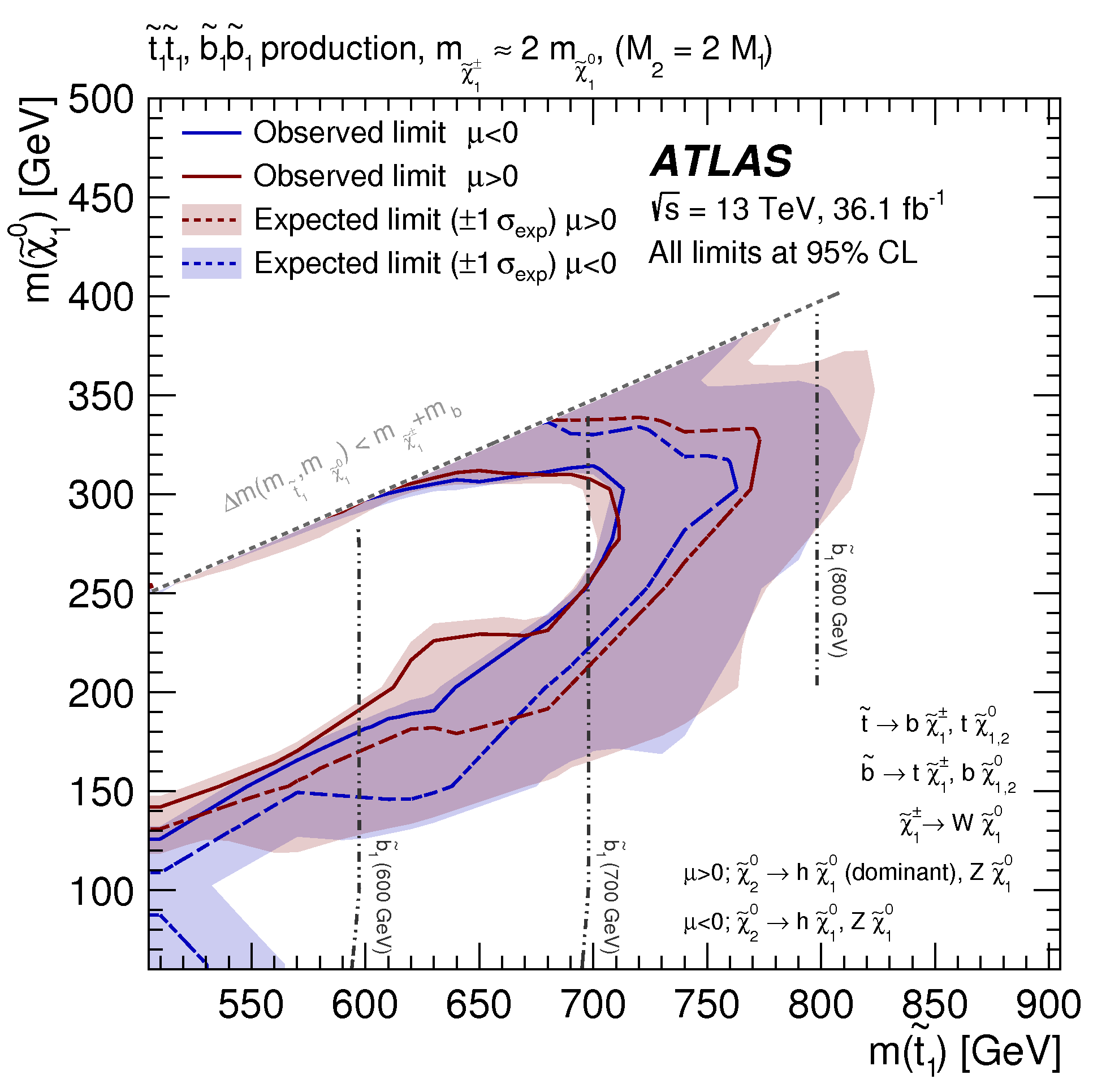}
\caption{Scalar top decay topologies for the final states with two leptons 
         and exclusion contours.
         }
\label{fig:stop}
\end{figure}

\begin{figure}[bp]
\centering
\includegraphics[width=0.24\textwidth]{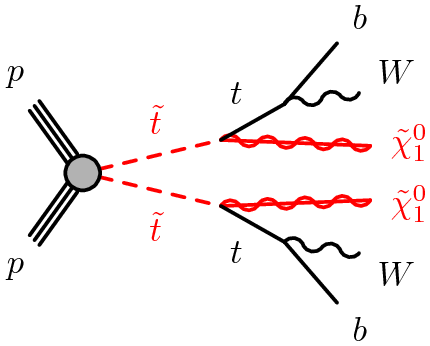}
\includegraphics[width=0.24\textwidth]{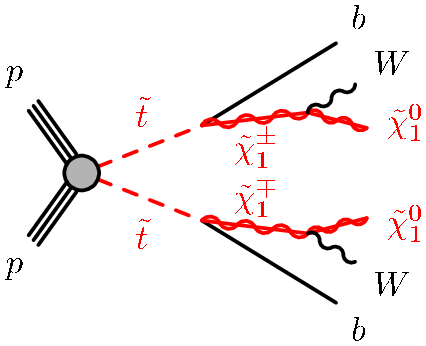}
\includegraphics[width=0.24\textwidth]{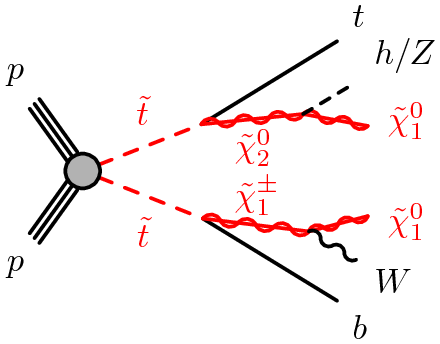}
\includegraphics[width=0.24\textwidth]{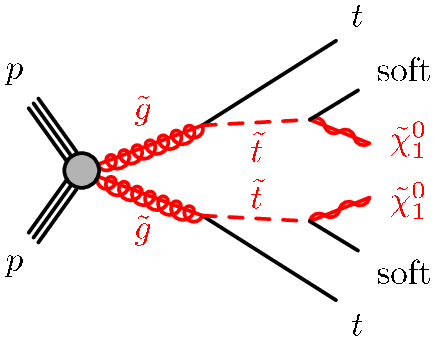}
\includegraphics[width=0.32\textwidth]{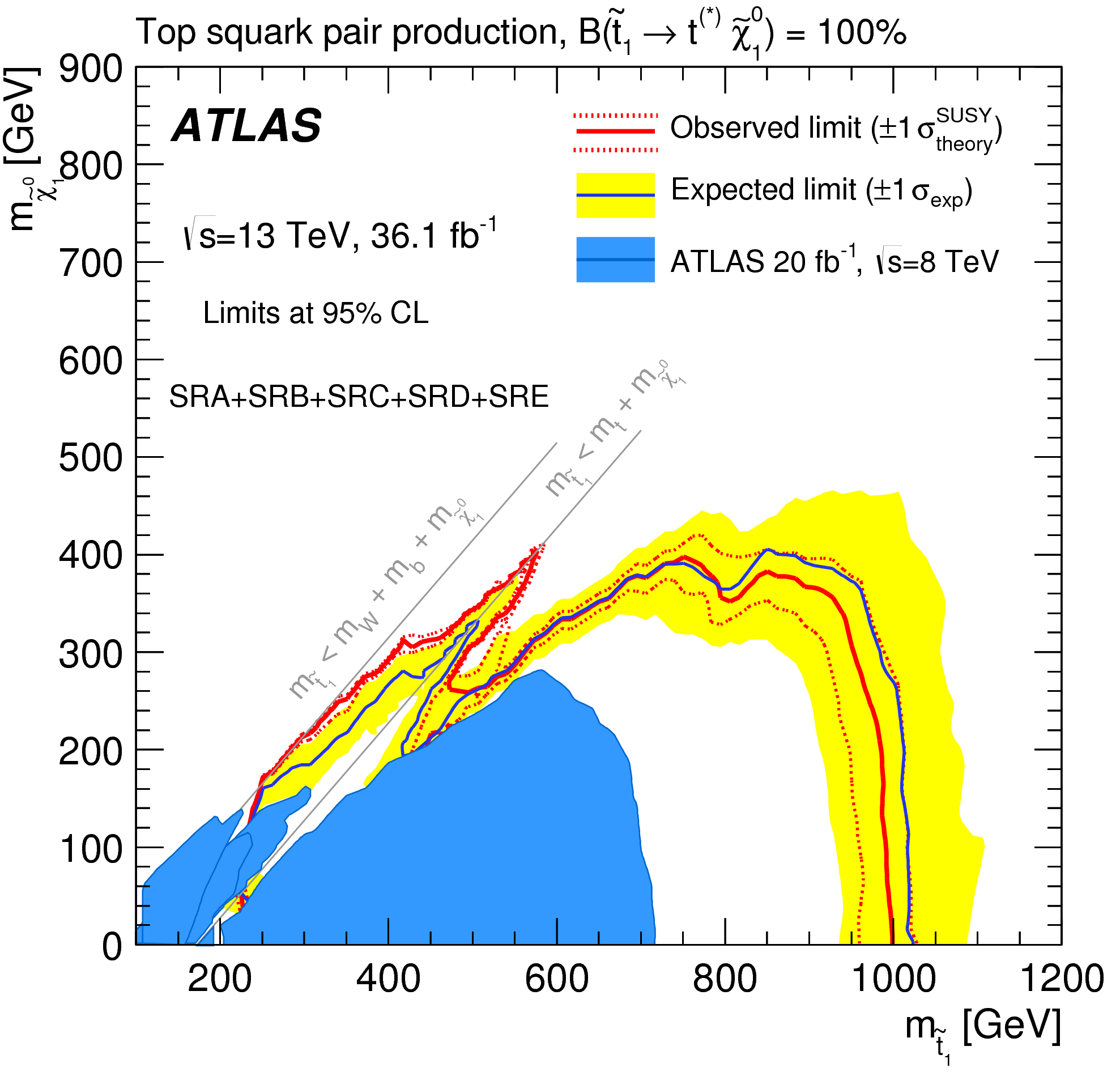}
\includegraphics[width=0.32\textwidth]{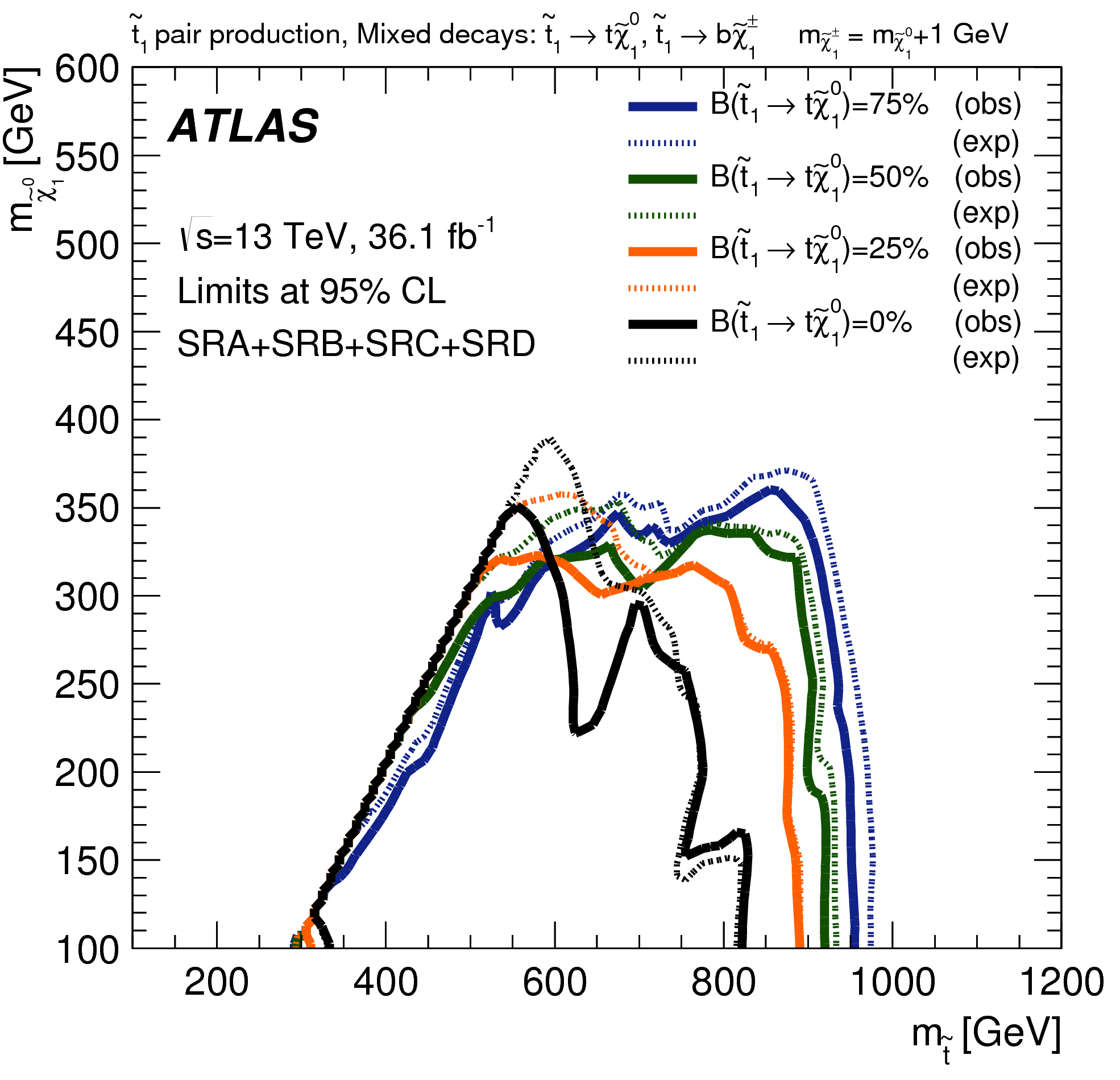}
\includegraphics[width=0.32\textwidth]{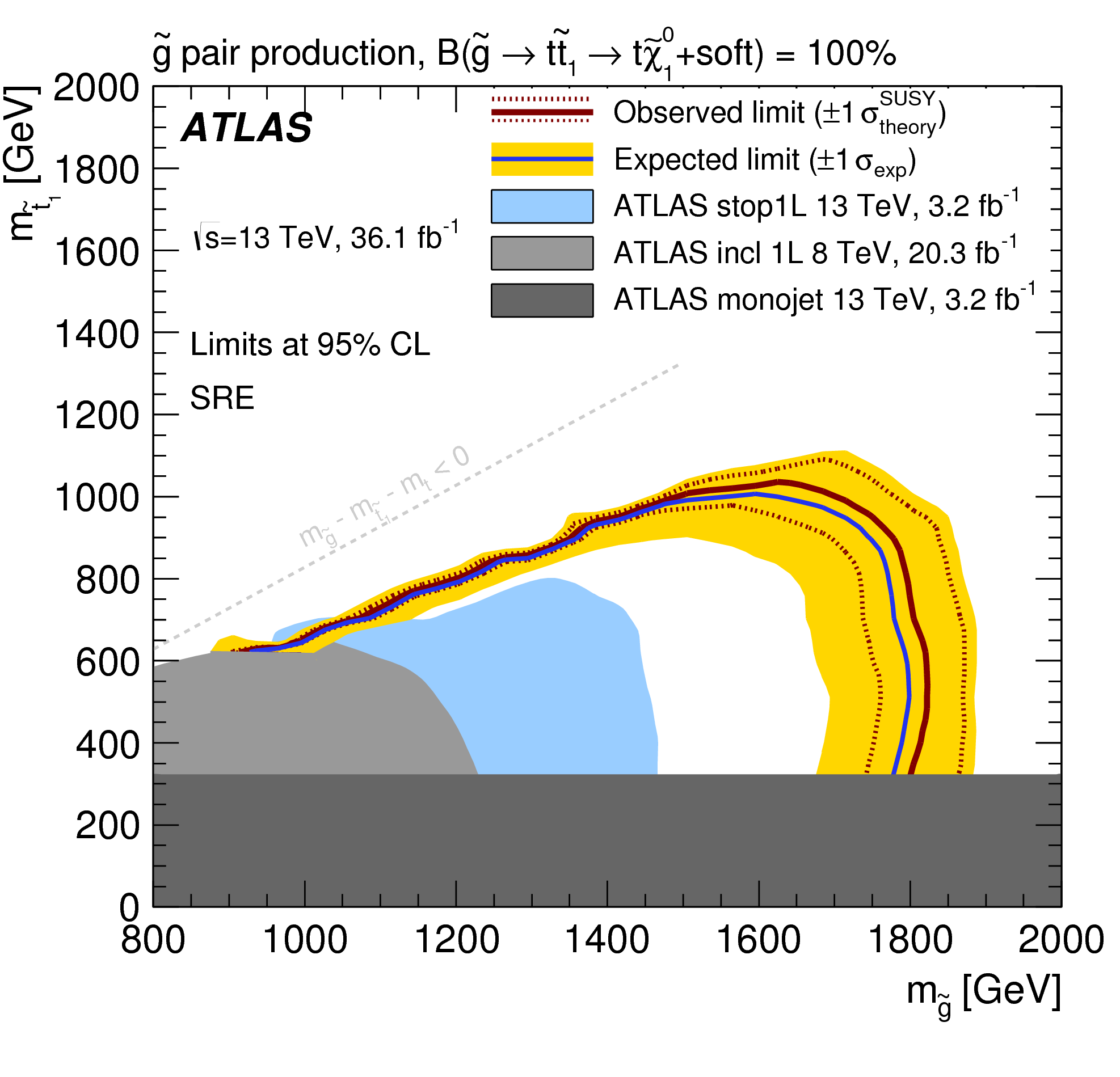}
\includegraphics[width=0.32\textwidth]{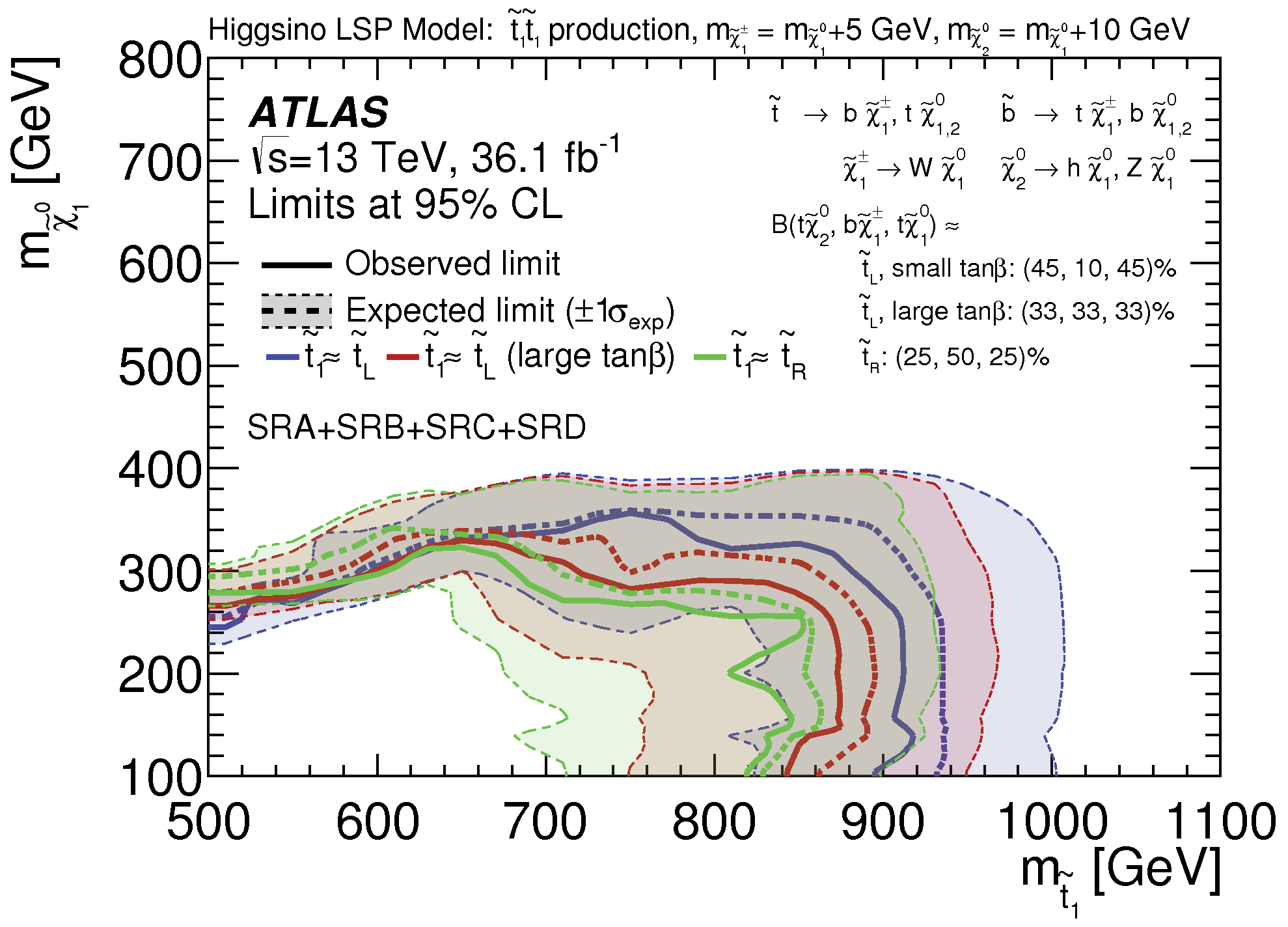}
\includegraphics[width=0.32\textwidth]{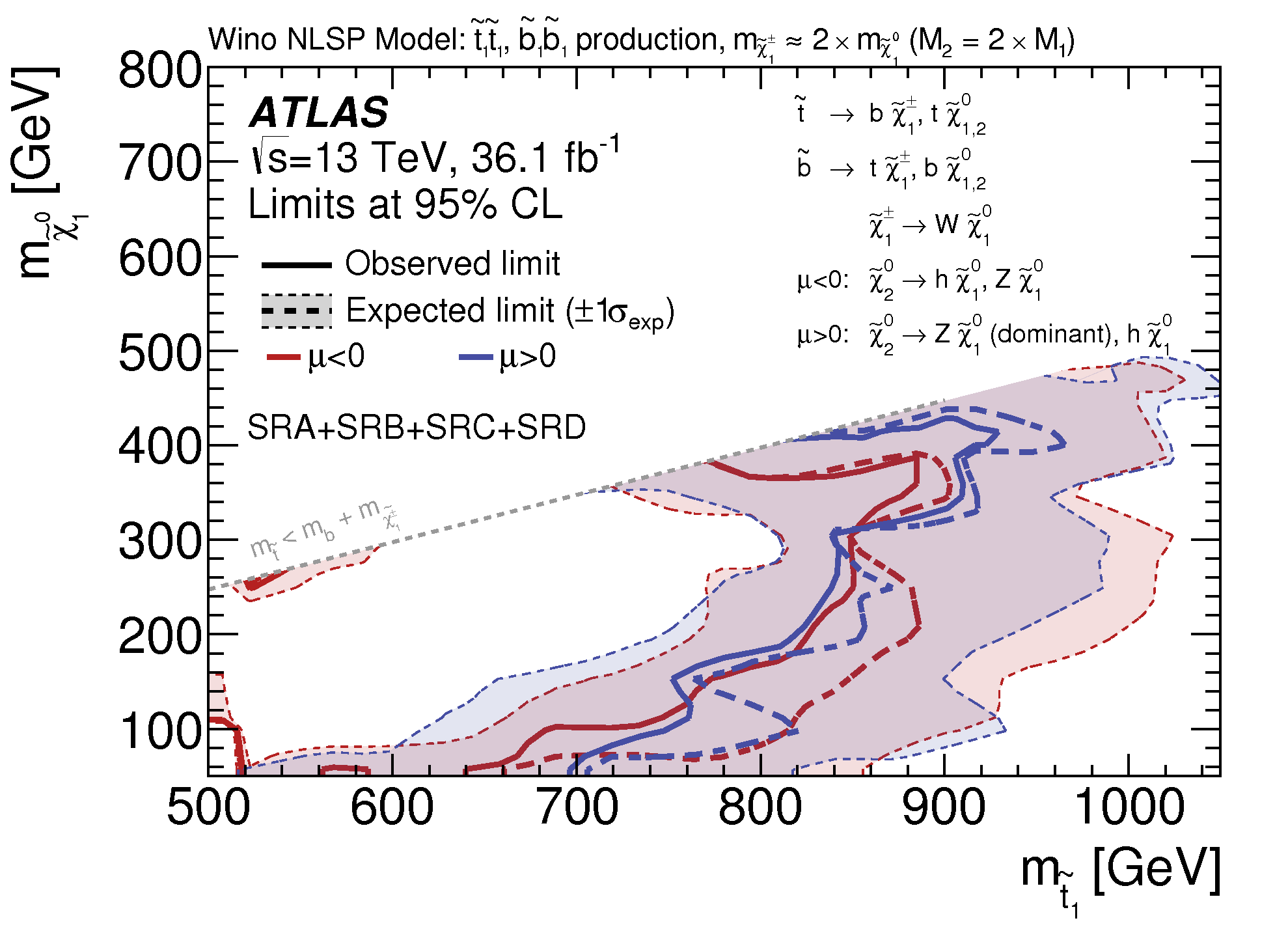}
\includegraphics[width=0.32\textwidth]{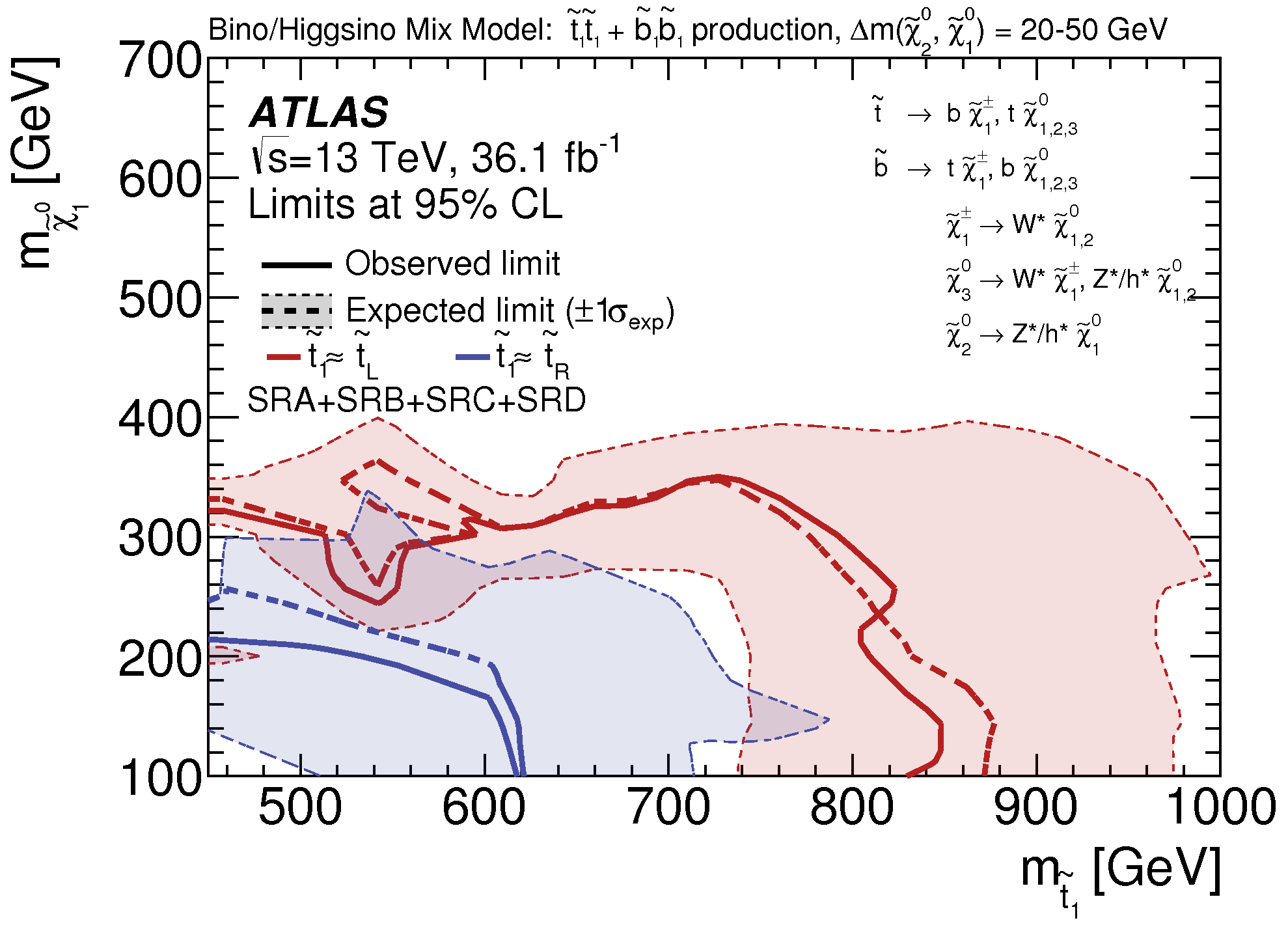}
\caption{Scalar top decay topologies for the final state with jets plus missing transverse momentum and exclusion contours.
         }
\label{fig:stop2}
\vspace*{-10mm}
\end{figure}

\begin{figure}[tp]
\centering
\includegraphics[width=0.24\textwidth]{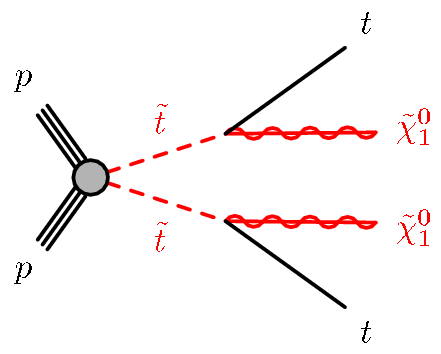}
\includegraphics[width=0.24\textwidth]{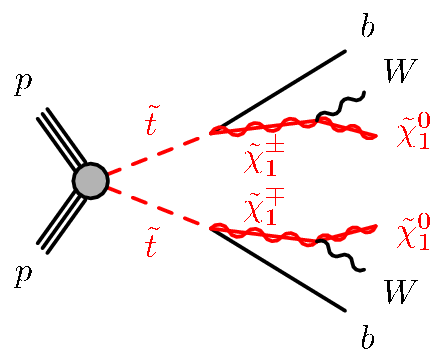}
\includegraphics[width=0.24\textwidth]{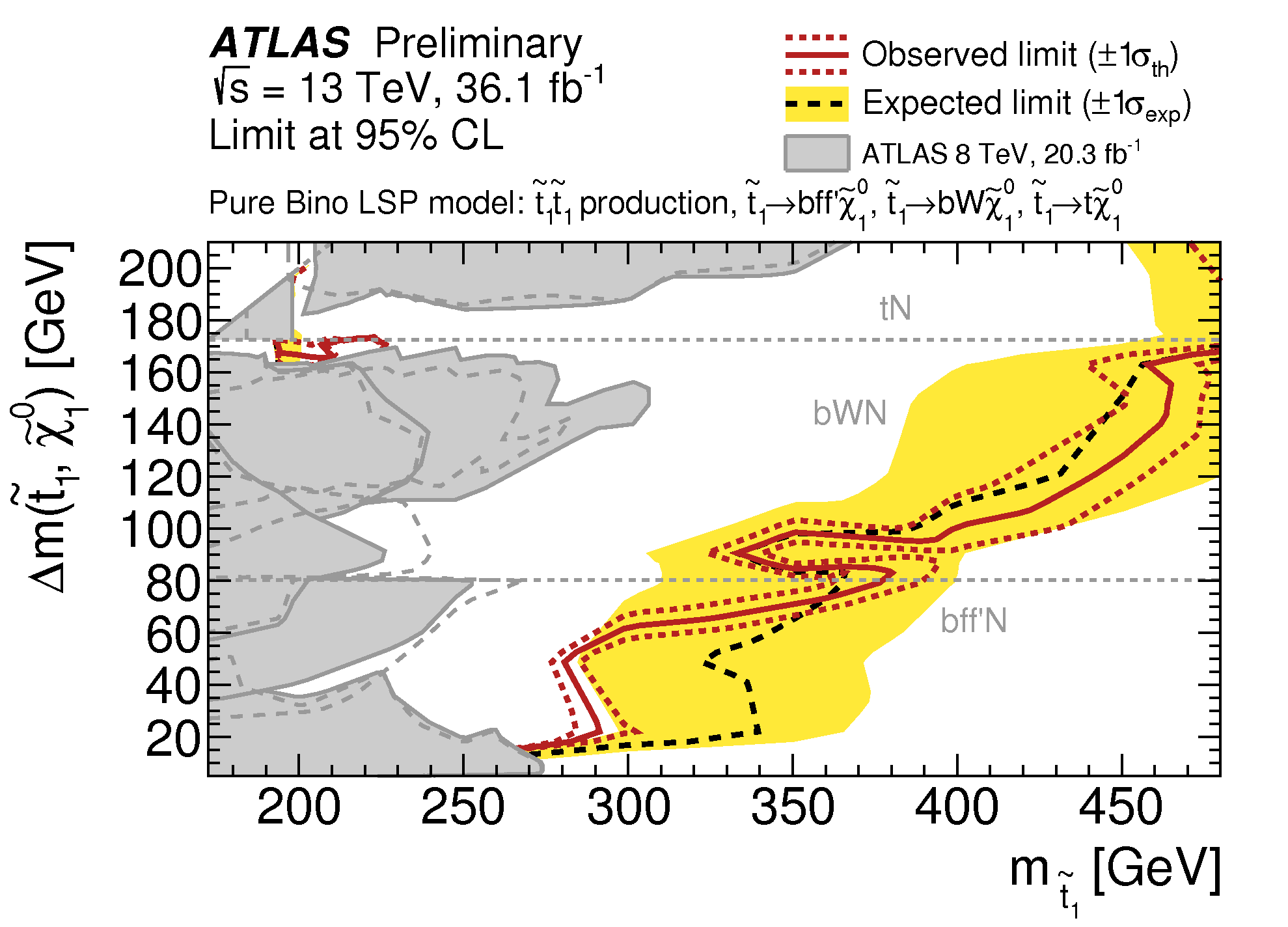}
\includegraphics[width=0.24\textwidth]{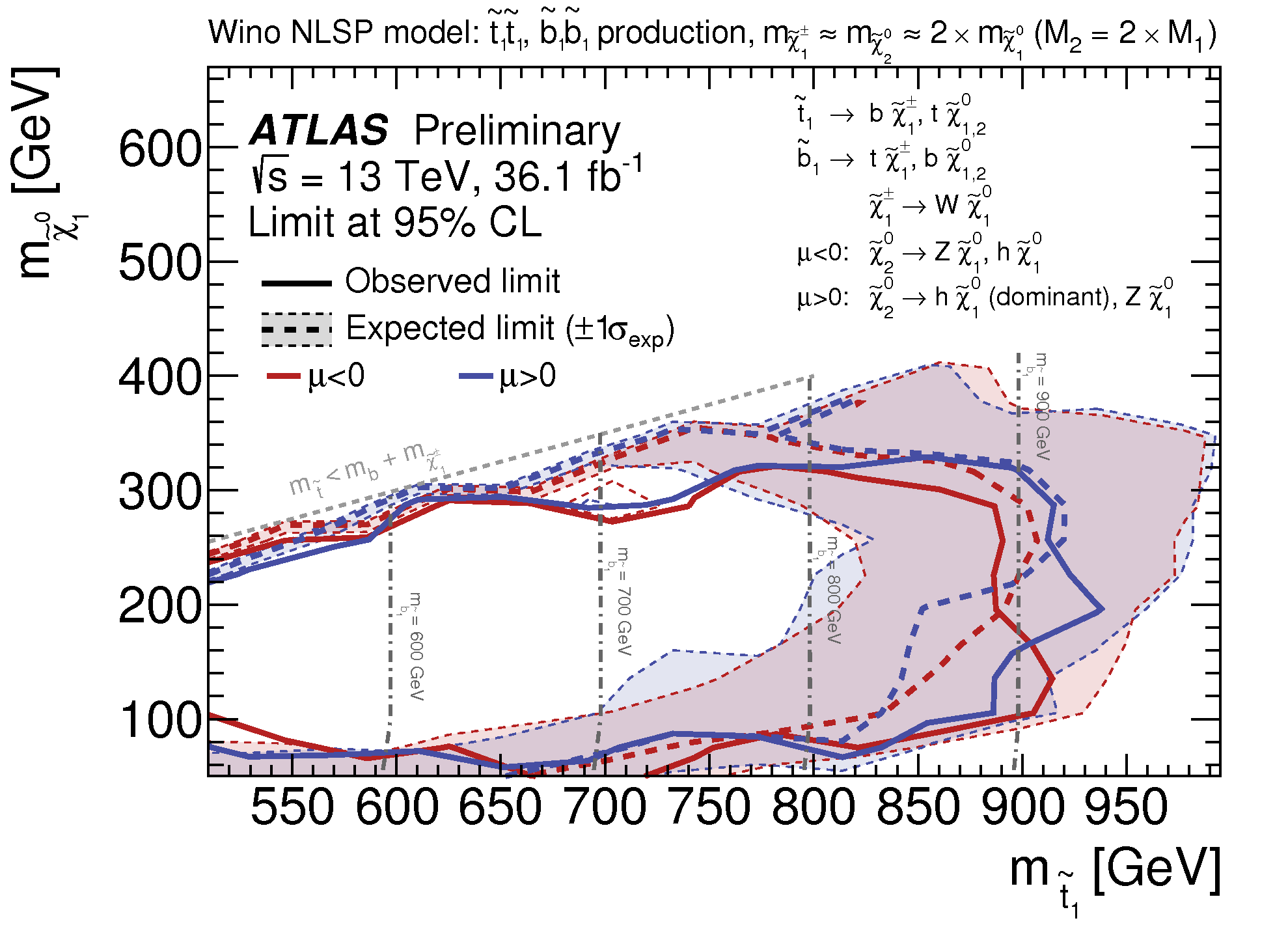}
\includegraphics[width=0.24\textwidth]{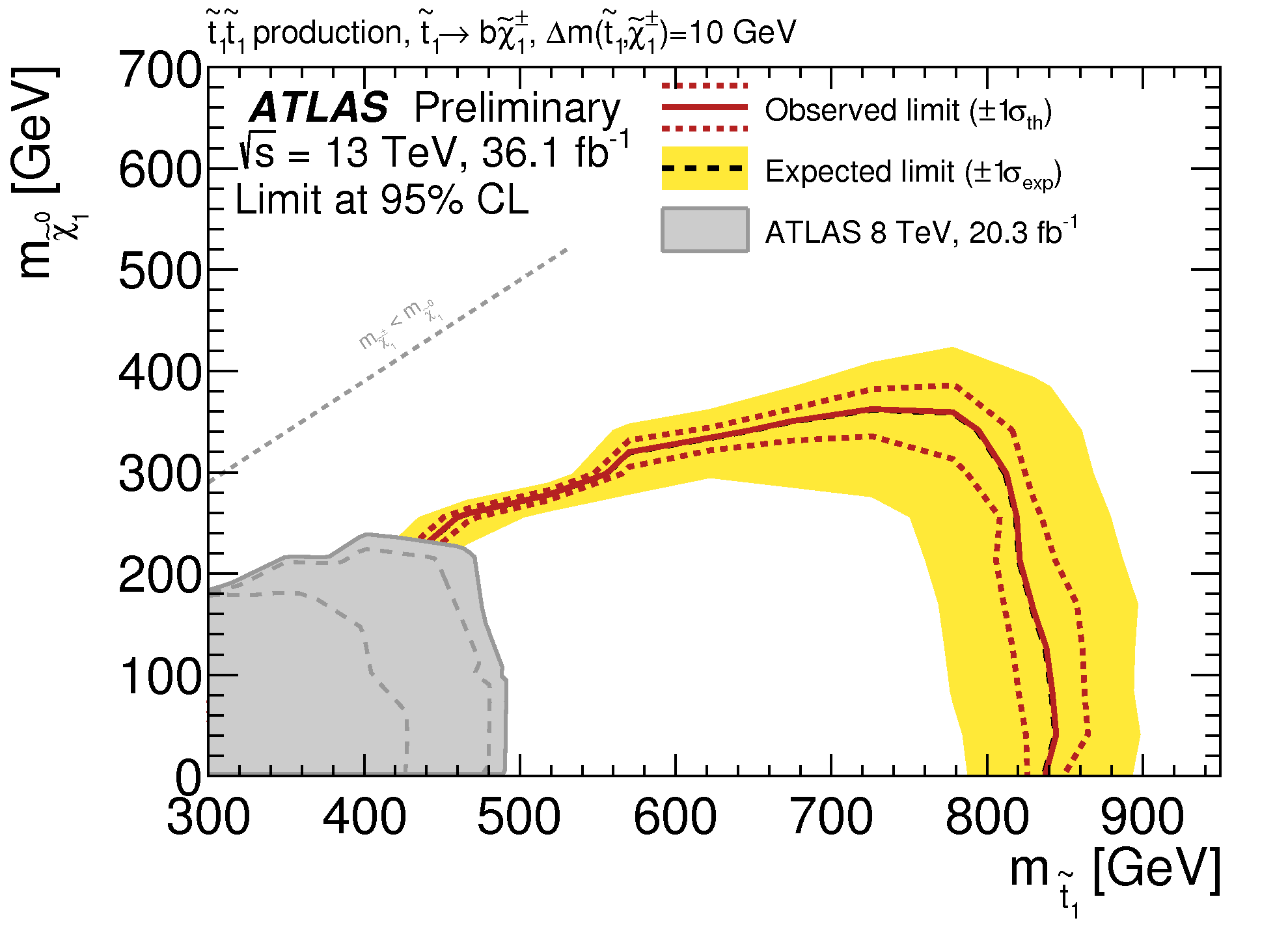}
\includegraphics[width=0.24\textwidth]{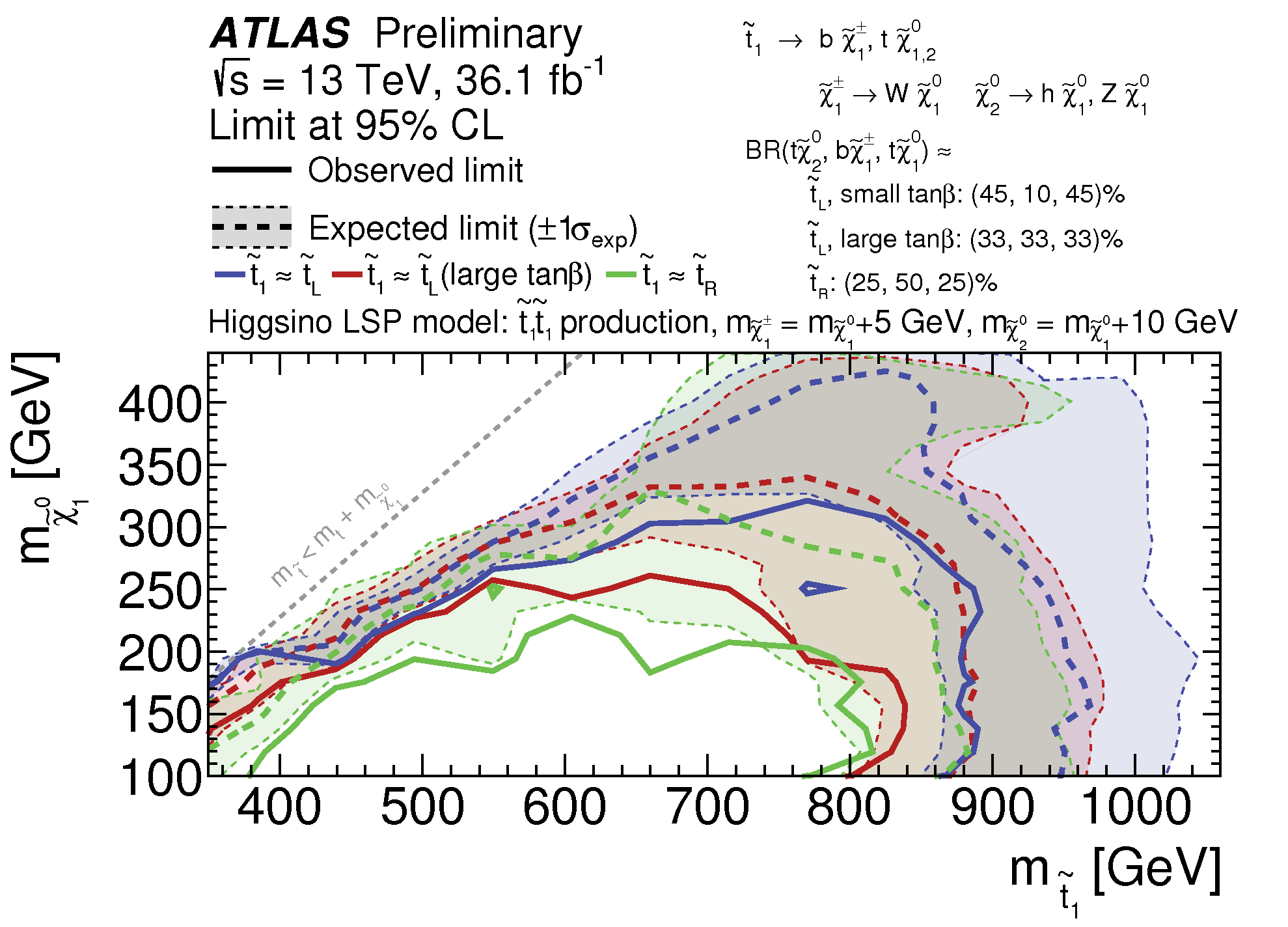}
\includegraphics[width=0.24\textwidth]{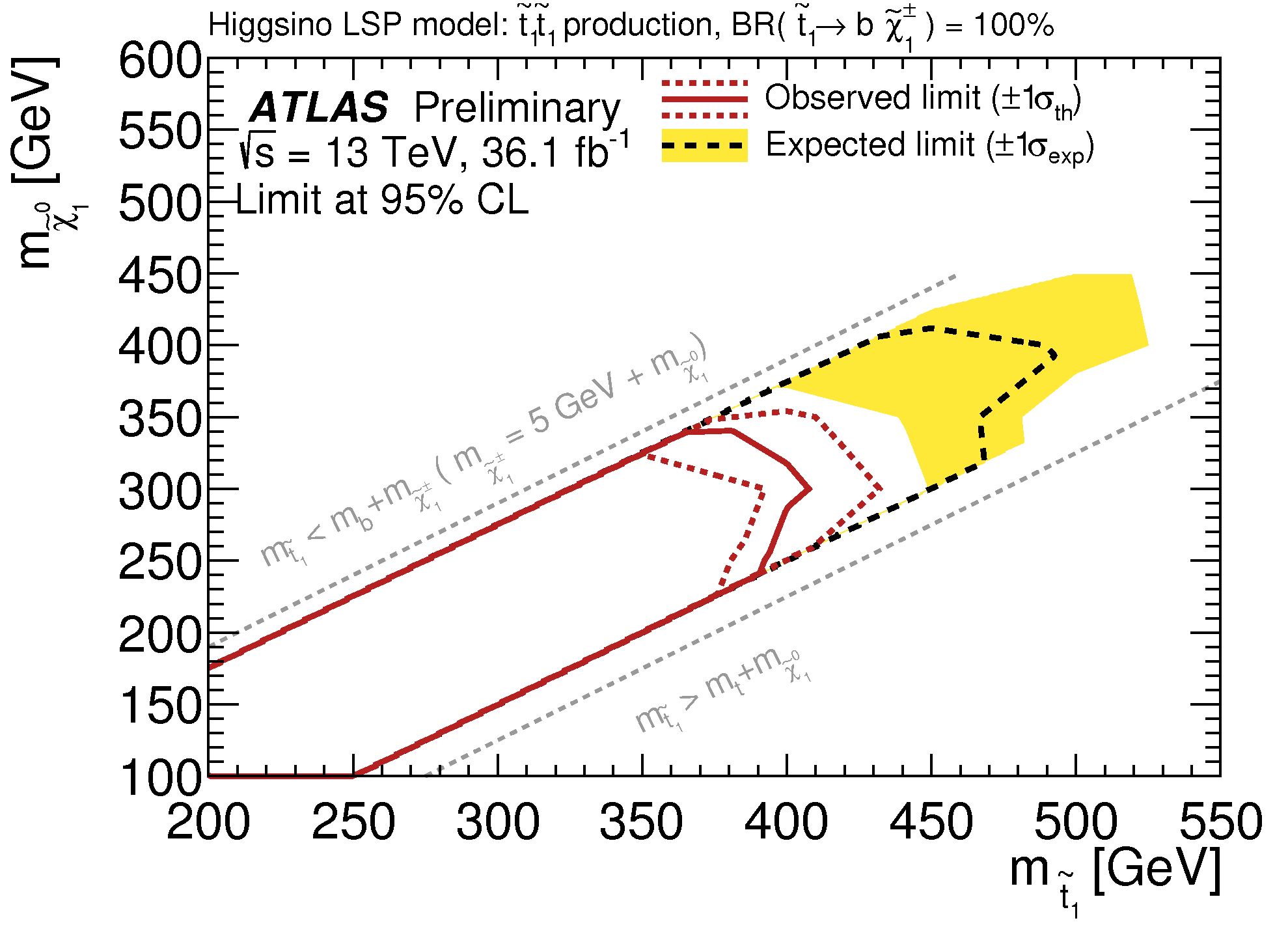}
\includegraphics[width=0.24\textwidth]{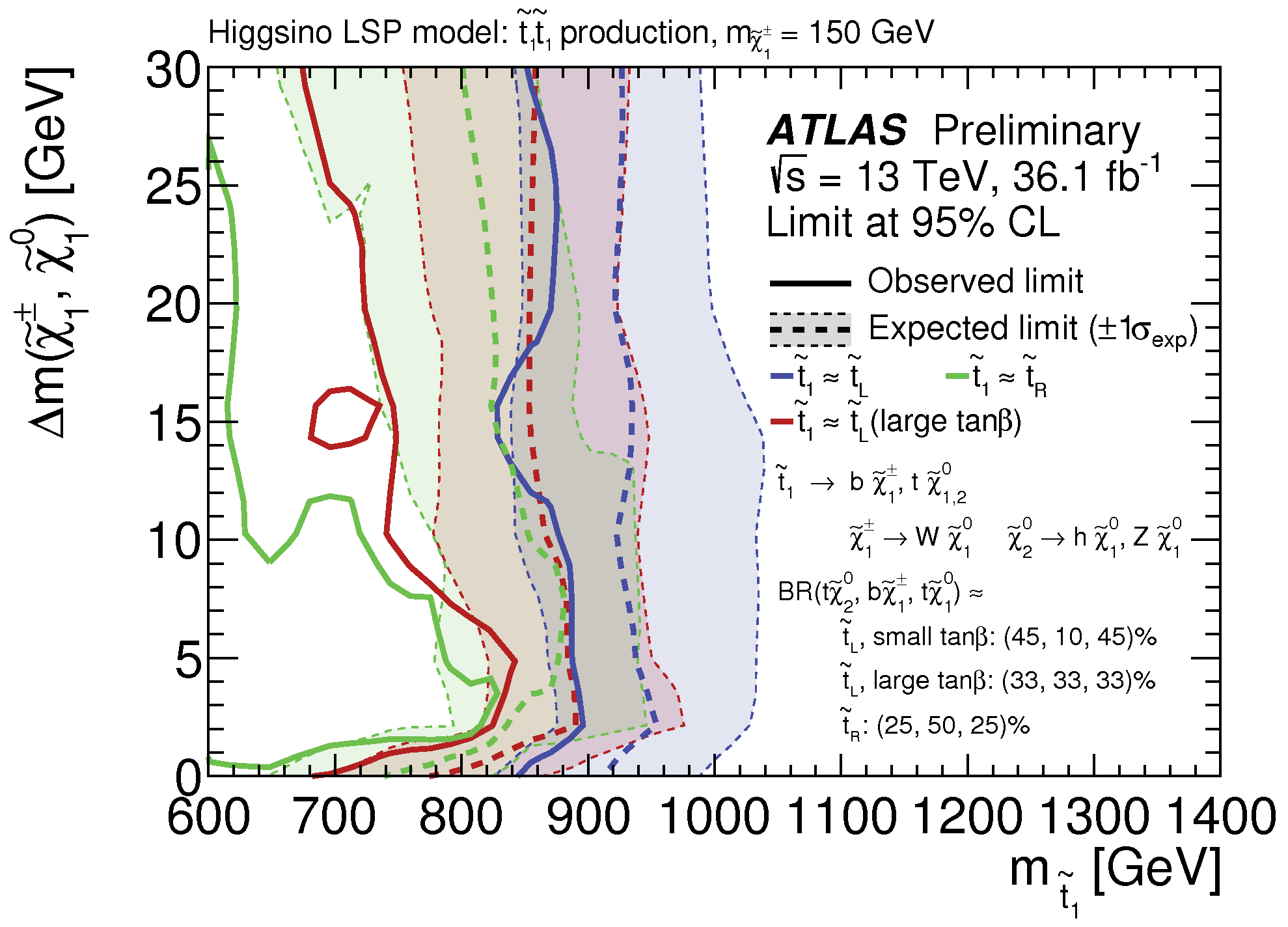}
\includegraphics[width=0.24\textwidth]{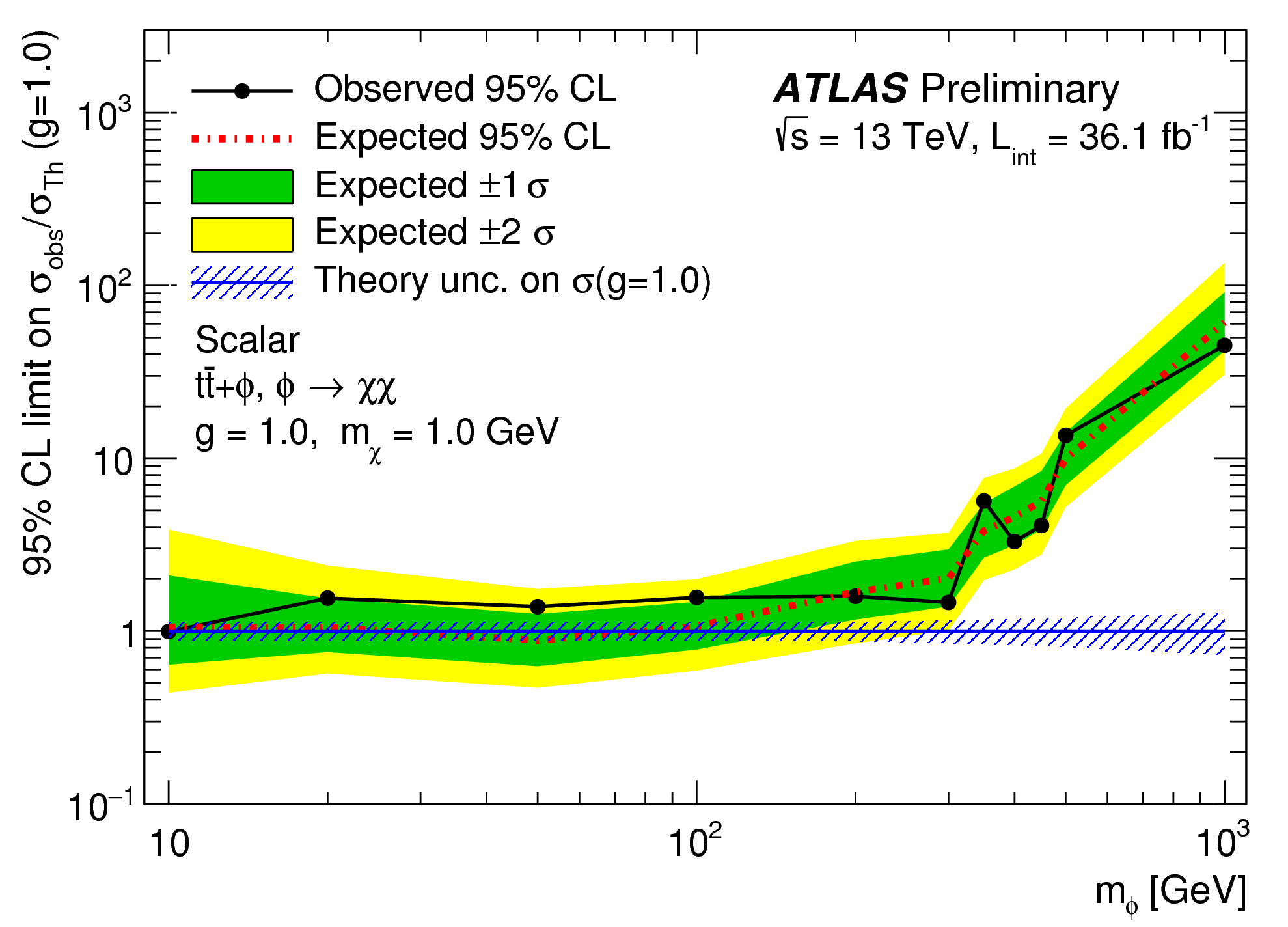}
\includegraphics[width=0.24\textwidth]{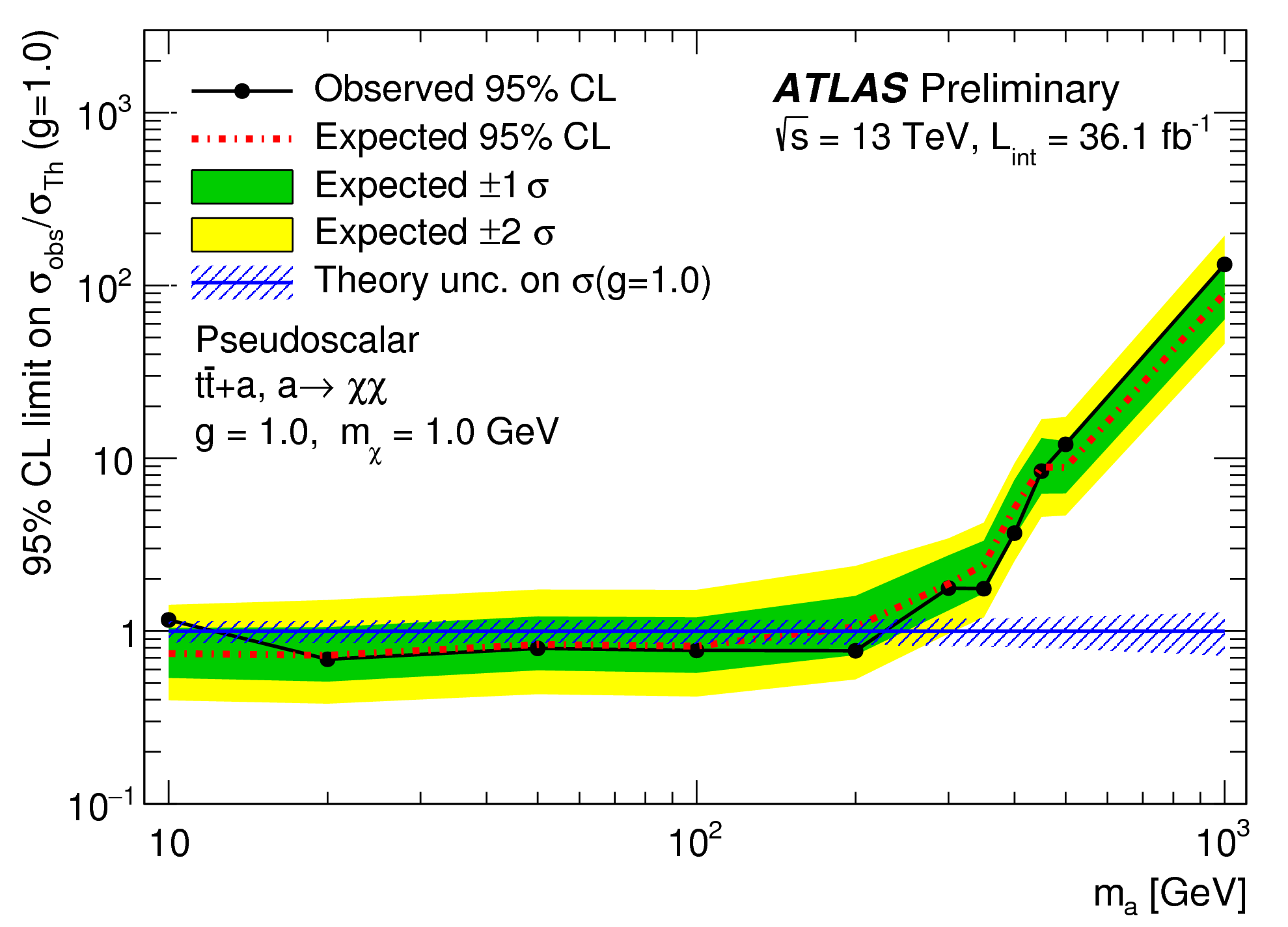}
\includegraphics[width=0.24\textwidth]{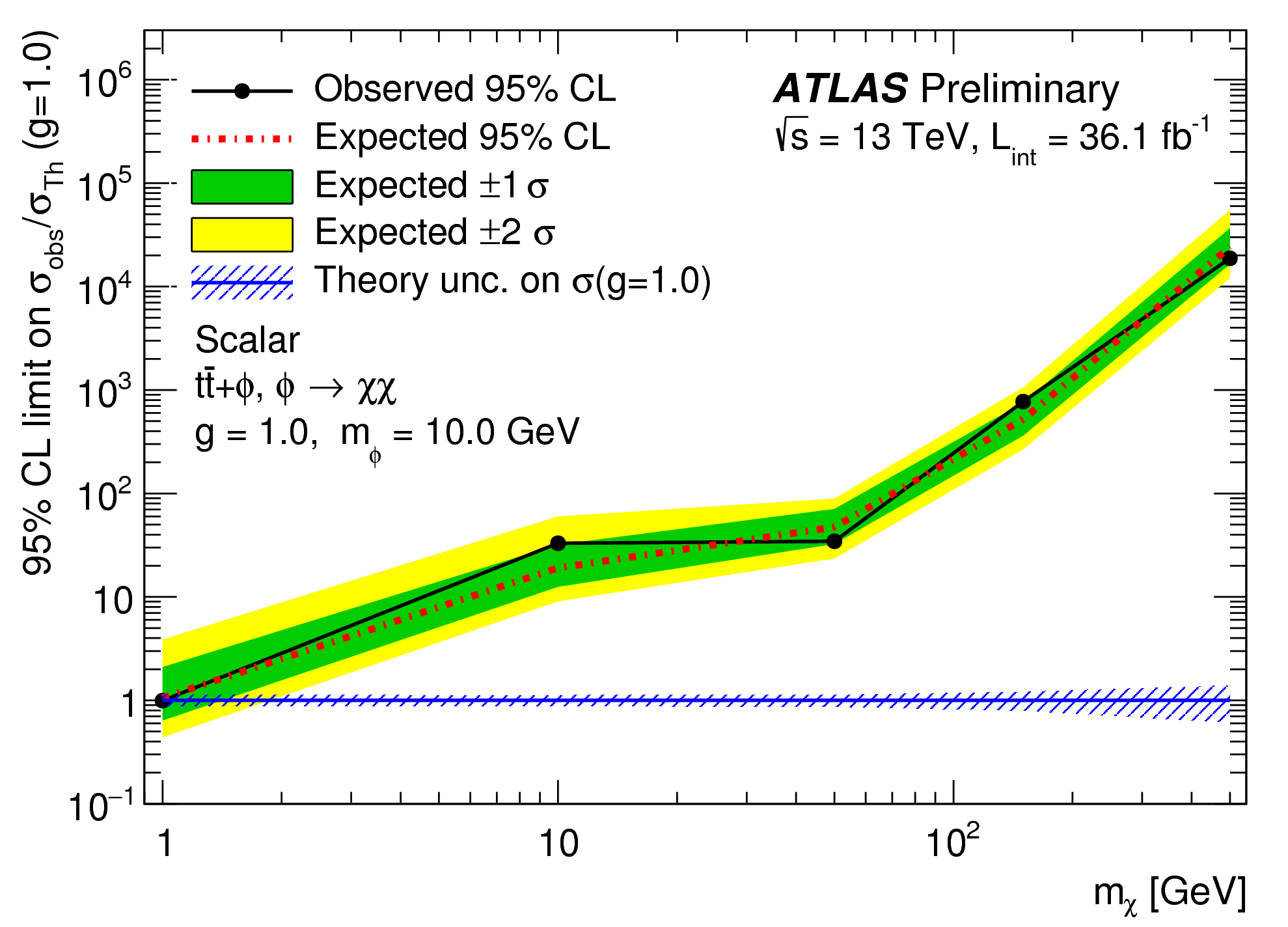}
\includegraphics[width=0.24\textwidth]{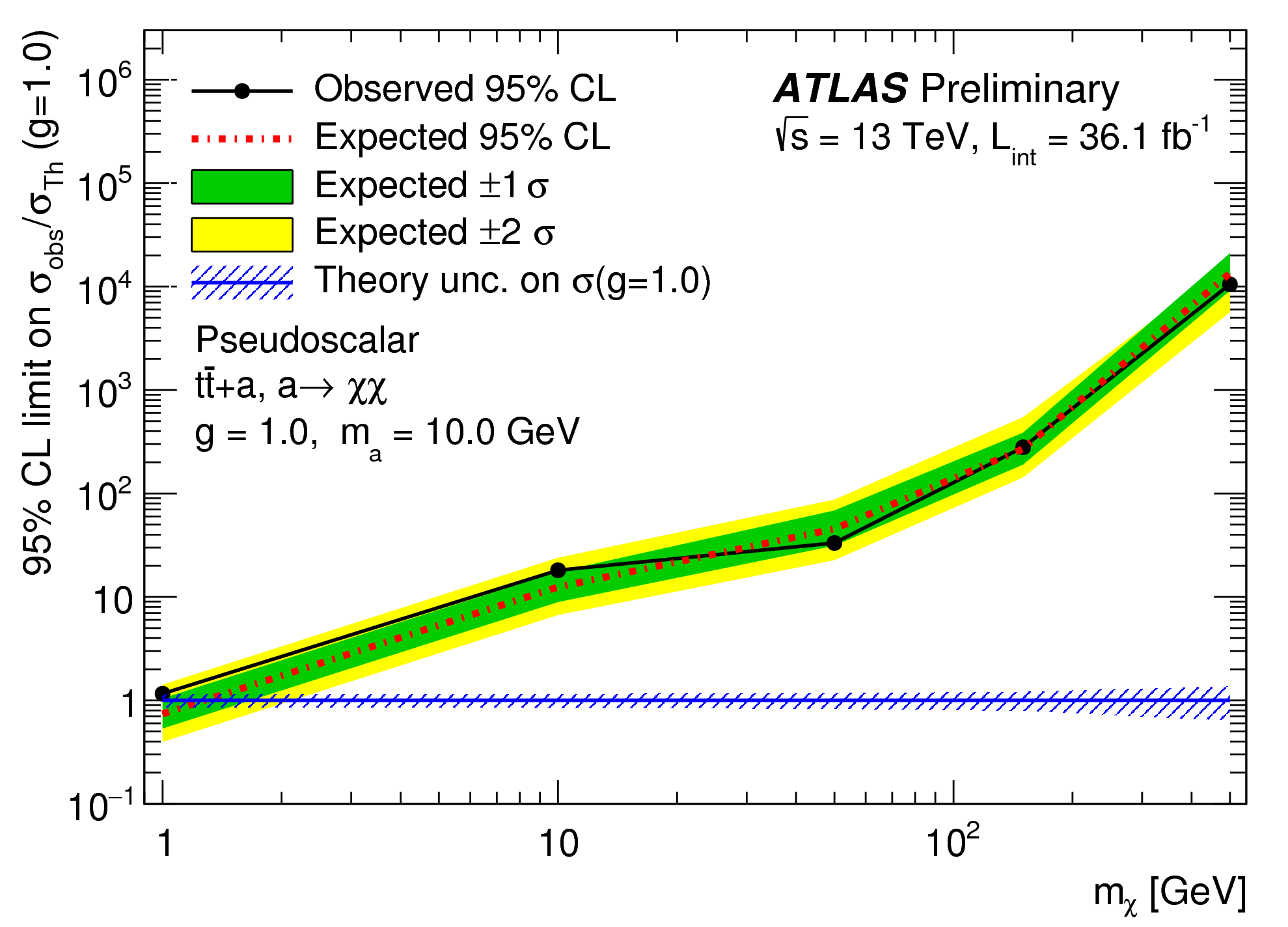}
\caption{Scalar top decay topologies for the final states with one isolated lepton, jets 
         and missing transverse momentum and exclusion contours.
         }
\label{fig:stop3}
\vspace*{-5mm}
\end{figure}

Searches for direct stop production were performed.
Feynman diagrams and results are summarized 
in Fig.~\ref{fig:stop}~\cite{Aaboud:2017nfd}
for final states with two leptons,
in Fig.~\ref{fig:stop2}~\cite{Aaboud:2017ayj}
for final states with jets plus missing transverse momentum,
and in Fig.~\ref{fig:stop3}~\cite{ATLAS-CONF-2017-037}
for final states with one isolated lepton, jets, and missing transverse momentum.

\section{Direct sbottom production}
\label{sec:bottom} 

\begin{figure}[h]
\centering
\includegraphics[width=0.24\textwidth]{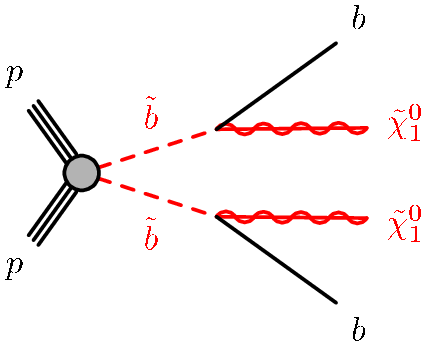}
\includegraphics[width=0.24\textwidth]{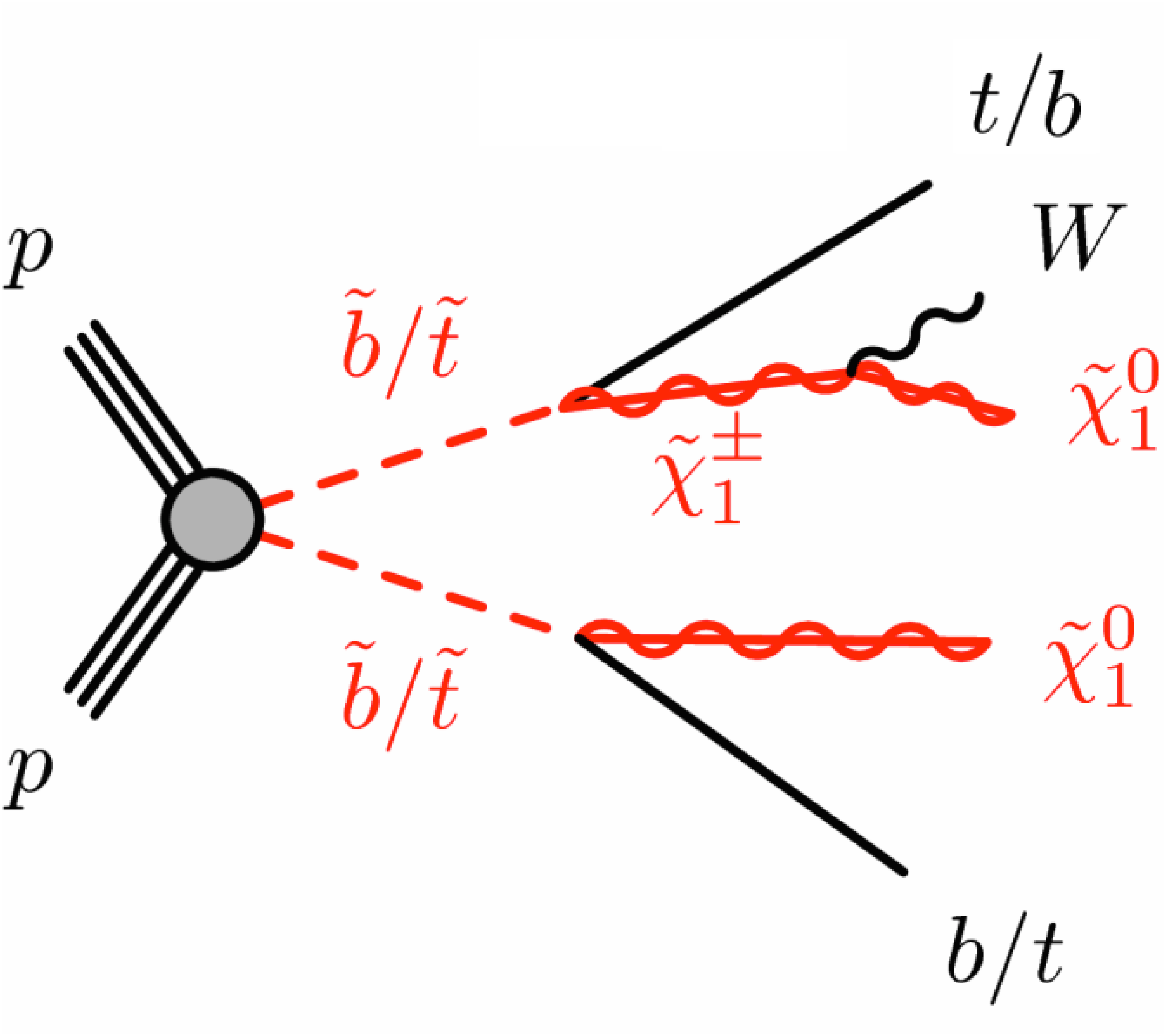}
\includegraphics[width=0.24\textwidth]{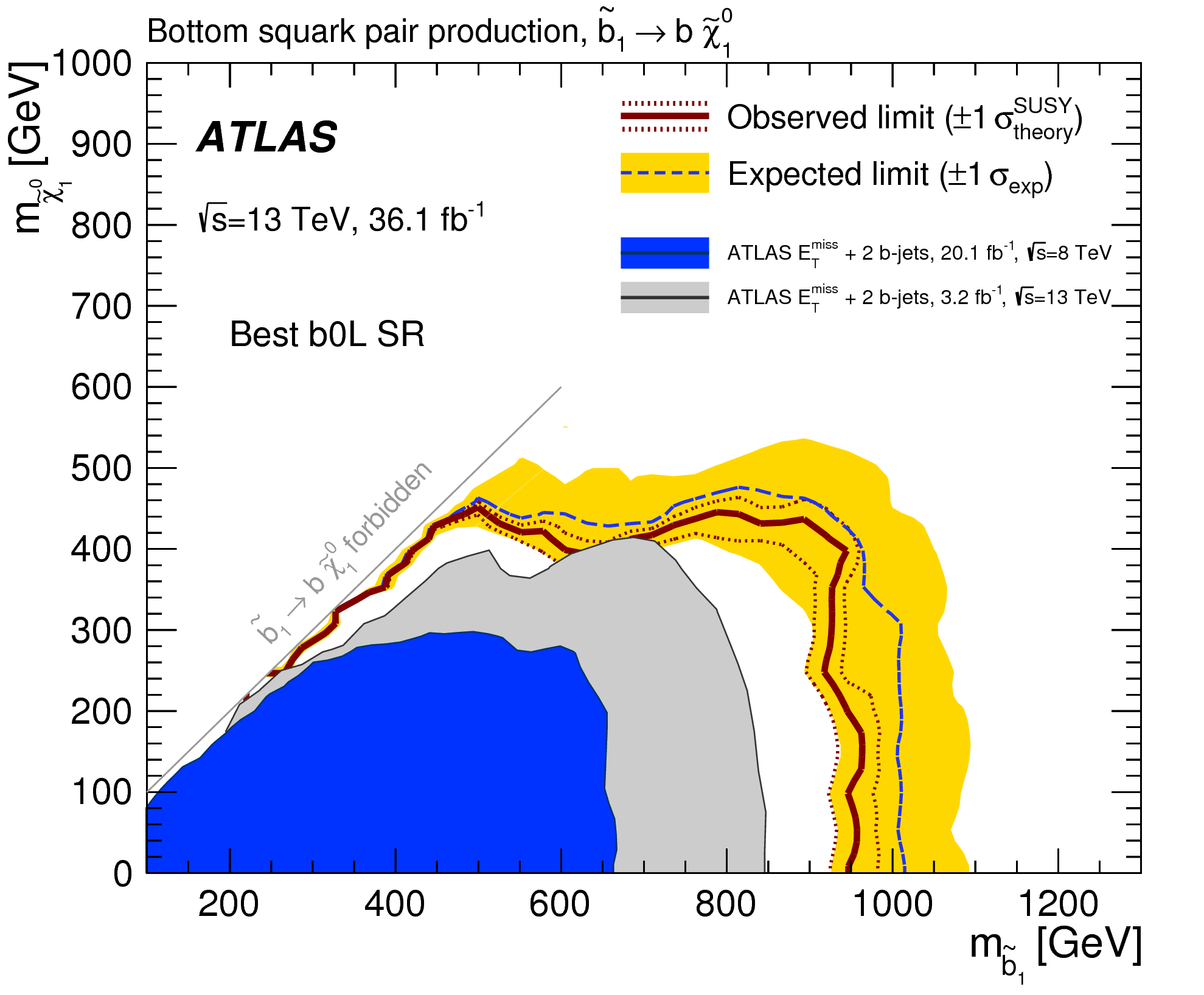}
\includegraphics[width=0.24\textwidth]{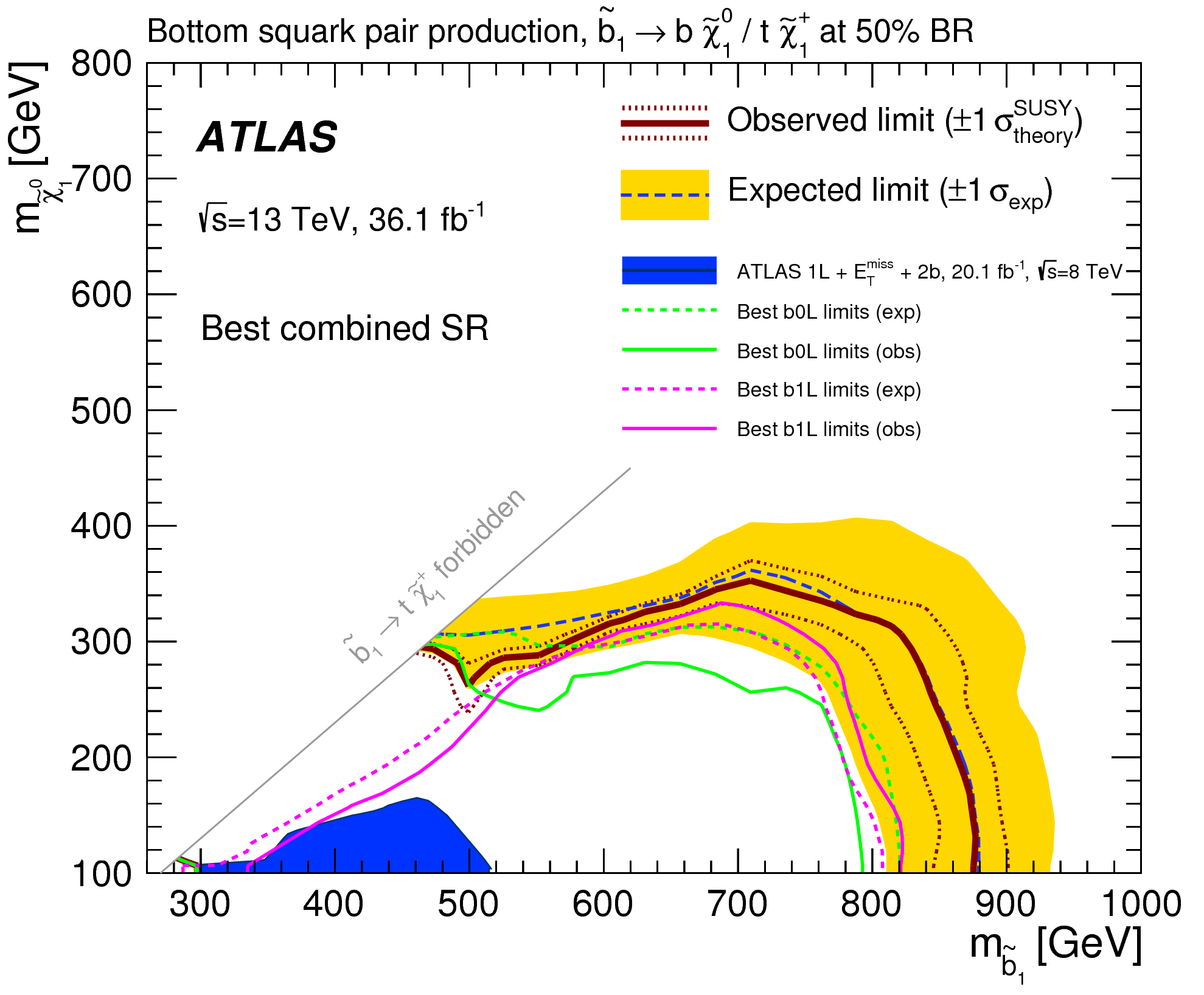}
\caption{Scalar bottom decay topologies and exclusion contours.
         }
\label{fig:sbottom}
\vspace*{-5mm}
\end{figure}

Searches for SUSY direct sbottom production were performed.
Feynman diagrams and results are summarized in Fig.~\ref{fig:sbottom}~\cite{Aaboud:2017wqg}.

\vspace*{-3mm}
\section{Small mass differences}
\label{sec:small}

\begin{figure}[h]
\centering
\includegraphics[width=0.24\textwidth]{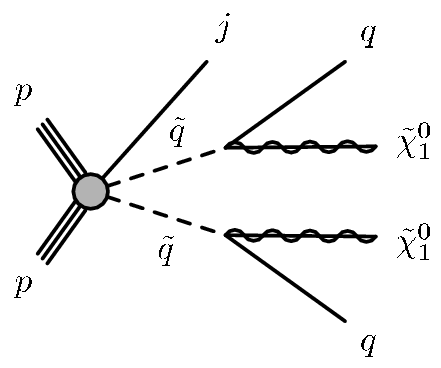}
\includegraphics[width=0.3\textwidth]{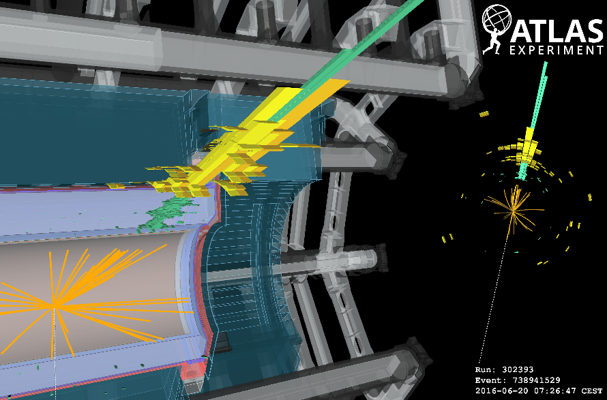}\\
\includegraphics[width=0.24\textwidth]{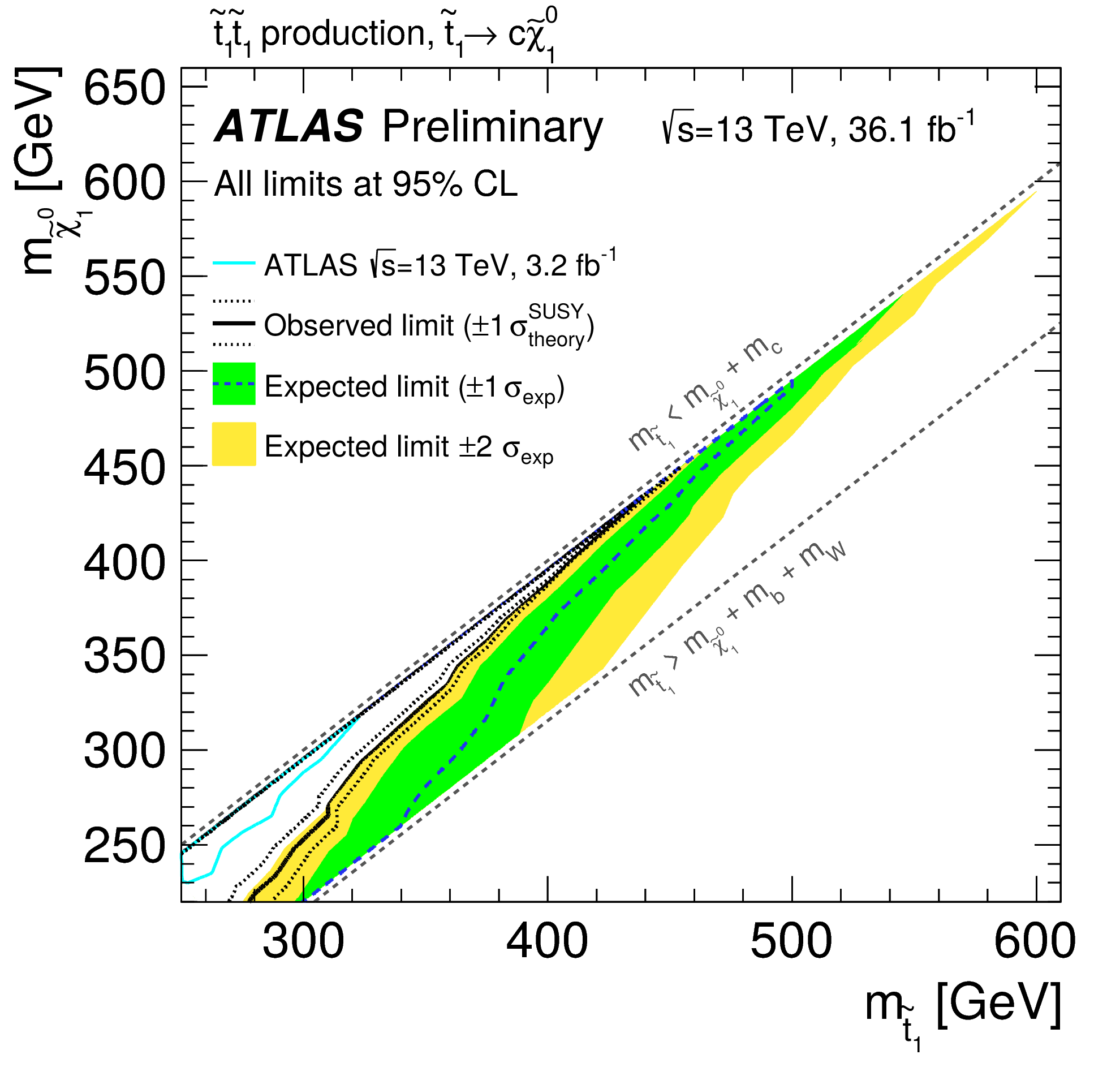}
\includegraphics[width=0.24\textwidth]{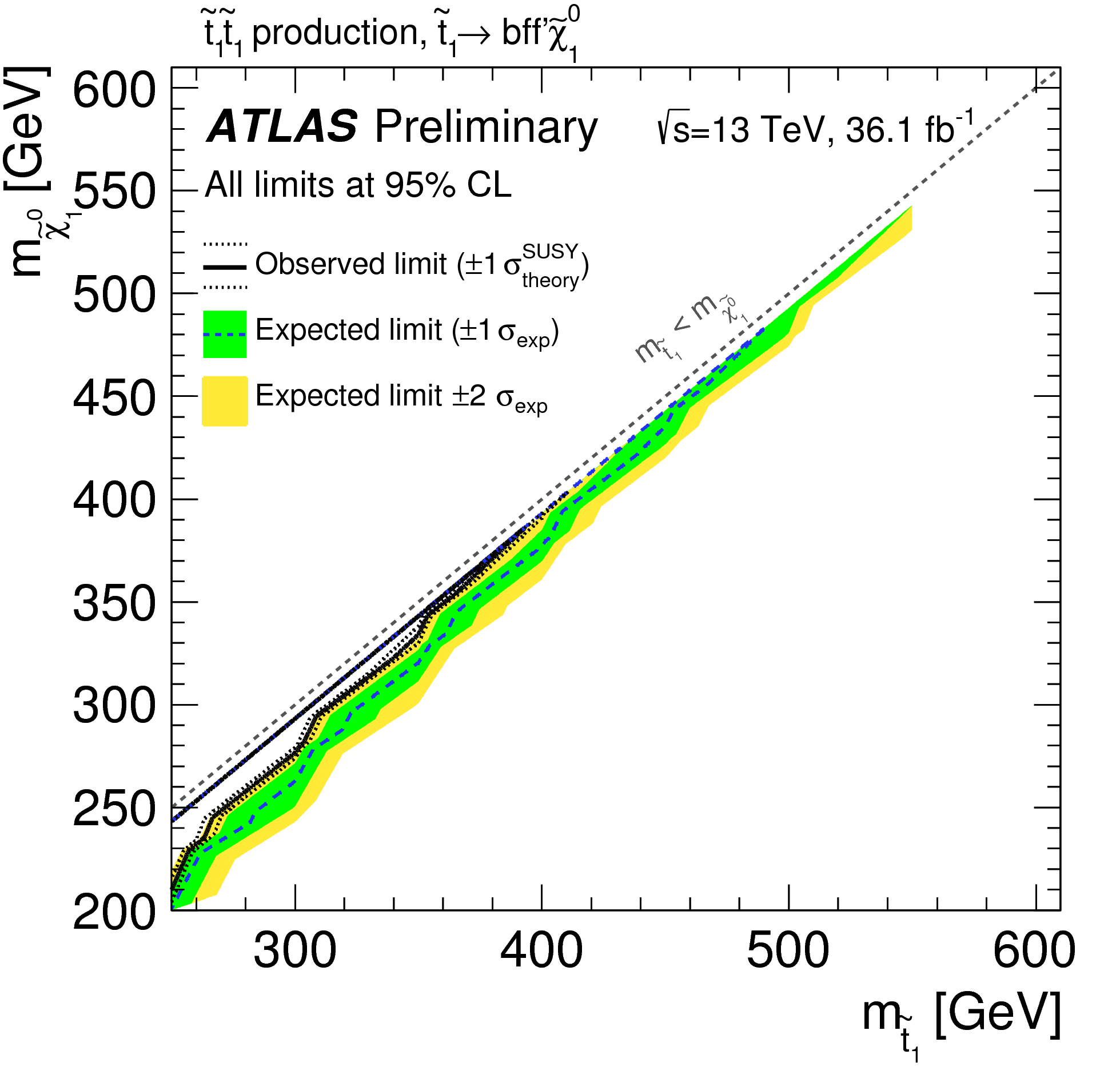}
\includegraphics[width=0.24\textwidth]{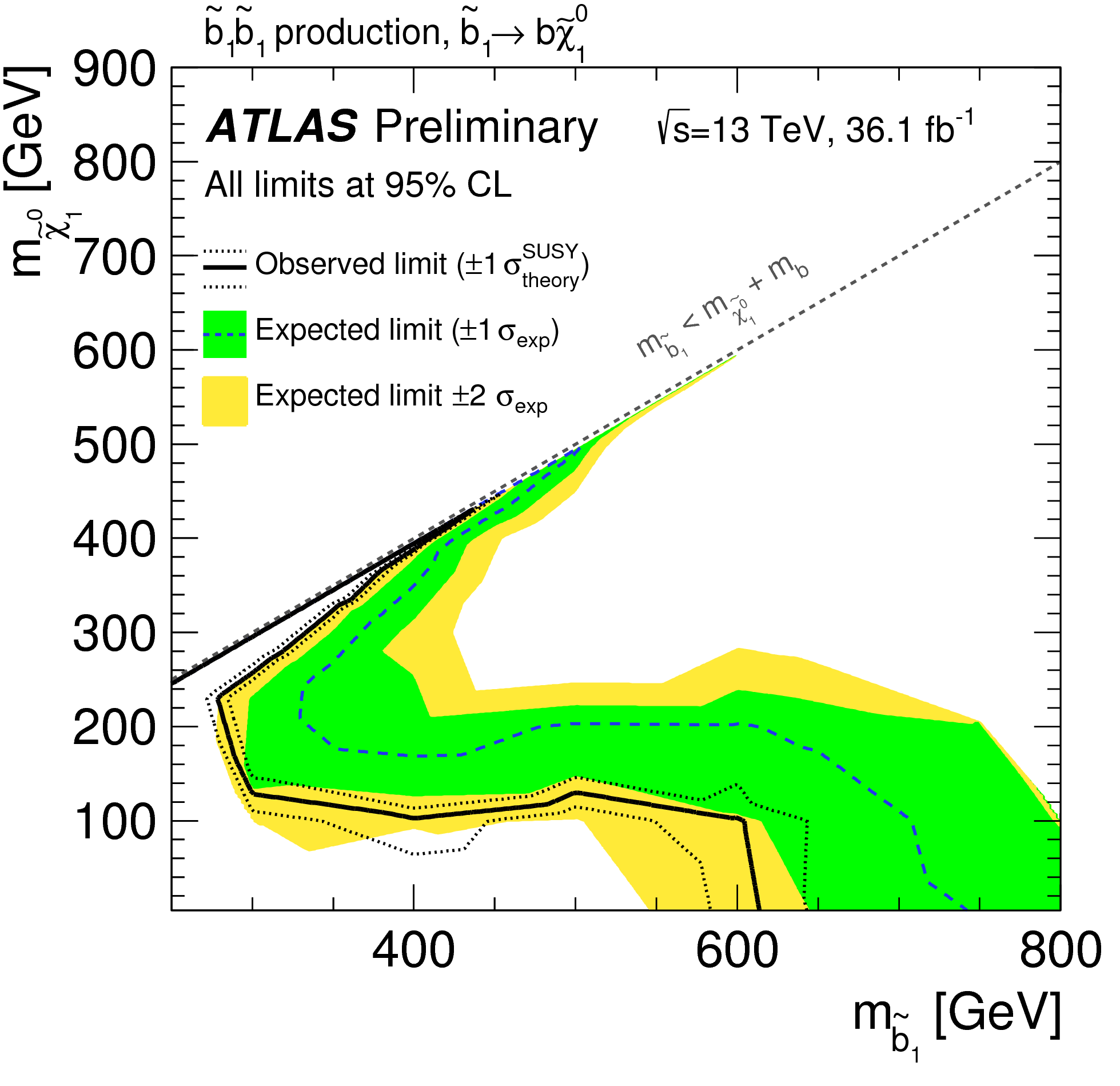}
\includegraphics[width=0.24\textwidth]{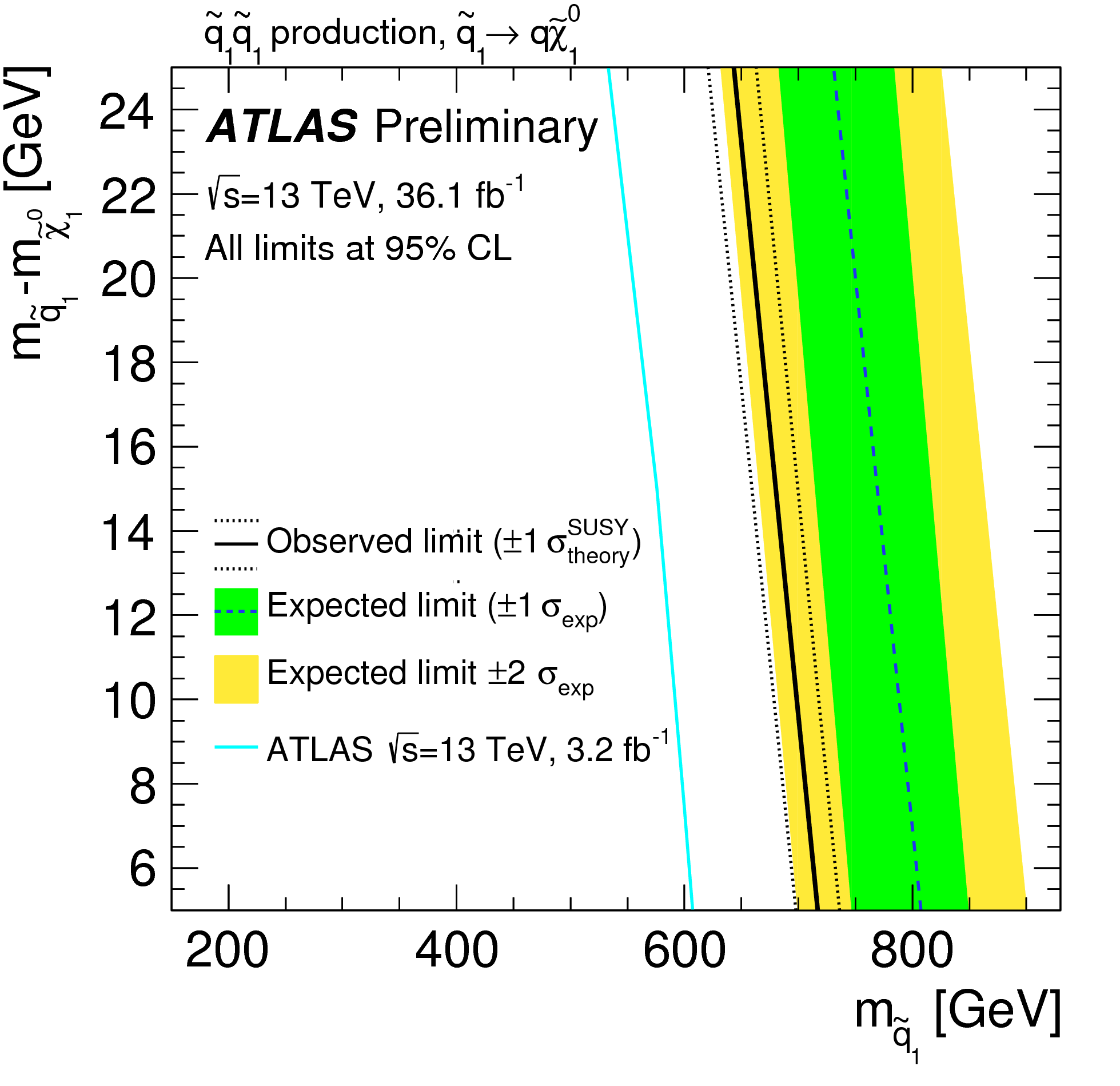}
\caption{Small mass difference of squarks and neutralinos decay topologies, 
         an event display of a typical monojet with missing energy candidate
         and exclusion contours.
         }
\label{fig:small}
\vspace*{-10mm}
\end{figure}

Searches for SUSY with a small mass difference of squarks and neutralinos were performed.
A Feynman diagram and results are summarized in Fig.~\ref{fig:small}~\cite{ATLAS-CONF-2017-060}.

\vspace*{-3mm}
\section{Electroweak direct production}
\label{sec:ew}

\begin{figure}[h]
\centering
\includegraphics[width=0.24\textwidth]{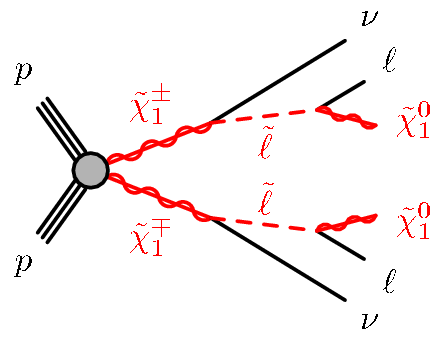}
\includegraphics[width=0.24\textwidth]{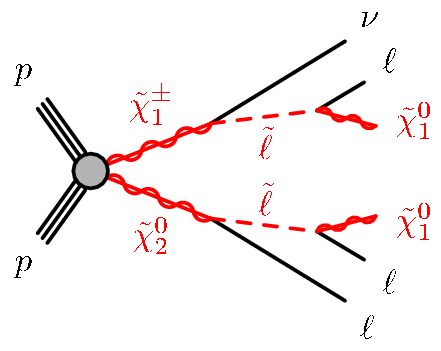}
\includegraphics[width=0.24\textwidth]{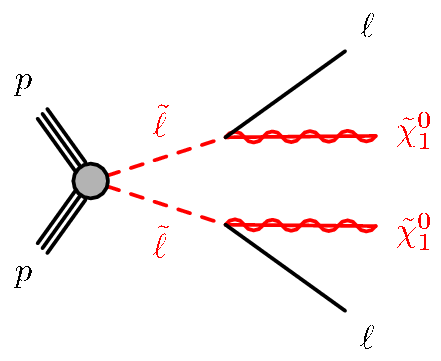}
\includegraphics[width=0.24\textwidth]{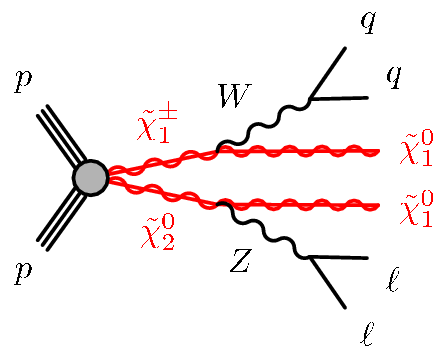}
\includegraphics[width=0.24\textwidth]{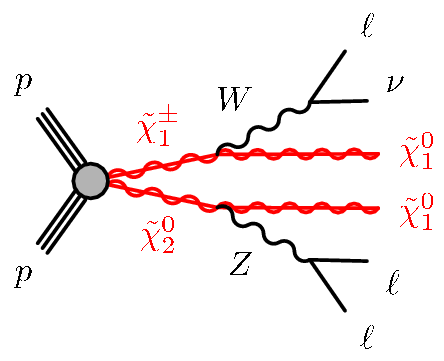}
\includegraphics[width=0.24\textwidth]{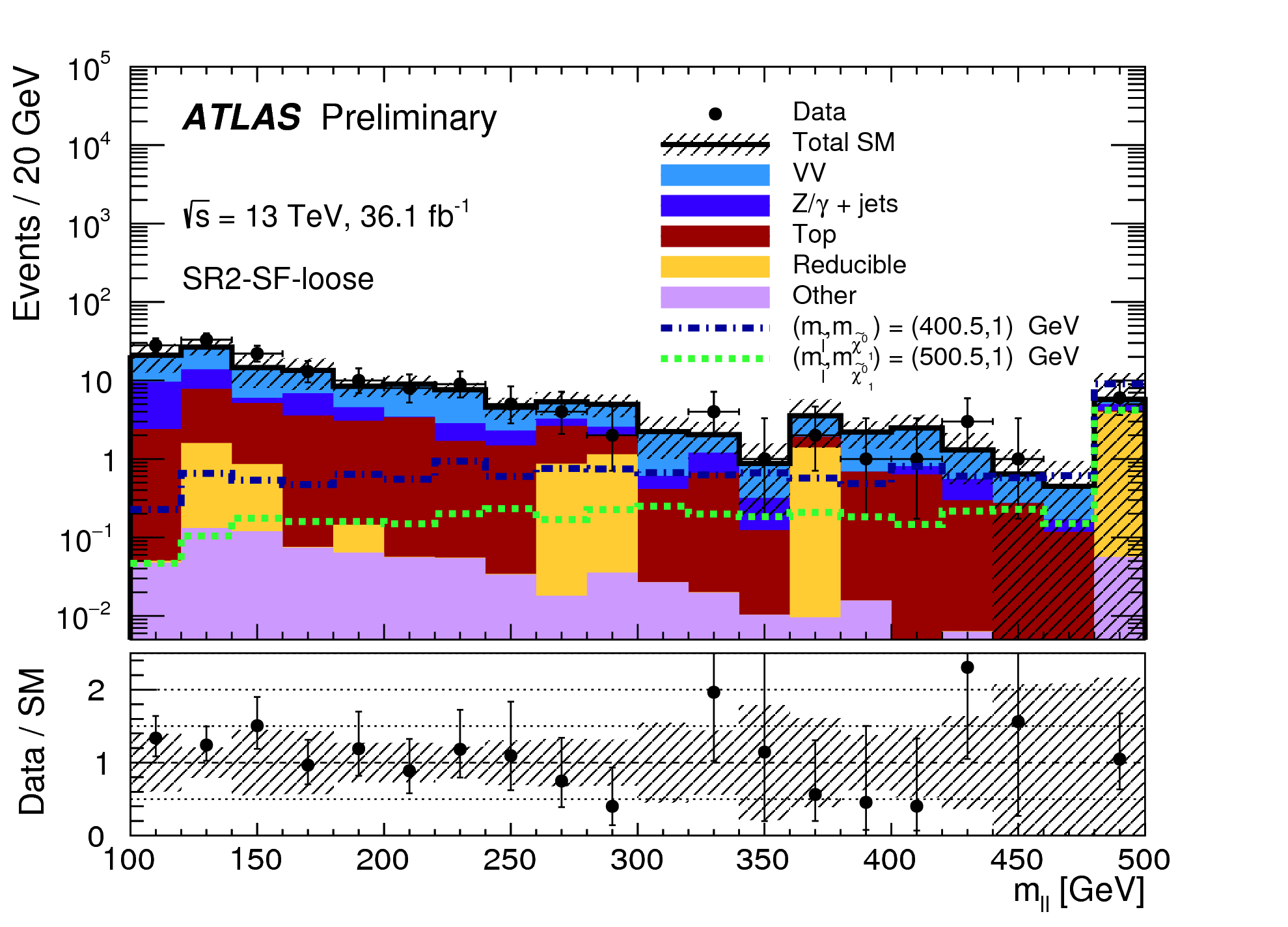}
\includegraphics[width=0.23\textwidth]{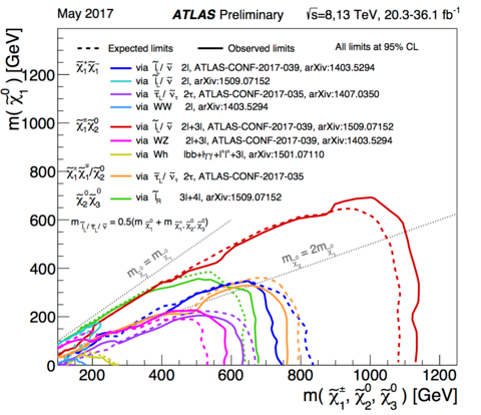}
\includegraphics[width=0.26\textwidth]{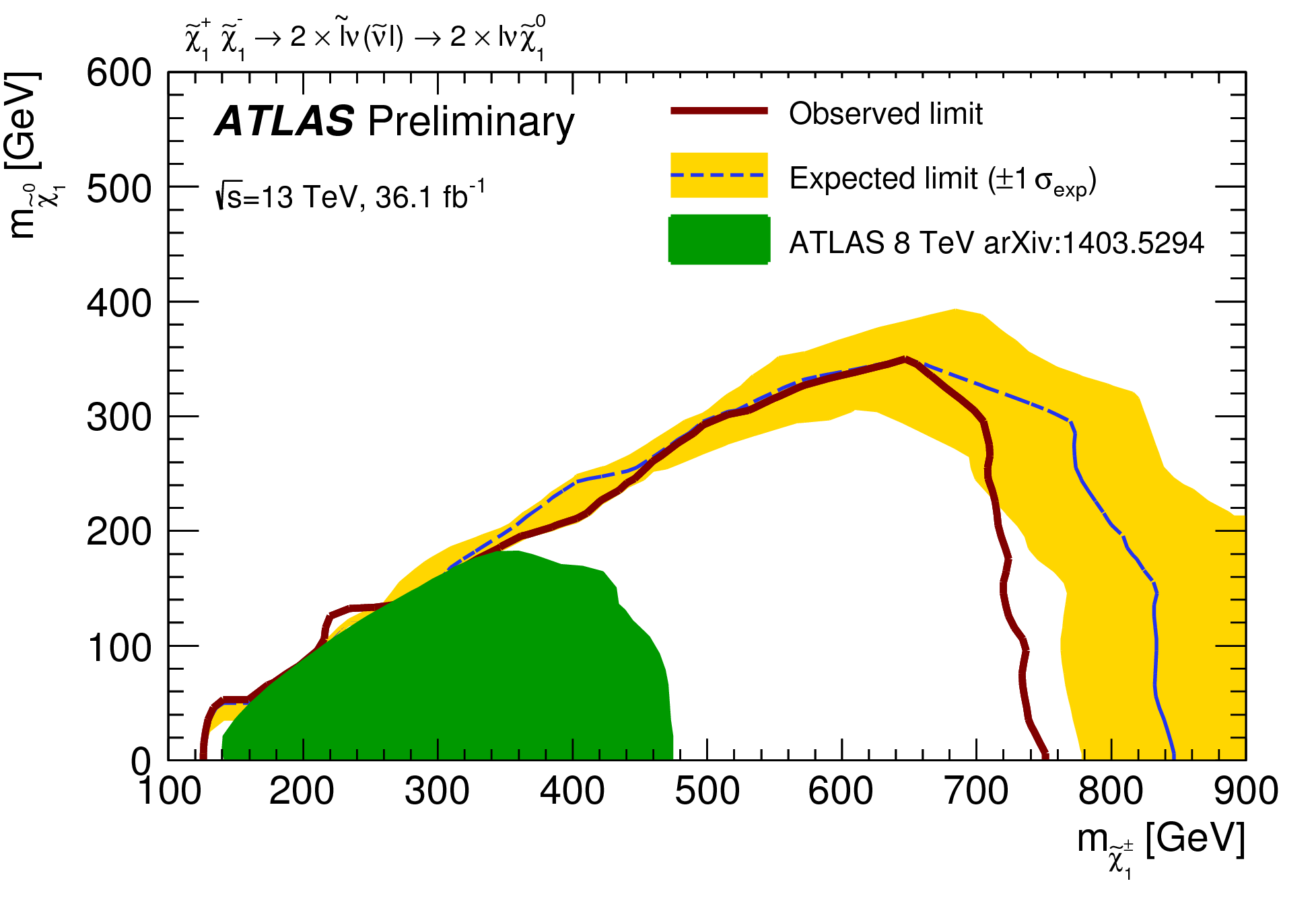}
\includegraphics[width=0.23\textwidth]{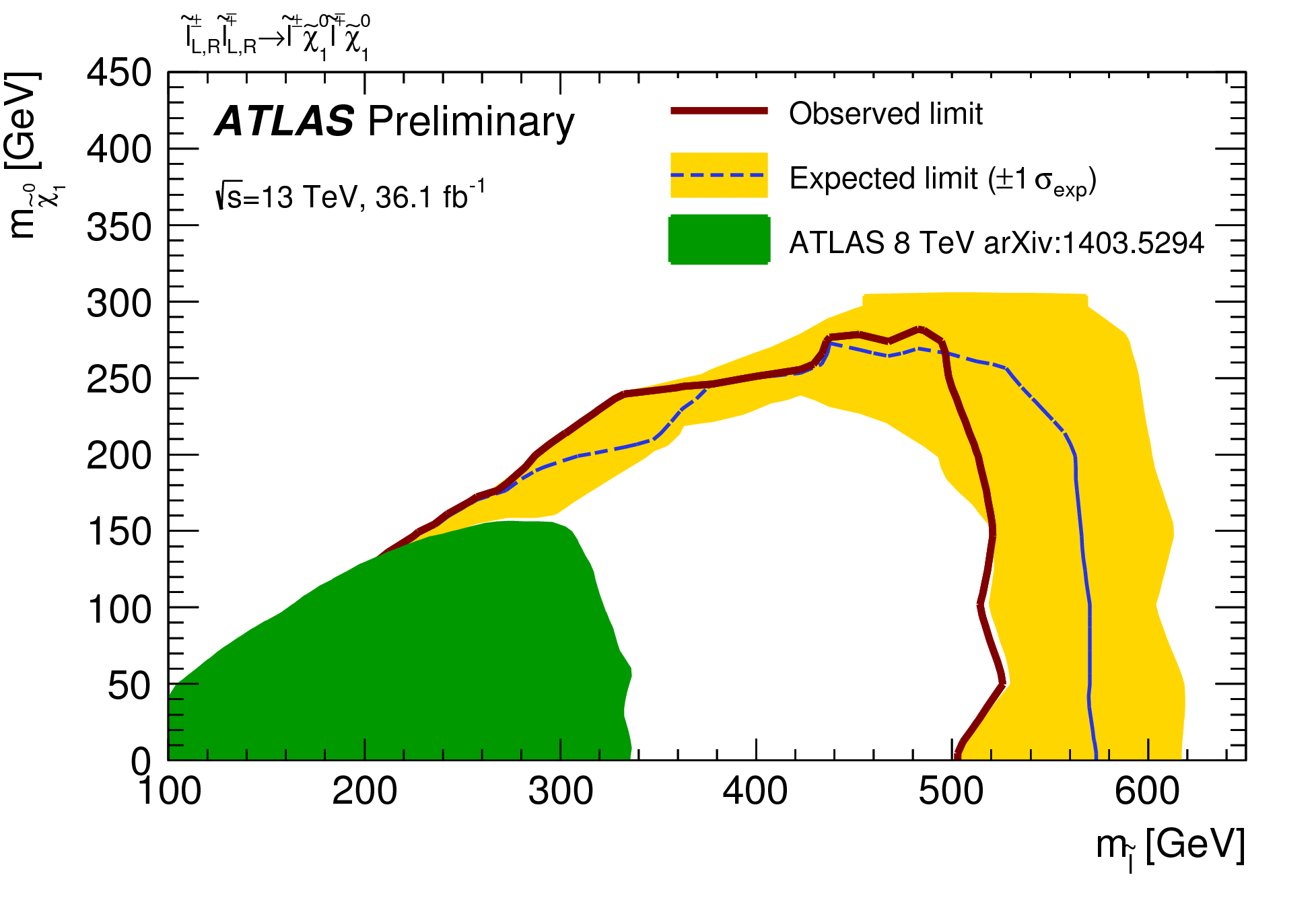}
\includegraphics[width=0.24\textwidth]{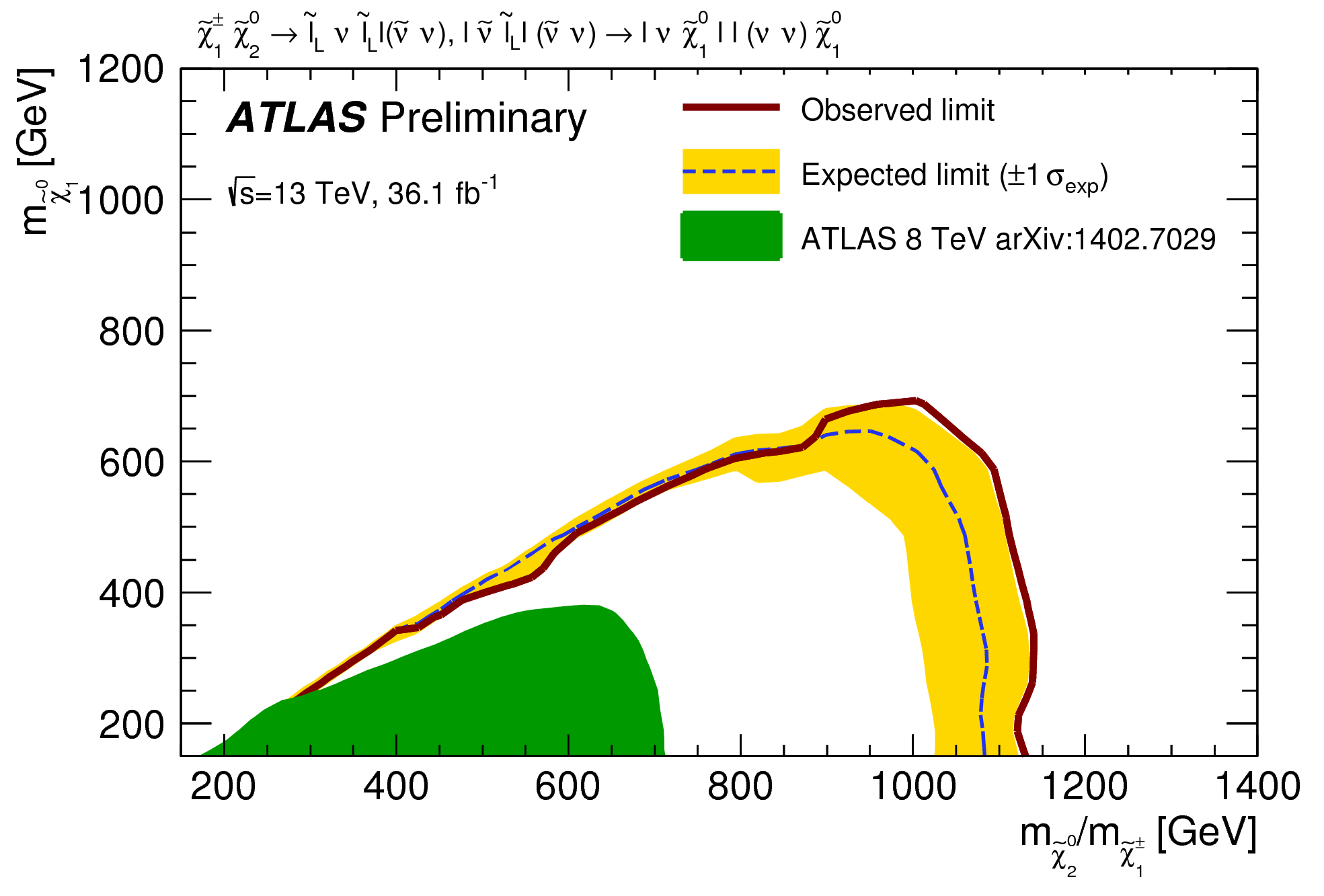}
\includegraphics[width=0.24\textwidth]{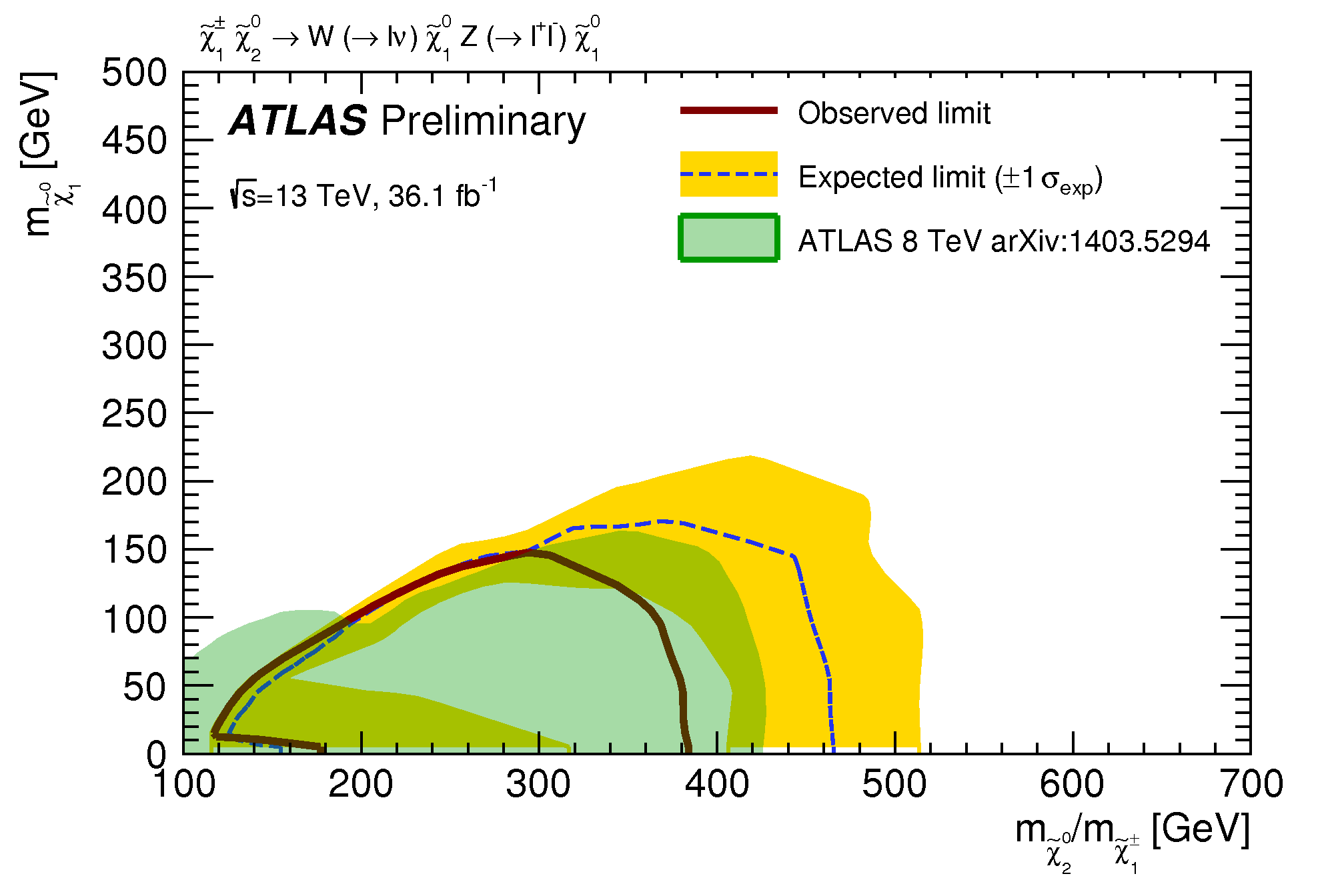}
\includegraphics[width=0.24\textwidth]{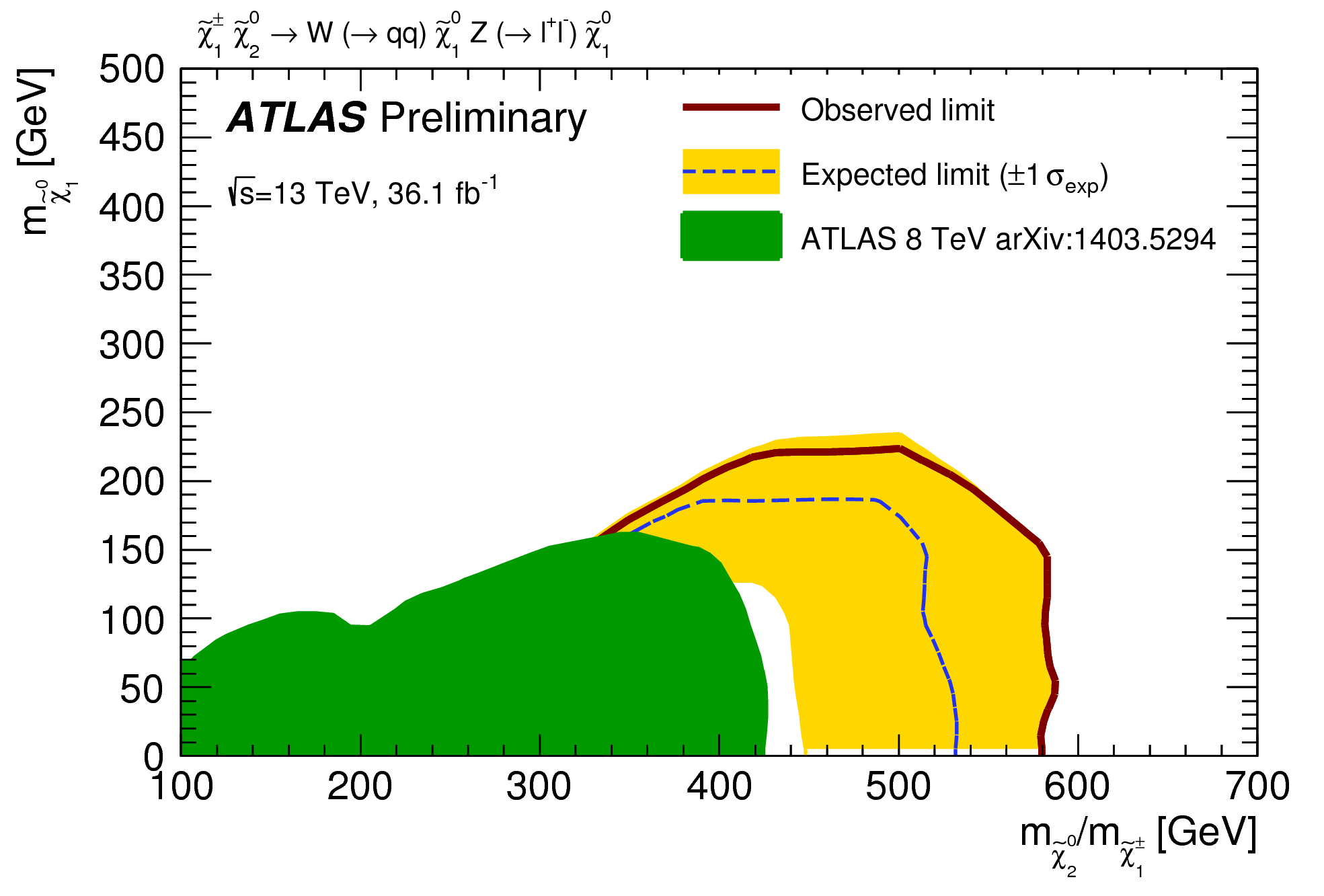}
\caption{Electroweak direct production with 2-leptons+0-jets, 2-leptons+jets, 3-leptons decay topologies, 
         2-lepton invariant mass distribution,         
         overview of inclusive searches
         and exclusion contours.
         }
\label{fig:ew}
\vspace{-10mm}
\end{figure}

Searches for SUSY electroweak direct production with 2-leptons+0-jets, 2-leptons+jets, 
and 3-leptons were performed.
Feynman diagrams, the 2-lepton invariant mass distribution for data, and the estimated SM backgrounds, an overview
and results are summarized in Fig.~\ref{fig:ew}~\cite{ATLAS-CONF-2017-039}.

\vspace*{-3mm}
\section{Photonic signatures}
\label{sec:photpnic} 

\begin{figure}[h]
\centering
\includegraphics[width=0.24\textwidth]{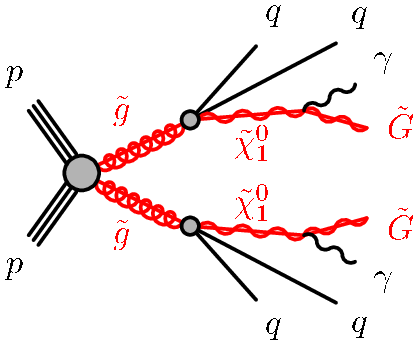}
\includegraphics[width=0.38\textwidth]{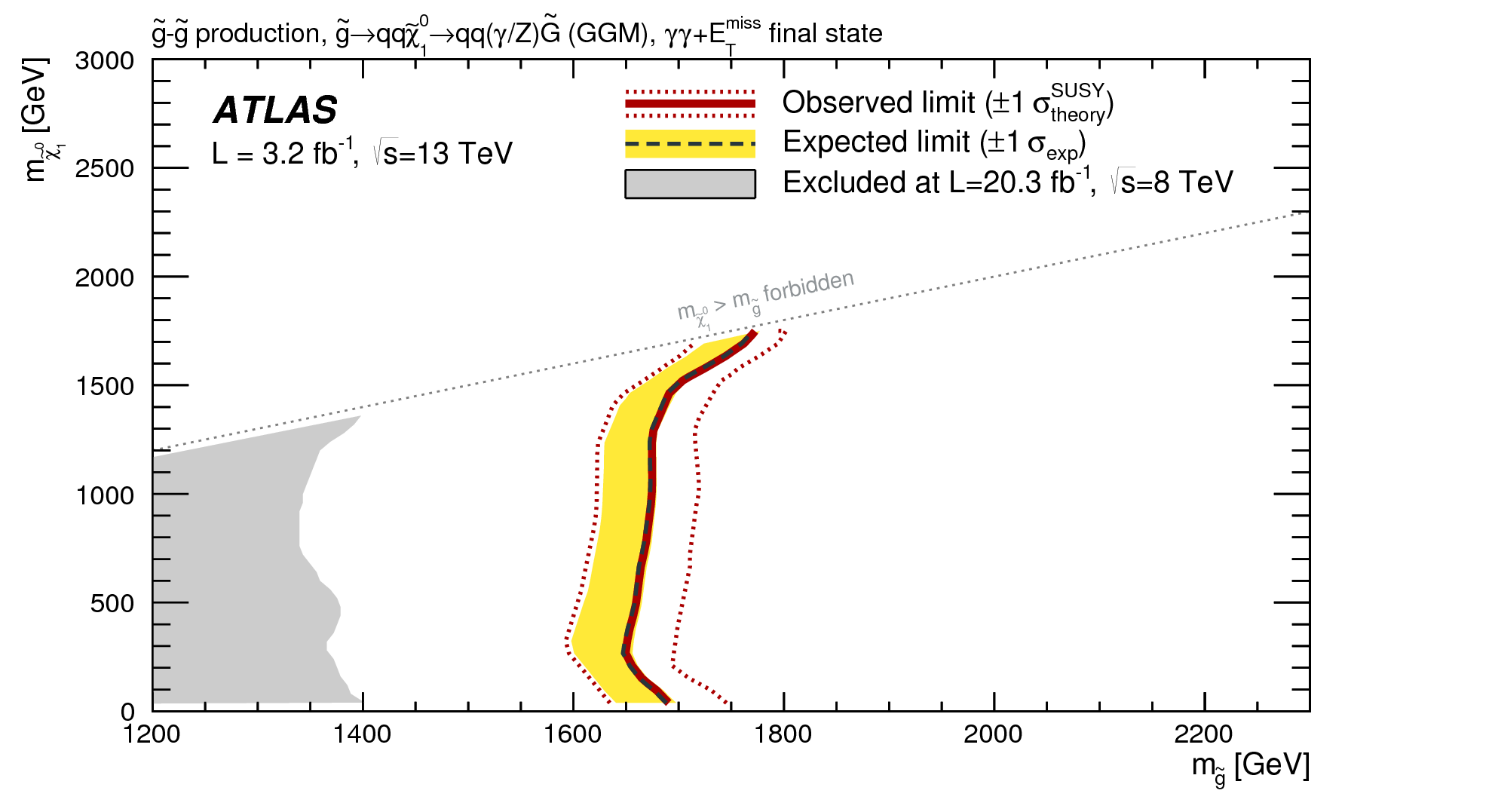}
\includegraphics[width=0.24\textwidth]{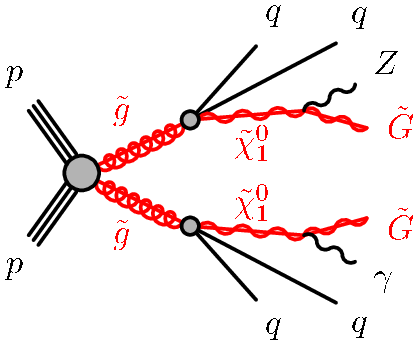}
\includegraphics[width=0.28\textwidth]{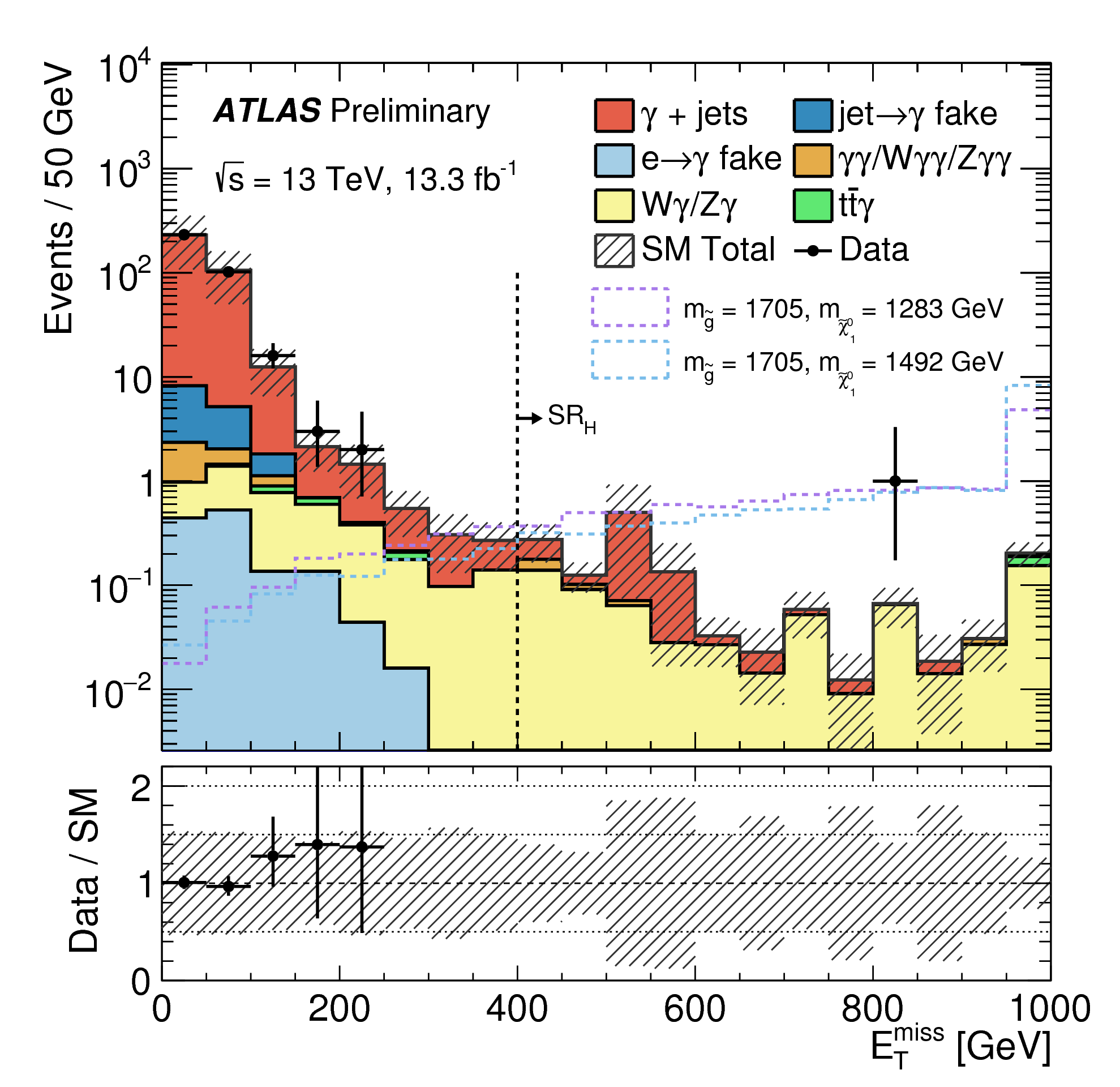}
\includegraphics[width=0.32\textwidth]{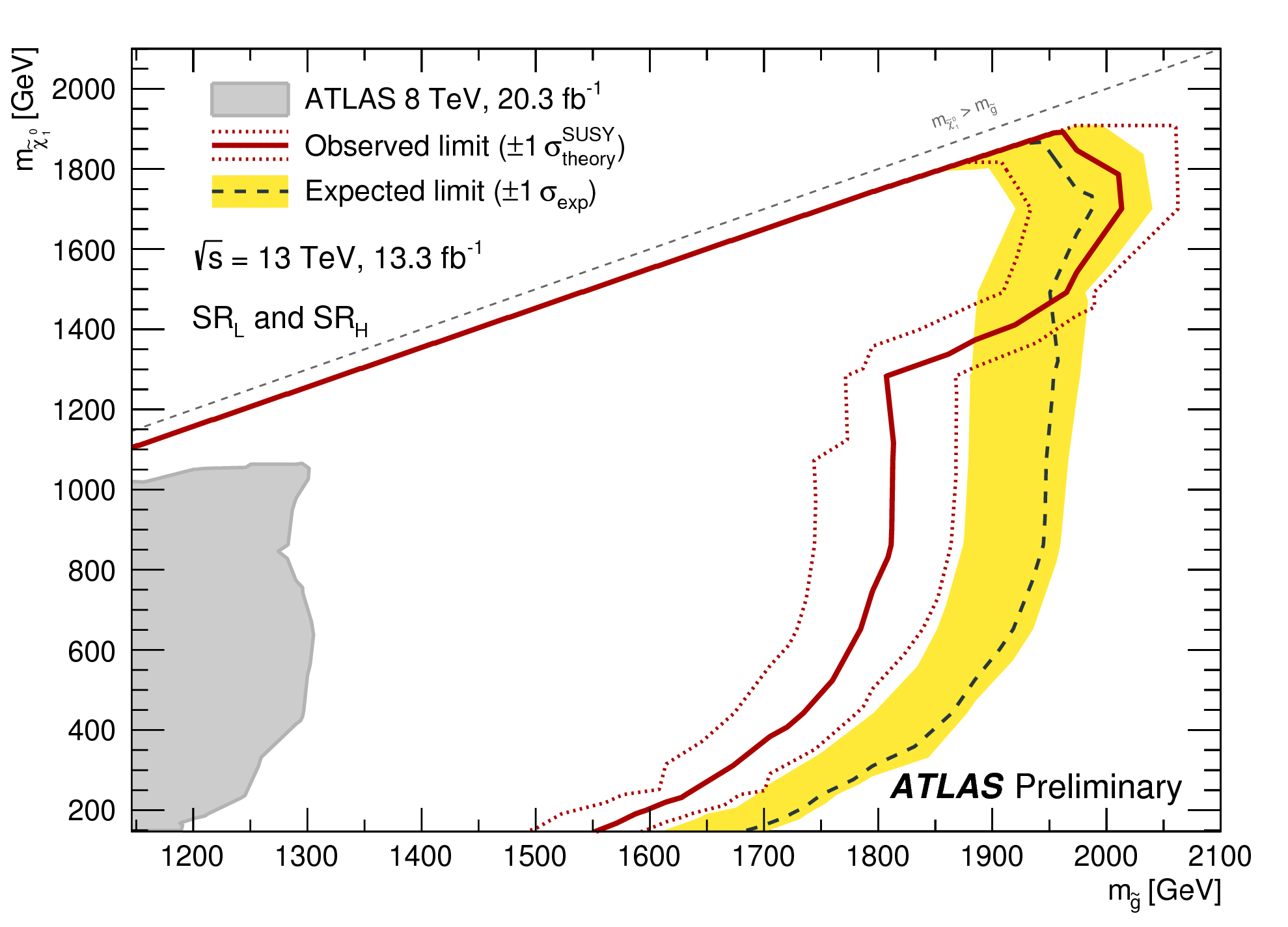}
\includegraphics[width=0.32\textwidth]{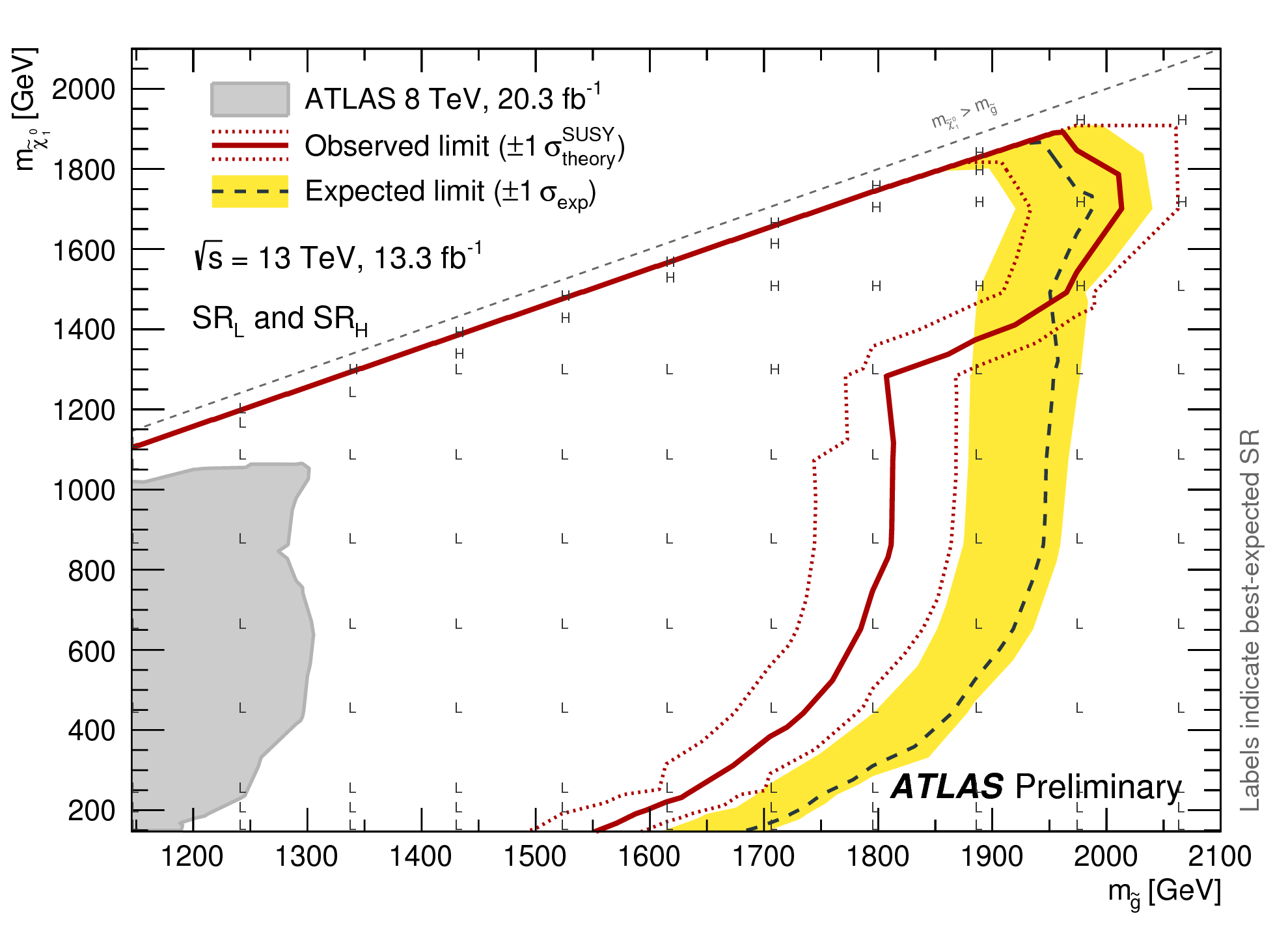}
\caption{Two photons and one photon decay topologies,
         missing transverse energy distribution and exclusion contours.
         }
\label{fig:photon}
\vspace{-10mm}
\end{figure}

Searches for SUSY with di-photon and missing energy were performed.
Feynman diagrams, missing transverse energy distribution,
and results are summarized in Fig.~\ref{fig:photon}
for the final states with two photons~\cite{ATLASCollaboration:2016wl},
and for the final states with one photon~\cite{ATLAS-CONF-2016-066}.

\clearpage

\section{Long-lived particles}
\label{sec:long} 

\begin{figure}[h]
\centering
\includegraphics[width=0.38\textwidth]{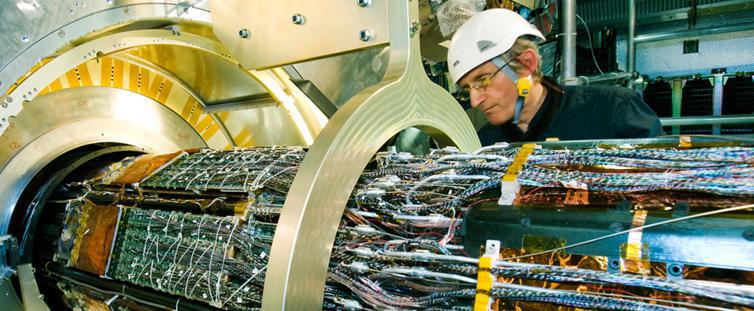}
\includegraphics[width=0.45\textwidth]{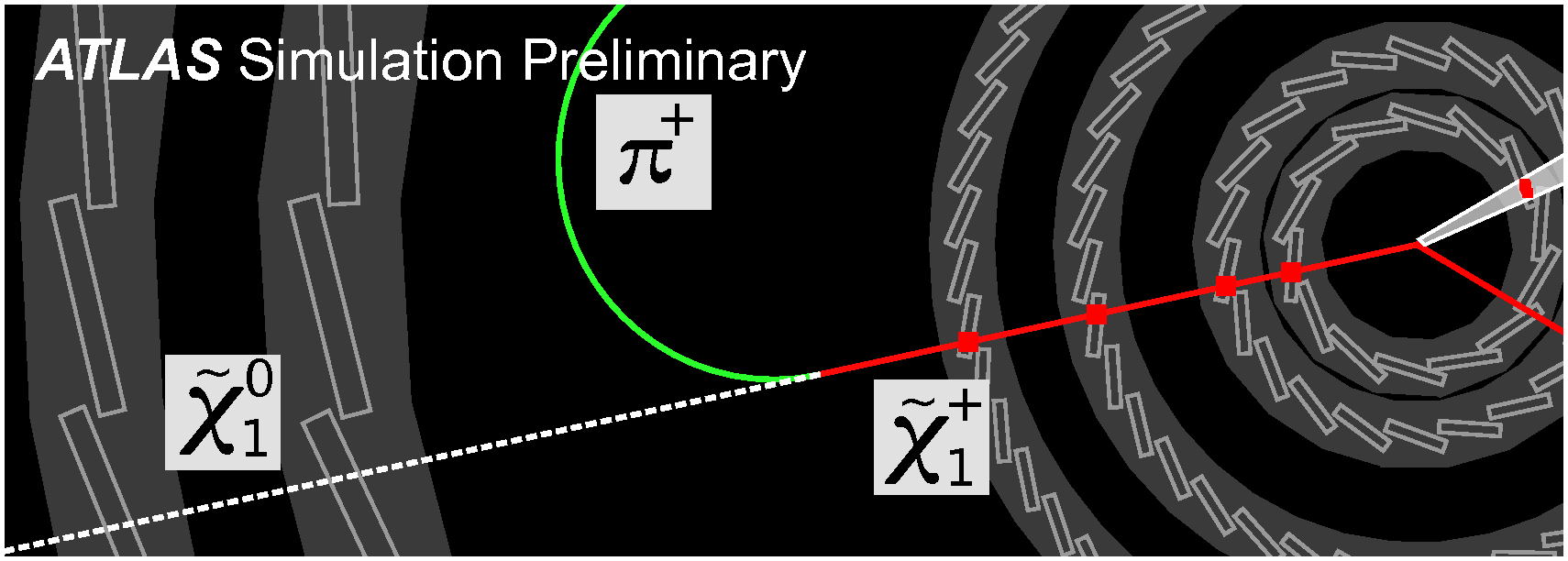}\\
\includegraphics[width=0.24\textwidth]{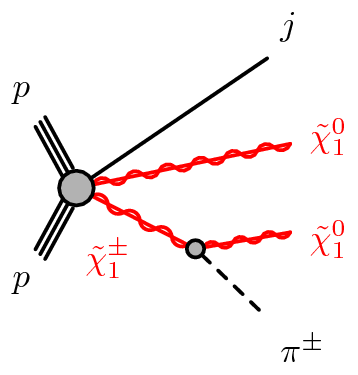}
\includegraphics[width=0.24\textwidth]{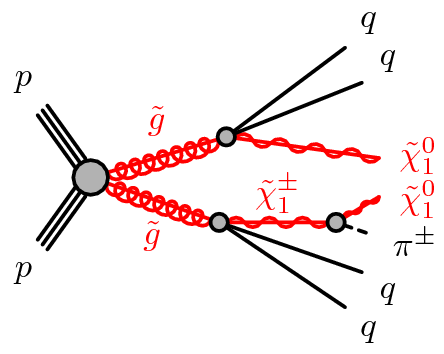}
\includegraphics[width=0.32\textwidth]{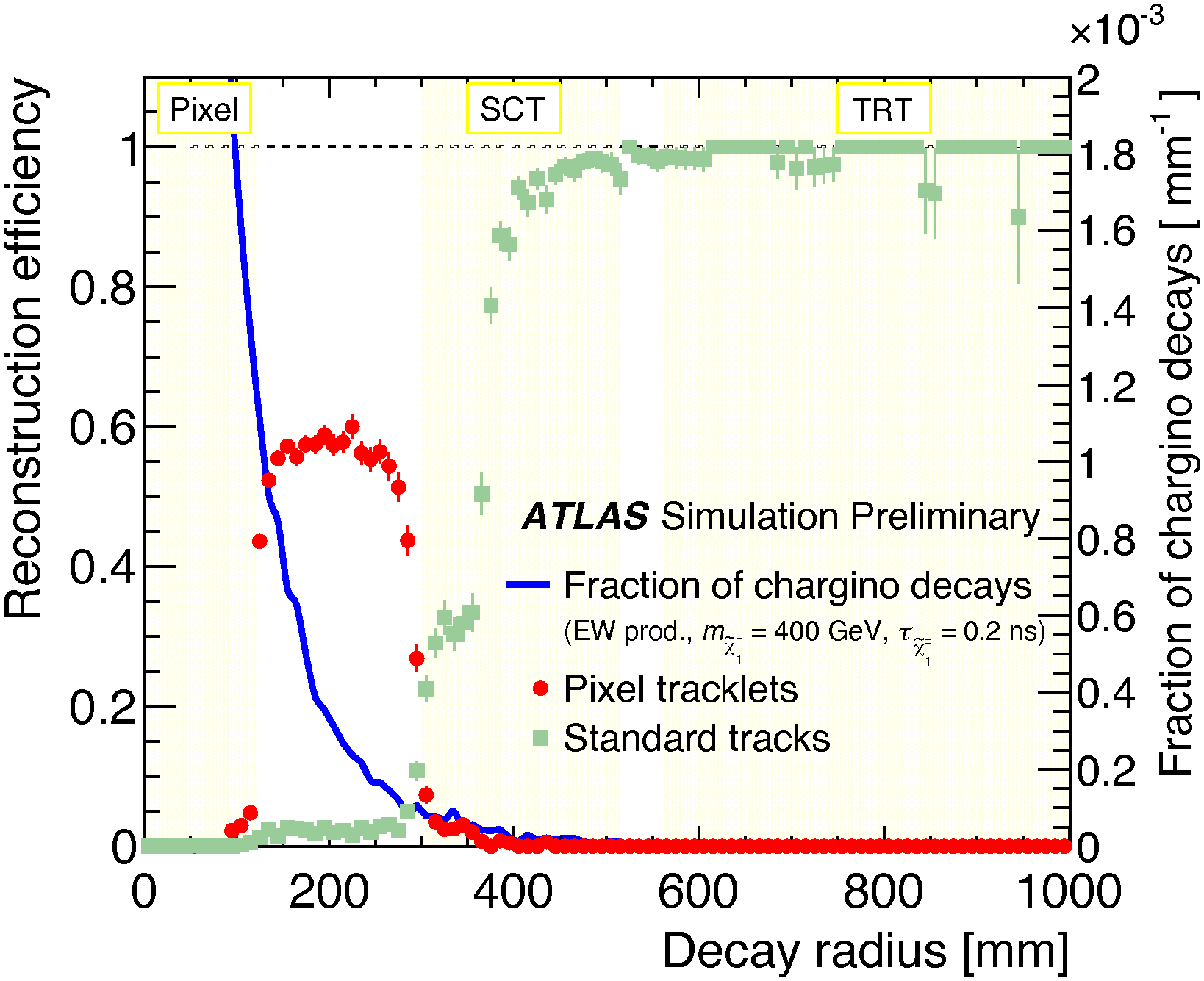}\\
\includegraphics[width=0.14\textwidth]{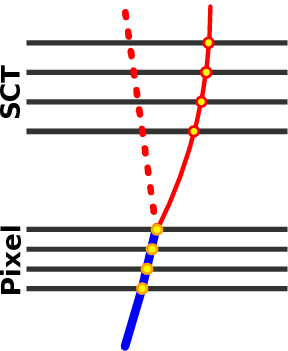}
\includegraphics[width=0.14\textwidth]{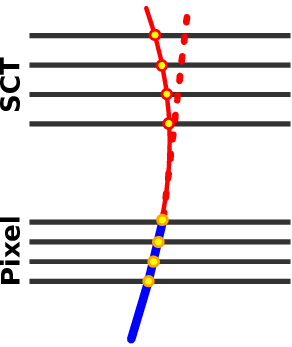}
\includegraphics[width=0.14\textwidth]{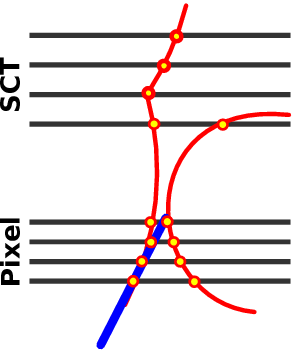}\\
\includegraphics[width=0.29\textwidth]{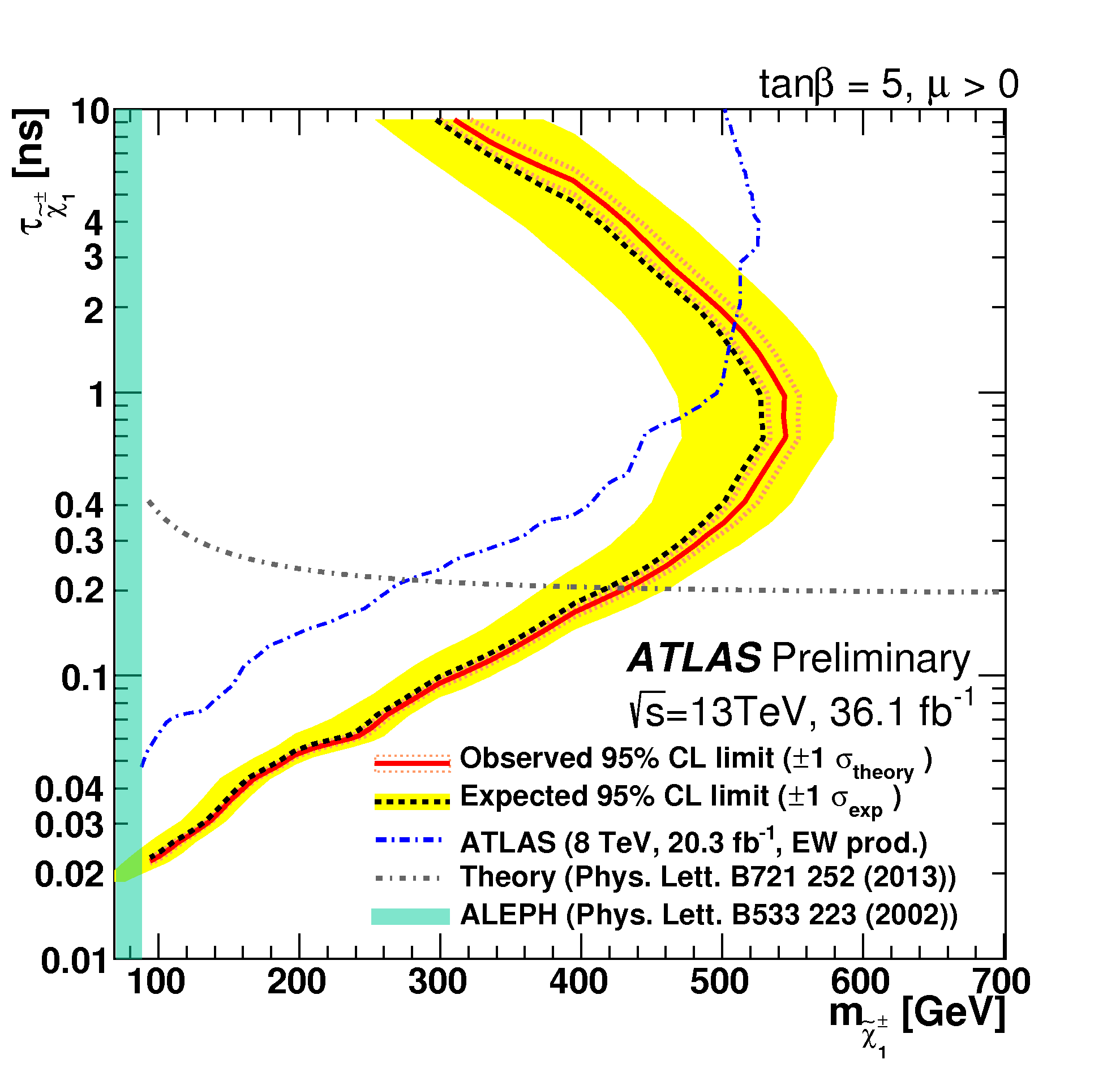}
\includegraphics[width=0.29\textwidth]{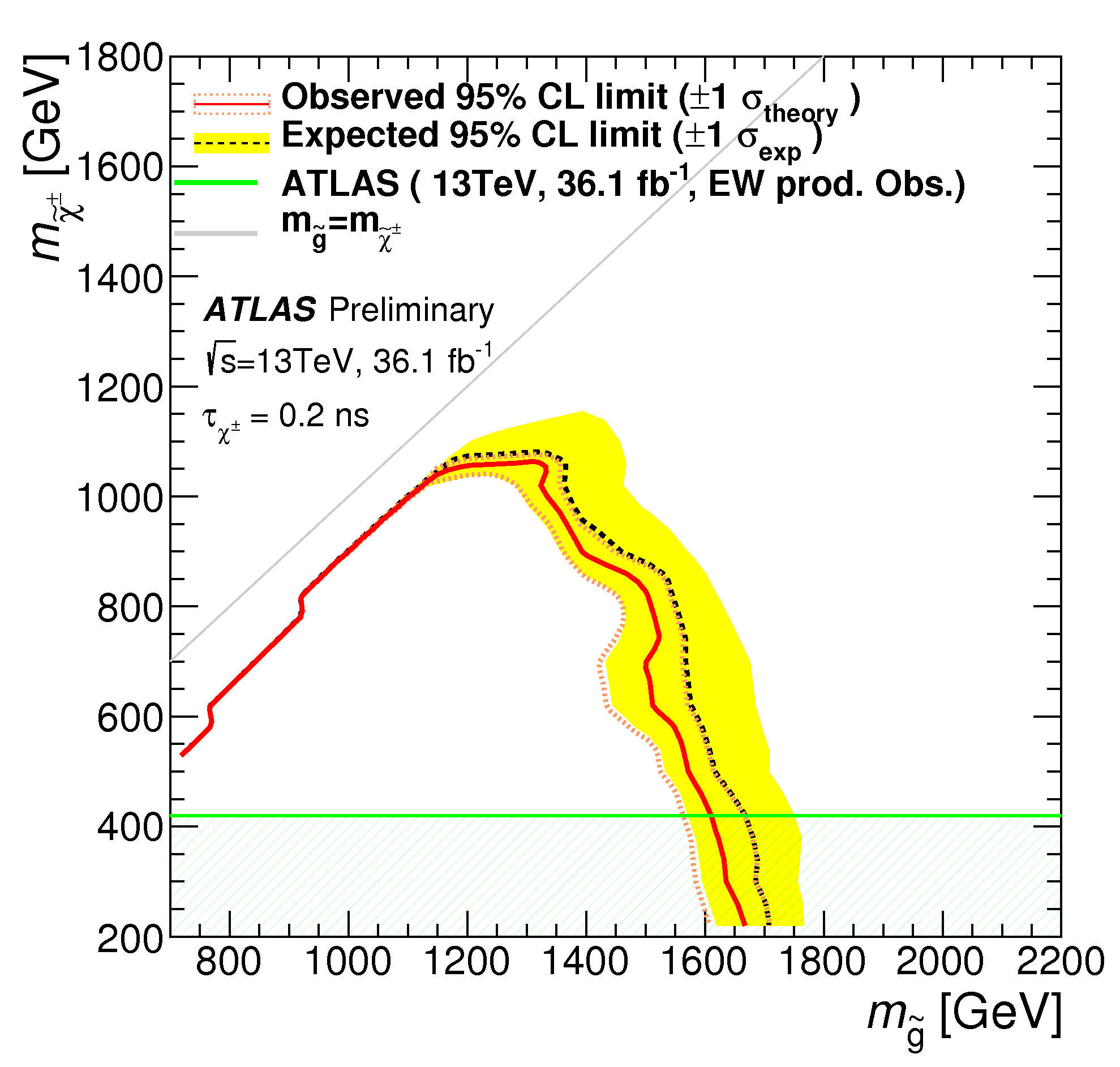}
\includegraphics[width=0.29\textwidth]{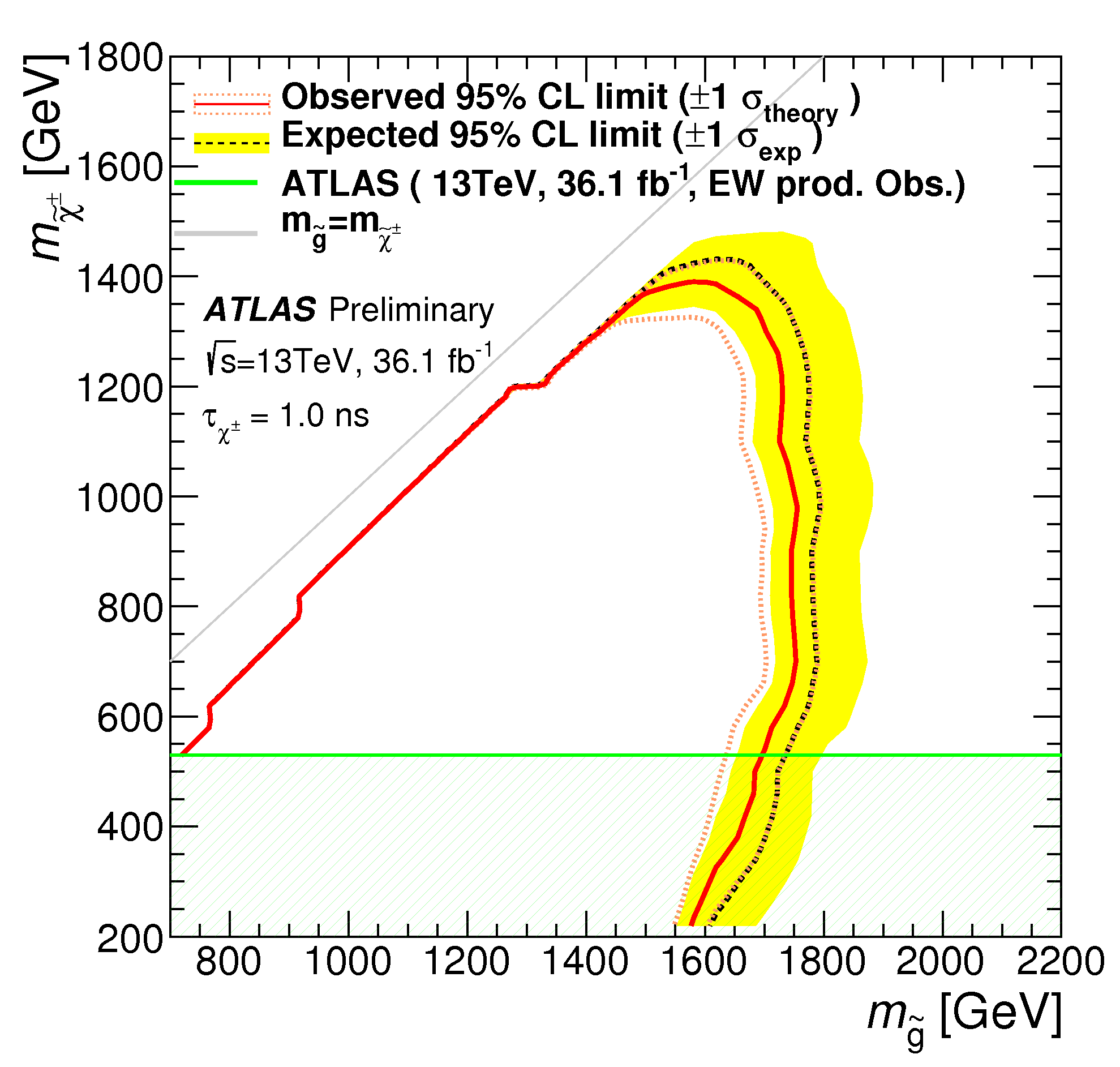}
\caption{Views of the ATLAS vertex detector, 
         long-lived SUSY particle decay topologies, 
         reconstruction efficiency as a function of the decay radius,
         sketches of the different background components to the search for pixel tracklets
         and exclusion contours.
         }
\label{fig:long}
\vspace{-5mm}
\end{figure}

Searches for SUSY particles with a long life time were performed.
Feynman diagrams, the reconstruction efficiency for such particles as a function of the decay radius
and results are summarized in Fig.~\ref{fig:long}~\cite{ATLAS-CONF-2017-017}.

\section{R-parity violation}
\label{sec:parity}

\begin{figure}[h]
\centering
\includegraphics[width=0.24\textwidth]{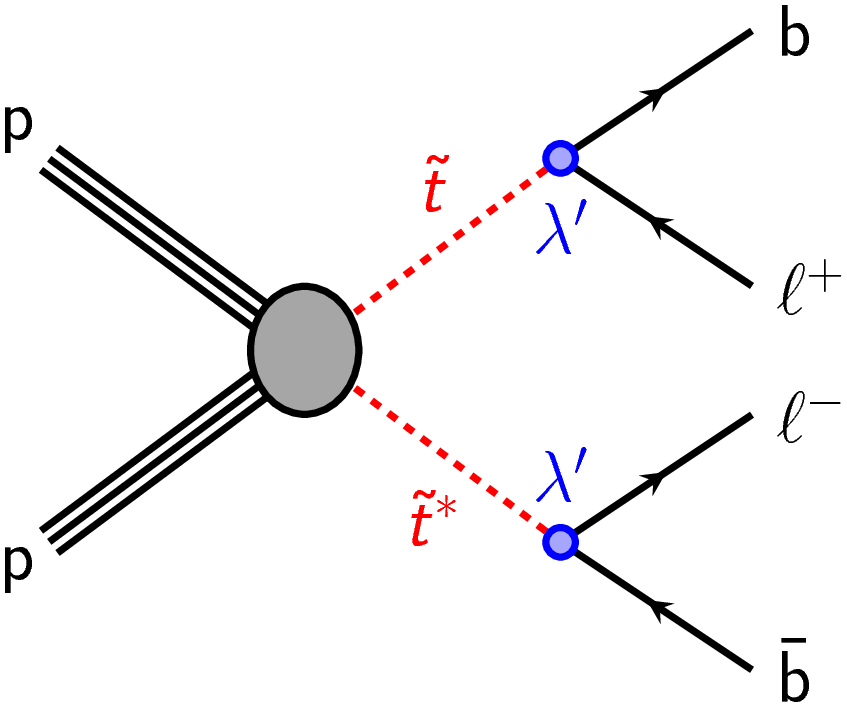}
\includegraphics[width=0.30\textwidth]{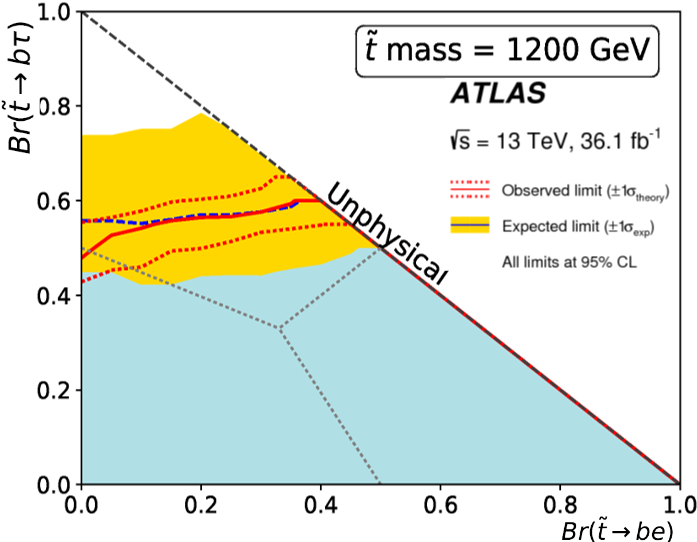}
\includegraphics[width=0.38\textwidth]{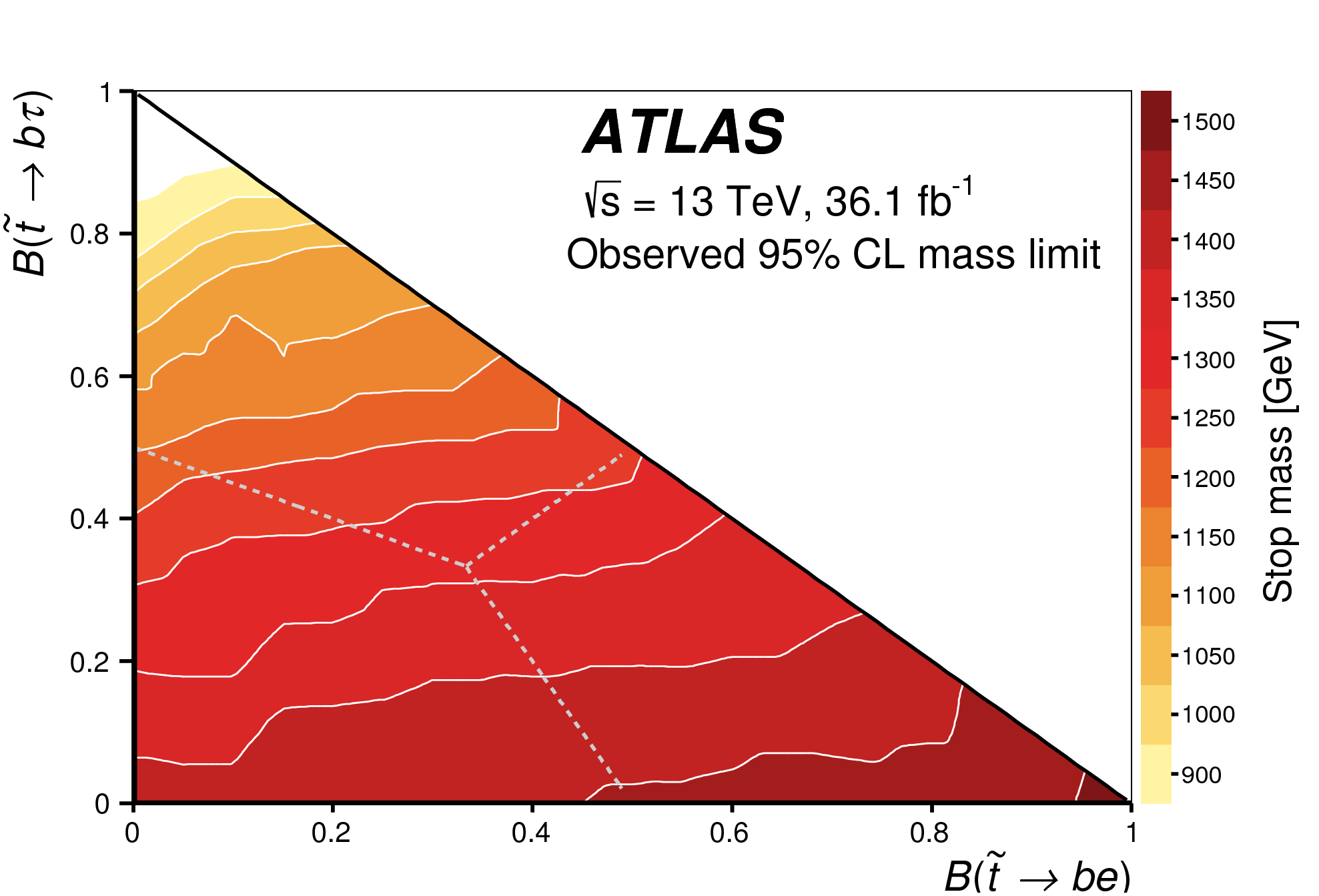}
\vspace*{-2mm}
\caption{R-parity-violating stop quark coupling to a final state with two leptons and two jets decay topologies
         and exclusion contours.
         }
\label{fig:rpv}
\vspace{-5mm}
\end{figure}

\begin{figure}[h]
\centering
\includegraphics[width=0.24\textwidth]{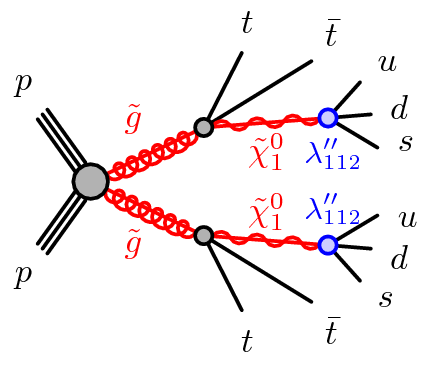}
\includegraphics[width=0.24\textwidth]{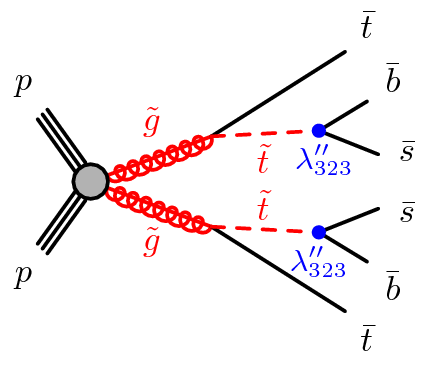}
\includegraphics[width=0.24\textwidth]{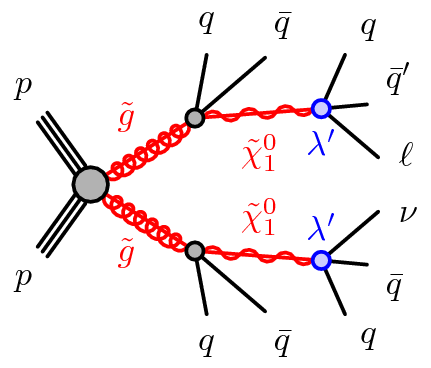}
\includegraphics[width=0.24\textwidth]{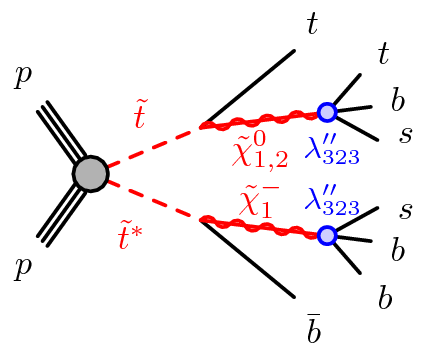}\\
\includegraphics[width=0.24\textwidth]{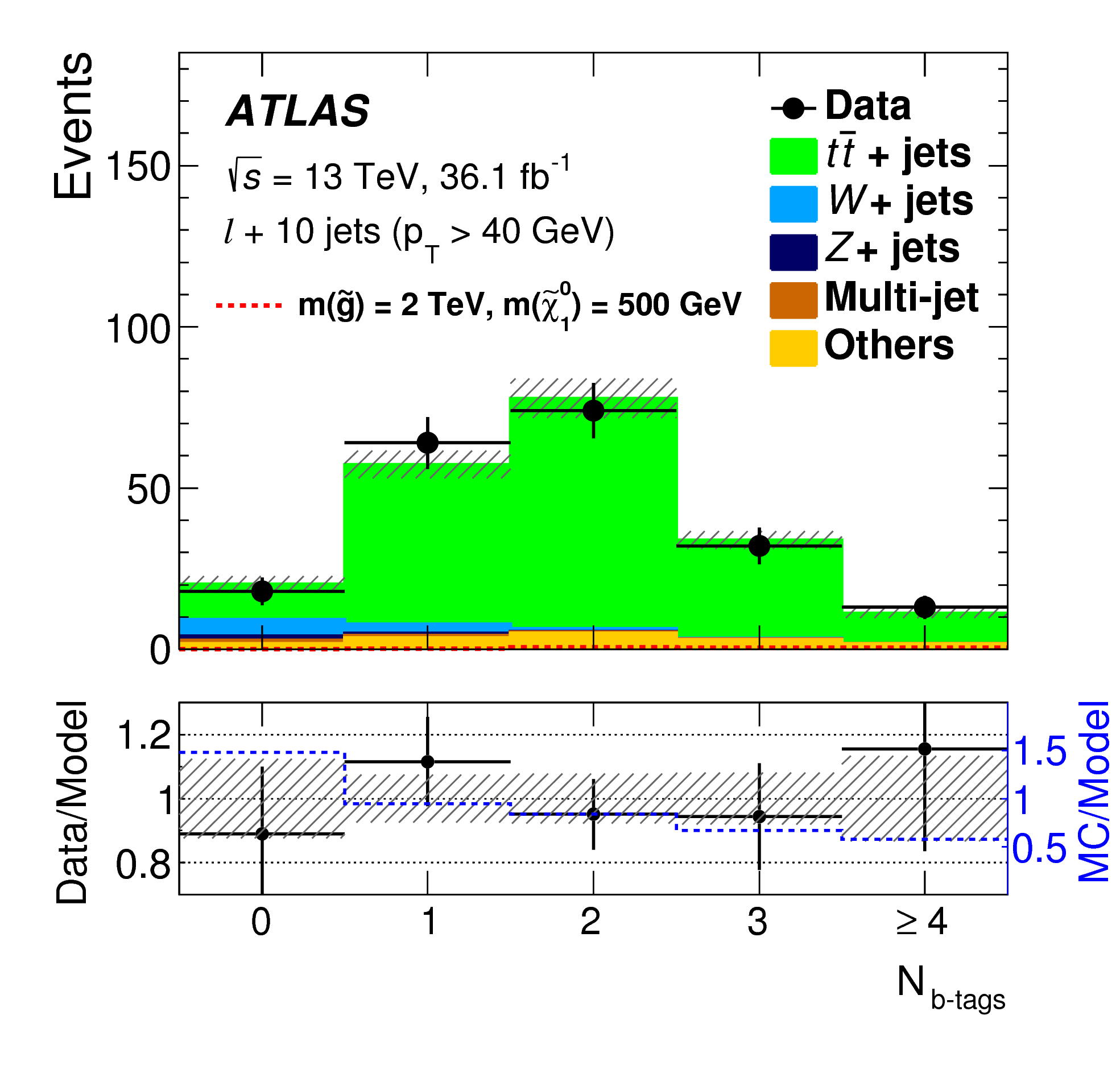}
\includegraphics[width=0.24\textwidth]{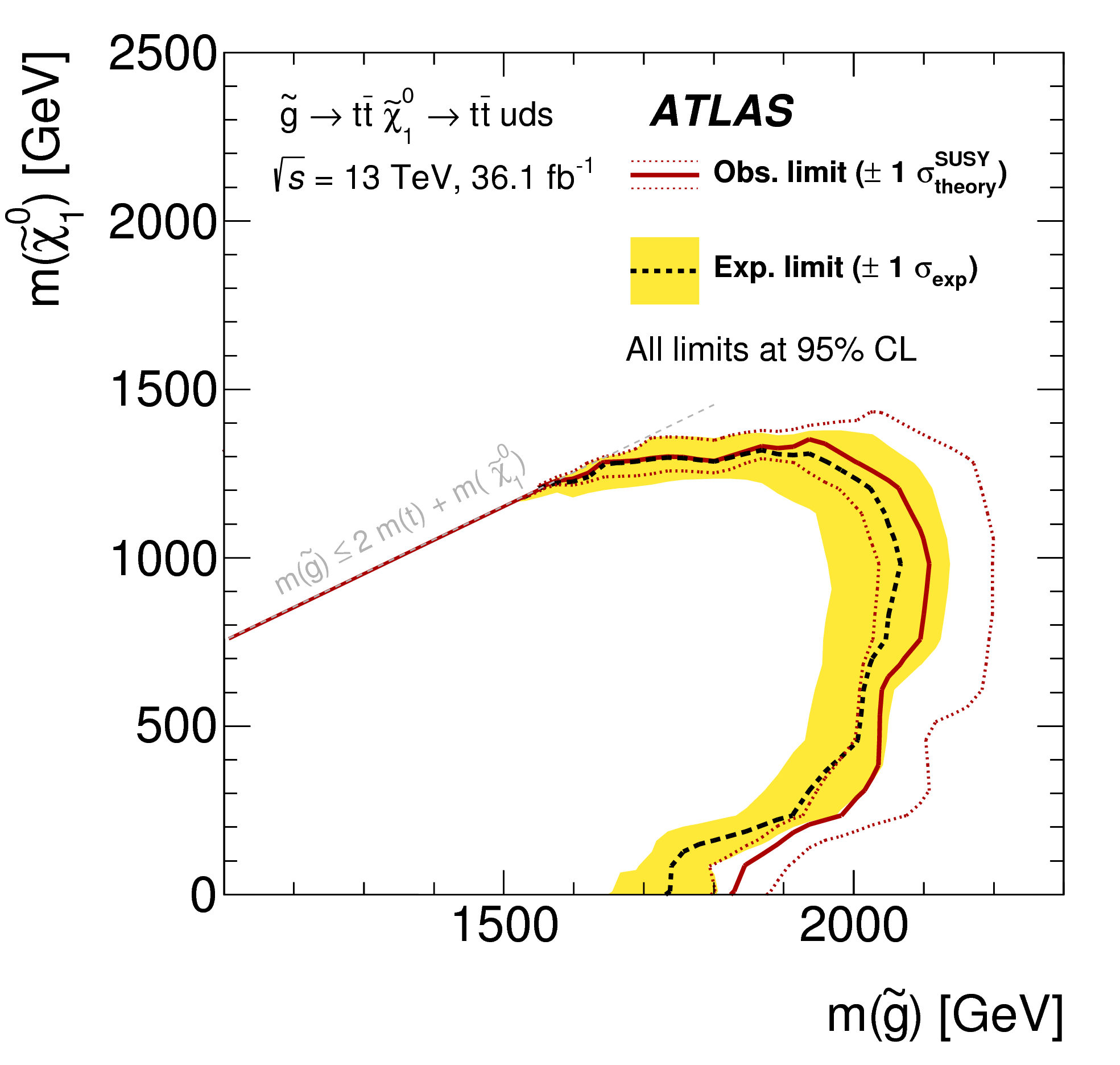}
\includegraphics[width=0.24\textwidth]{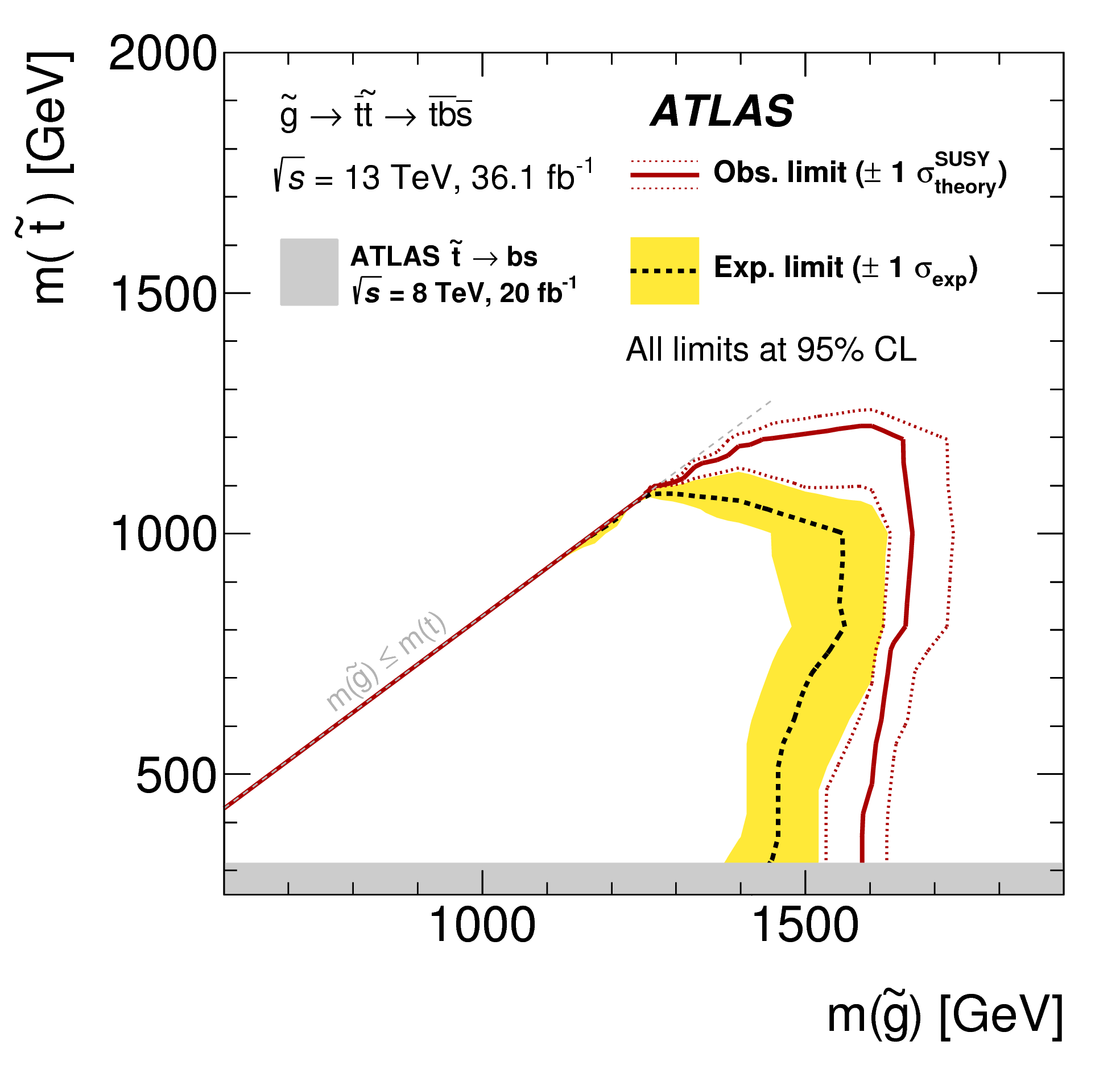}
\includegraphics[width=0.24\textwidth]{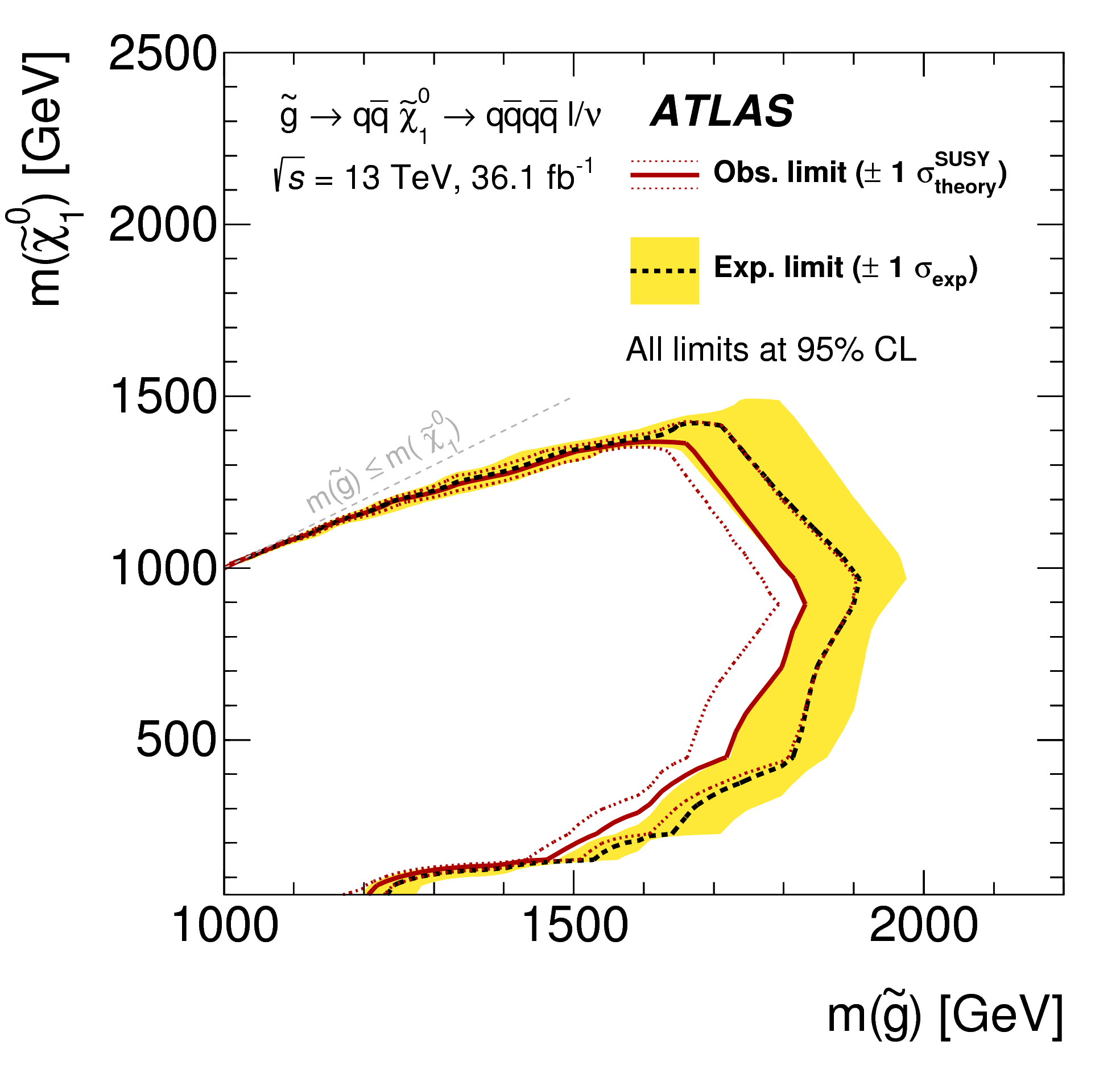}
\vspace*{-2mm}
\caption{R-parity-violating SUSY particle decay topologies with high jet multiplicities decay topologies 
         and exclusion contours.
         }
\label{fig:rpv2}
\end{figure}

Searches for SUSY particles with R-parity violation were performed.
An example is presented for an R-parity-violating stop quark coupling 
to a final state with two leptons and two jets, at least one of which 
is identified as a b-jet. 
Feynman diagrams and results are summarized in Fig.~\ref{fig:rpv}~\cite{Aaboud:2017opj}.
Figure~\ref{fig:rpv2}~\cite{Aaboud:2017faq} shows the corresponding results for searches with a lepton 
plus high (up to 12) jet multiplicity in the final states.

\section{Displaced vertices}
\label{sec:displaced}

\begin{figure}[h]
\centering
\includegraphics[width=0.24\textwidth]{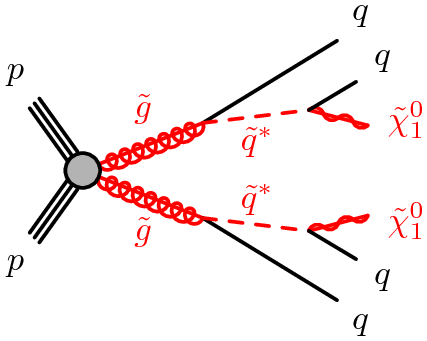}
\includegraphics[width=0.24\textwidth]{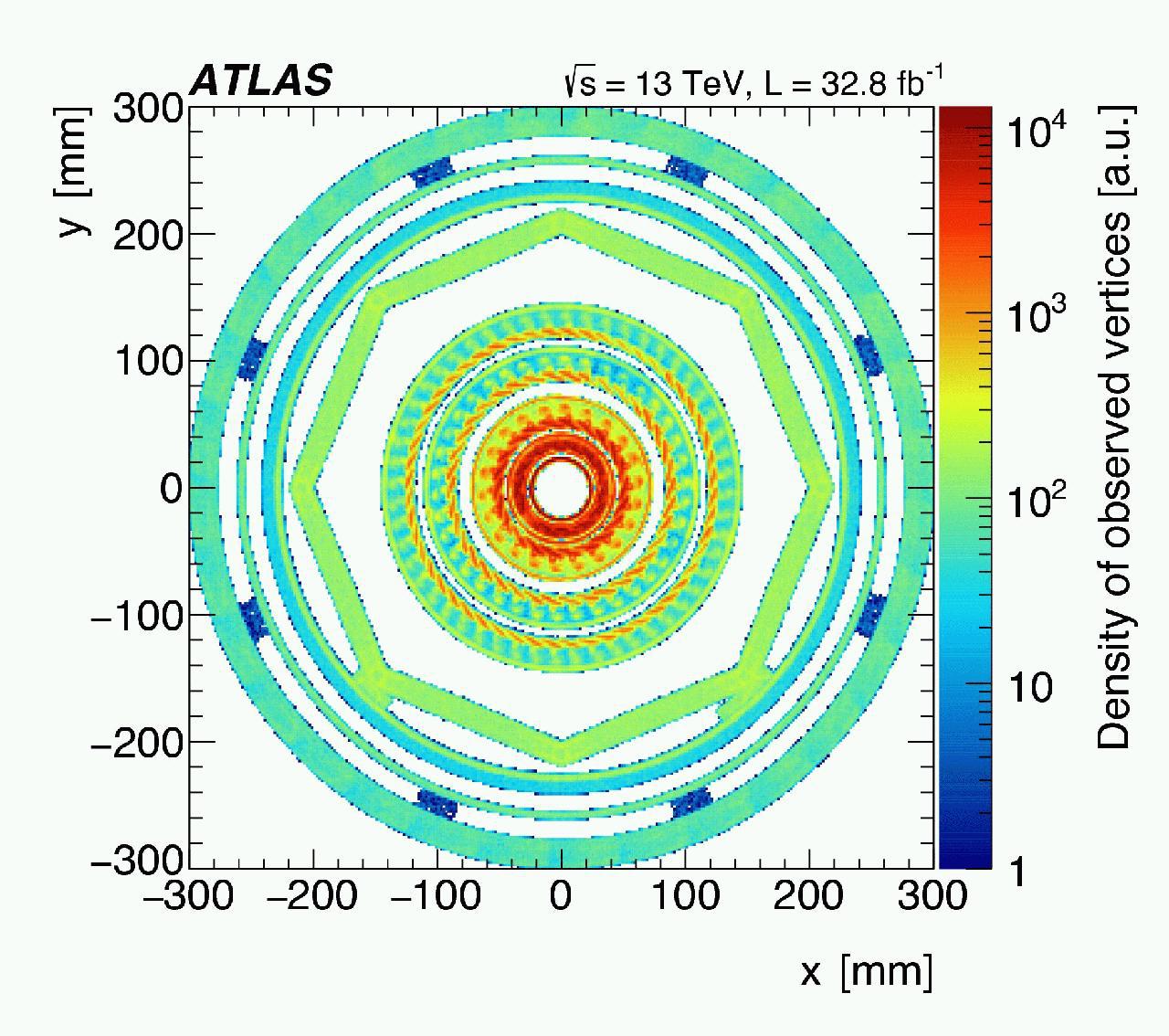}
\includegraphics[width=0.24\textwidth]{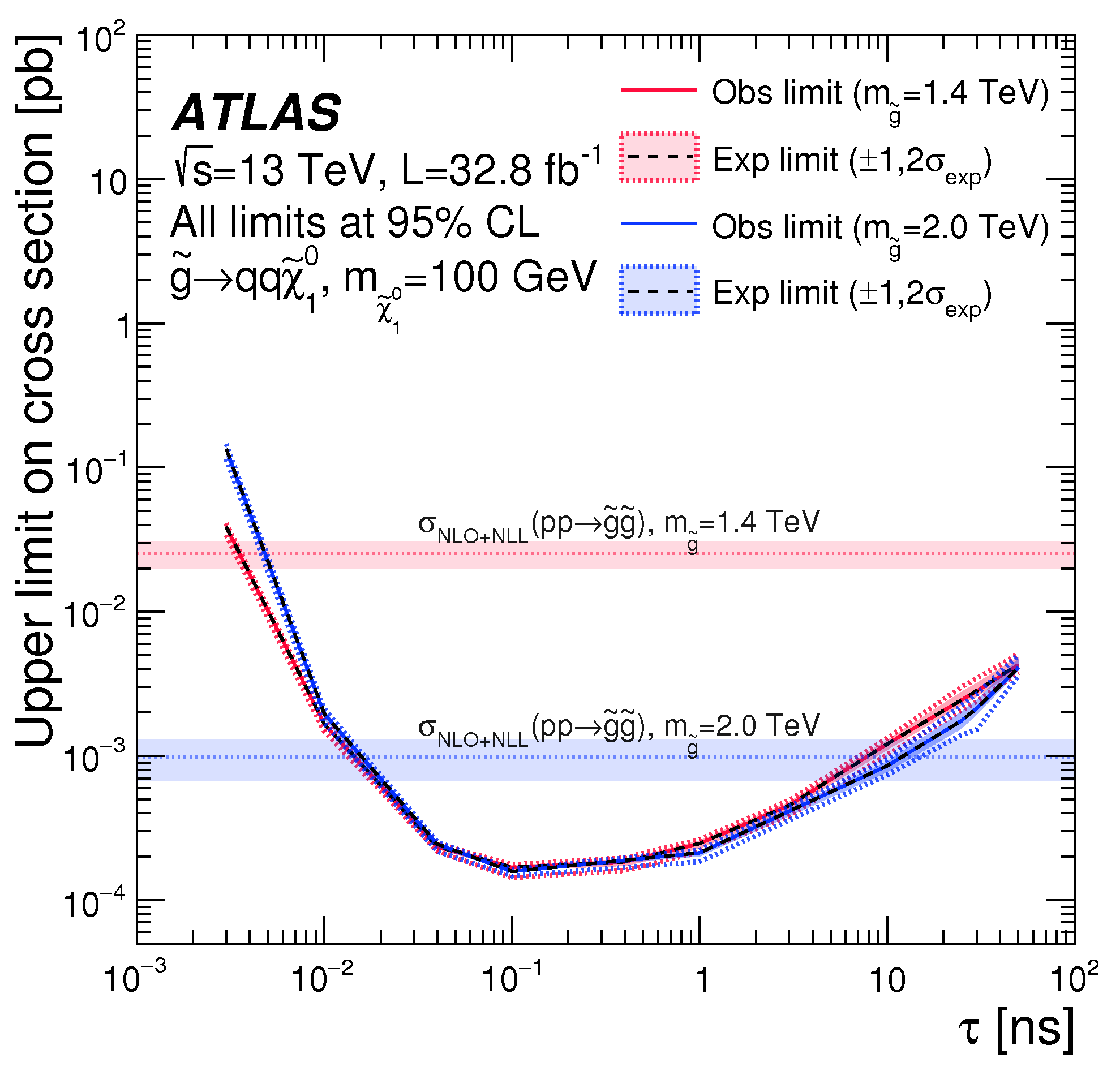}
\includegraphics[width=0.24\textwidth]{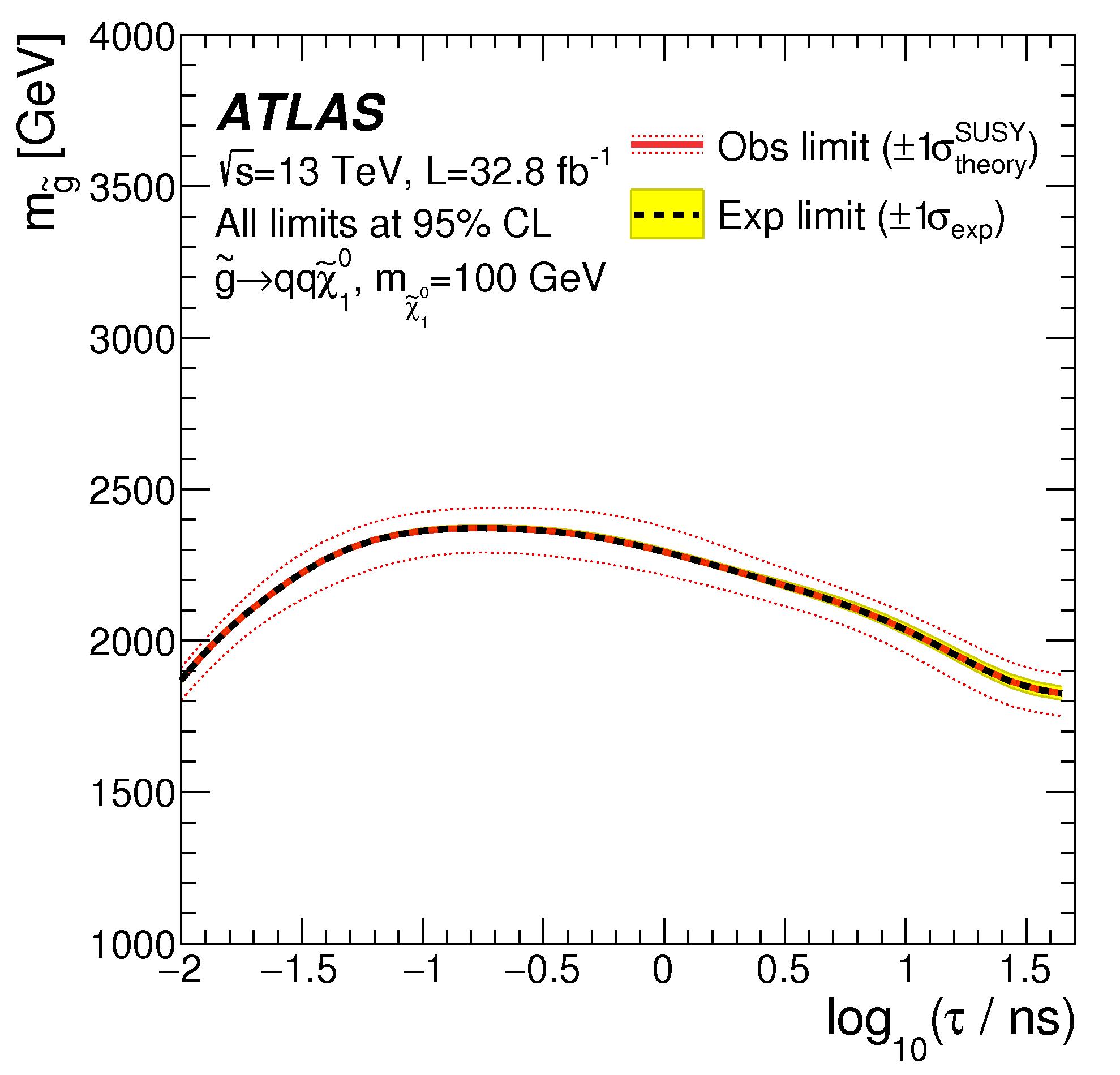}
\includegraphics[width=0.24\textwidth]{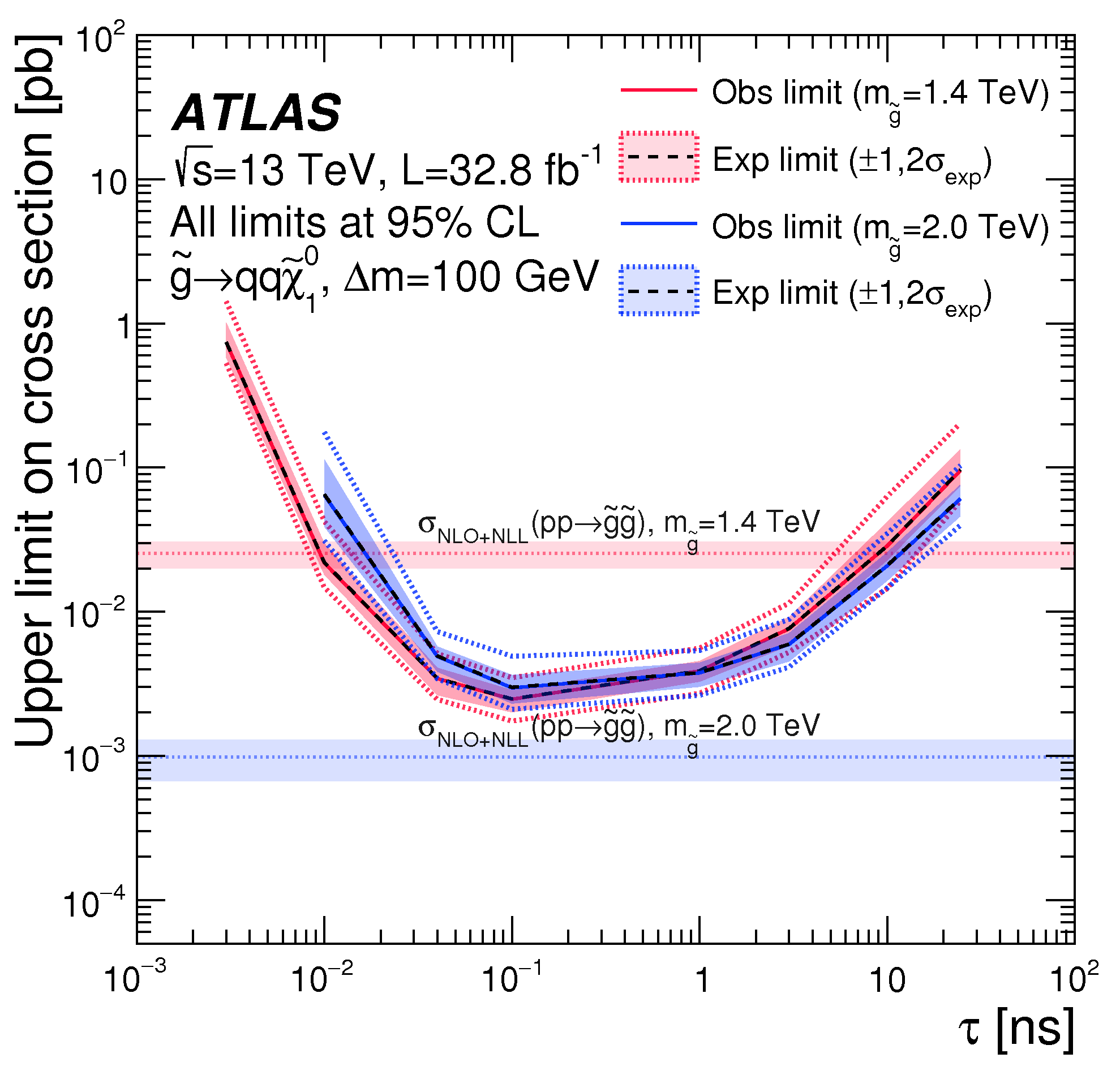}
\includegraphics[width=0.24\textwidth]{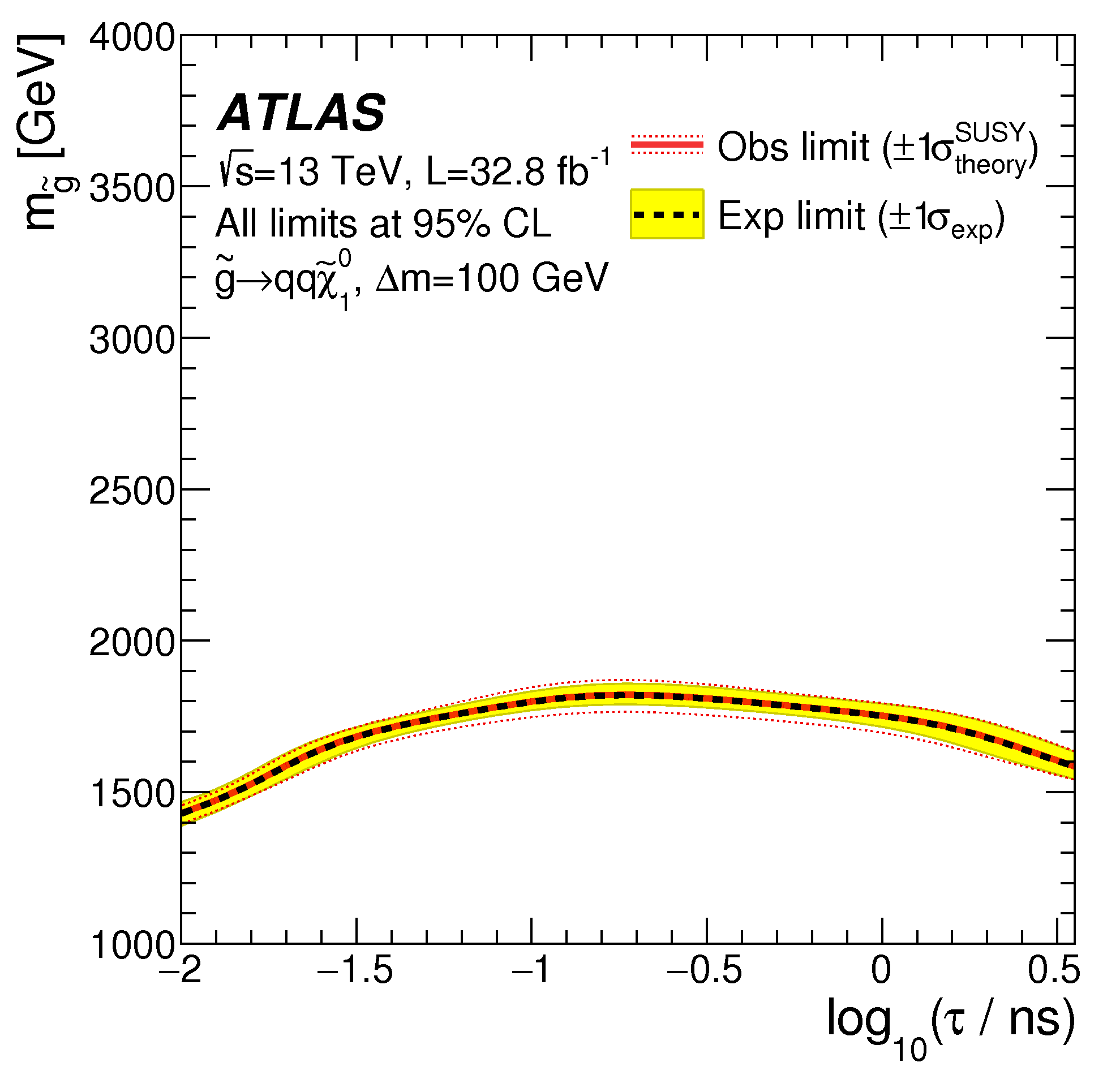}
\includegraphics[width=0.24\textwidth]{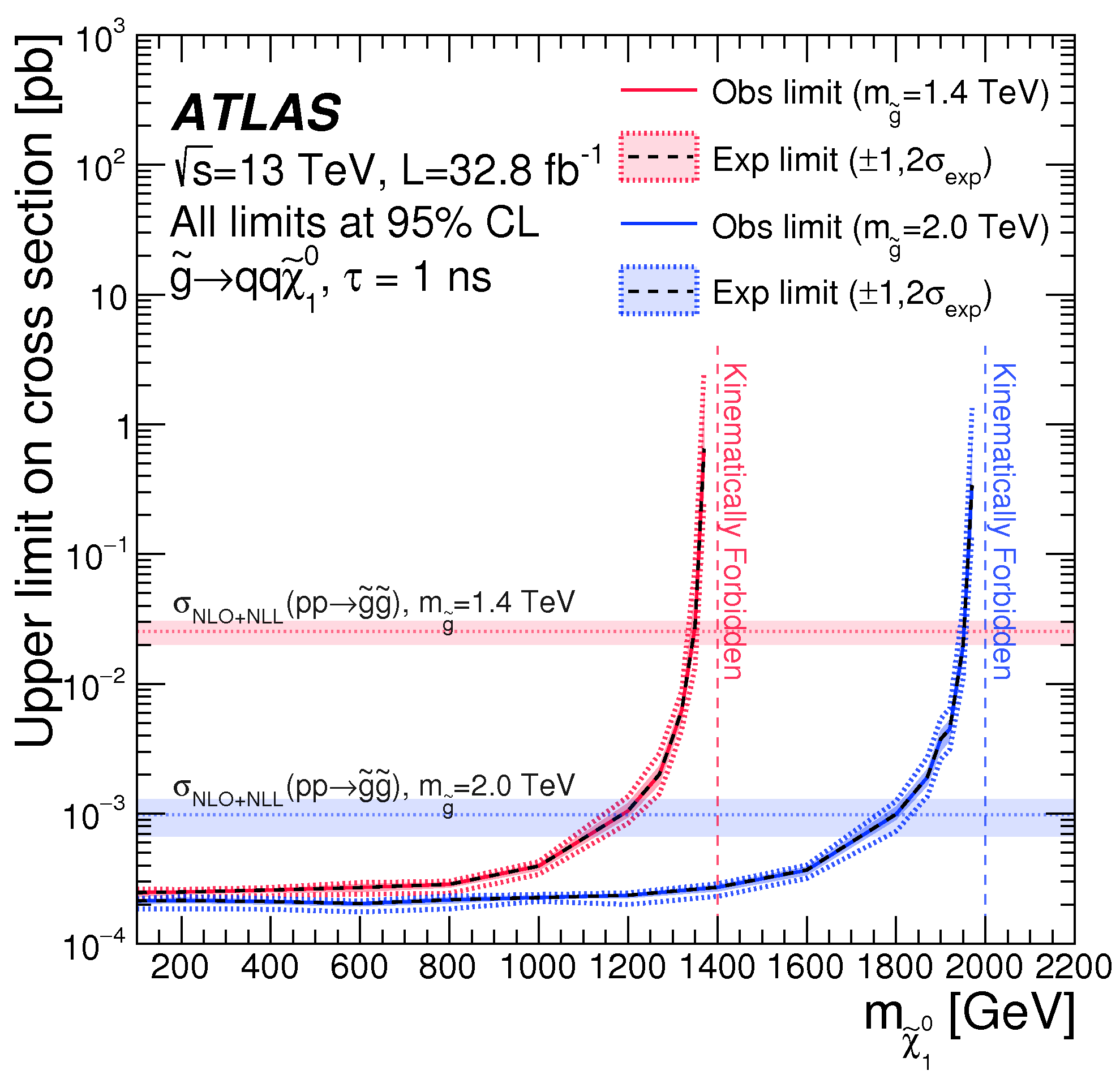}
\includegraphics[width=0.24\textwidth]{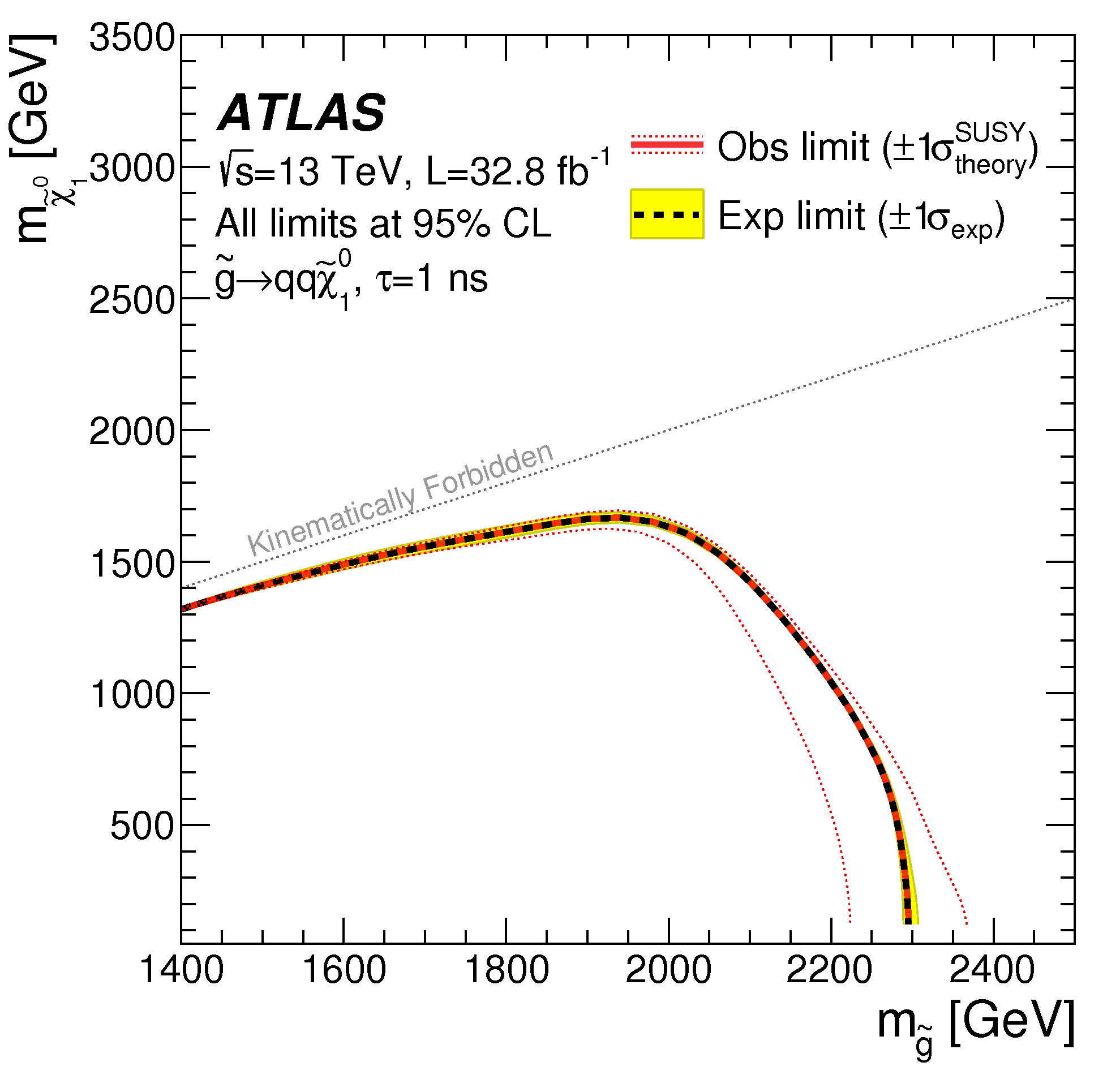}
\vspace*{-2mm}
\caption{Massive particles in events with displaced vertices  
         and large missing transverse momentum decay topologies,
         densities of observed vertices in the inner ATLAS detector
         and exclusion contours.
         }
\label{fig:displaced}
\vspace*{-5mm}
\end{figure}

In split SUSY scenarios, the gluinos are long-lived enough to hadronise to R-hadrons 
that can give rise to displaced vertices when they decay.
Searches for such massive particles were performed in events with displaced vertices 
and large missing transverse momentum.
A Feynman diagram, densities of vertices in the inner ATLAS detector 
and results are summarized in Fig.~\ref{fig:displaced}~\cite{Aaboud:2017iio}.

\section{pMSSM}
\label{sec:pmssm}

\begin{figure}[h]
\centering
\includegraphics[width=0.3\textwidth]{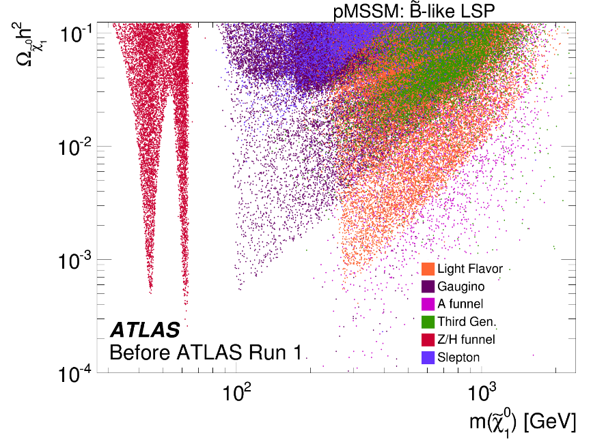}
\includegraphics[width=0.3\textwidth]{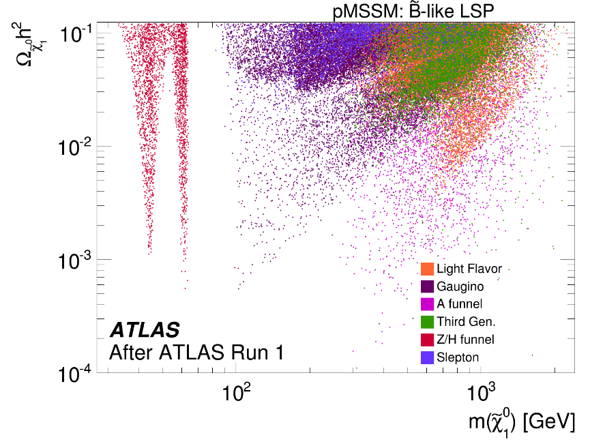}
\includegraphics[width=0.38\textwidth]{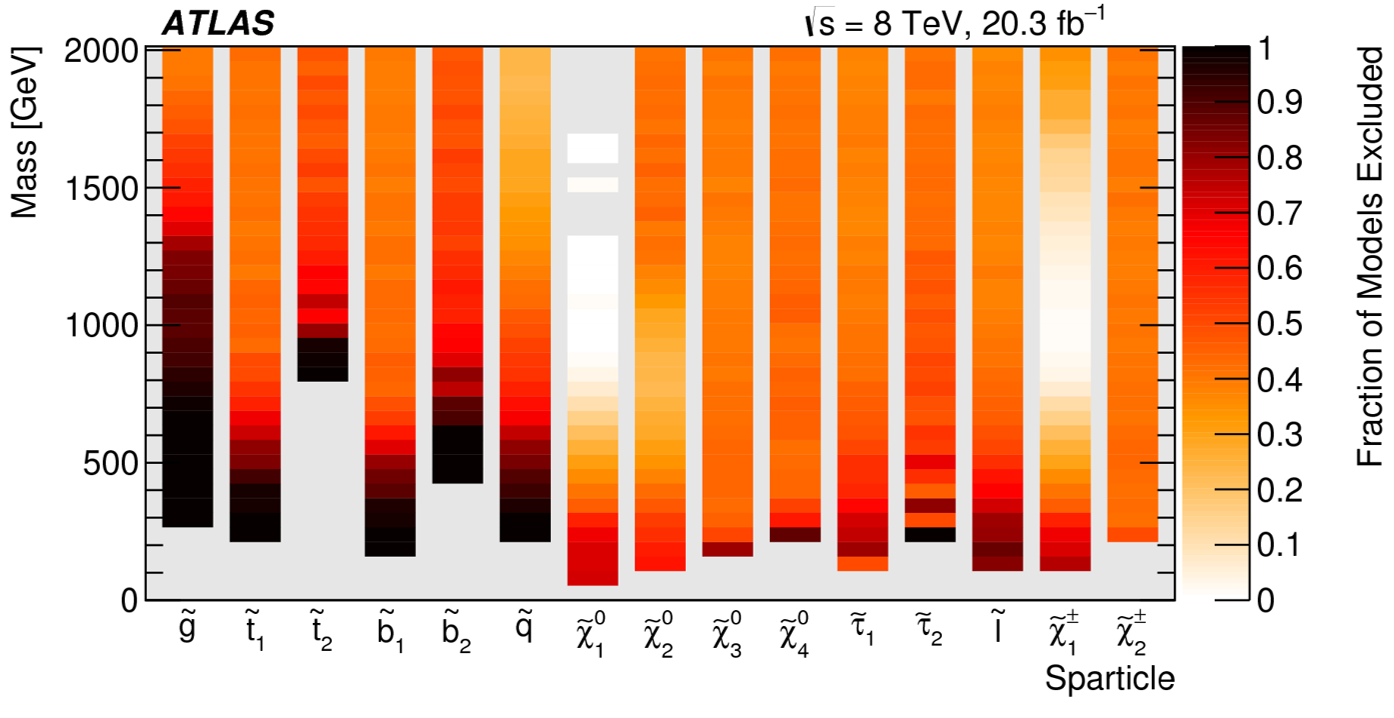}
\vspace*{-2mm}
\caption{pMSSM interpretation of the search results from LHC Run-1 data taking.
         Density of pMSSM points on the plane of relic density versus LSP mass.
         Fraction of model points excluded, with colour coding representing the 
         fraction of model points excluded for each sparticle.
        }
\label{fig:pMSSM}
\vspace*{-5mm}
\end{figure}

The search result for supersymmetry after LHC Run-1 are interpreted in the phenomenological MSSM
(pMSSM)~\cite{Aaboud:2016wna}. A general parameter scan was performed with 19 parameters.
Feynman diagrams and results are summarized in Fig.~\ref{fig:pMSSM}~\cite{Aaboud:2016wna}.

\section{Prospects}
\label{sec:prospects}

\begin{figure}[h]
\centering
\includegraphics[width=0.6\textwidth]{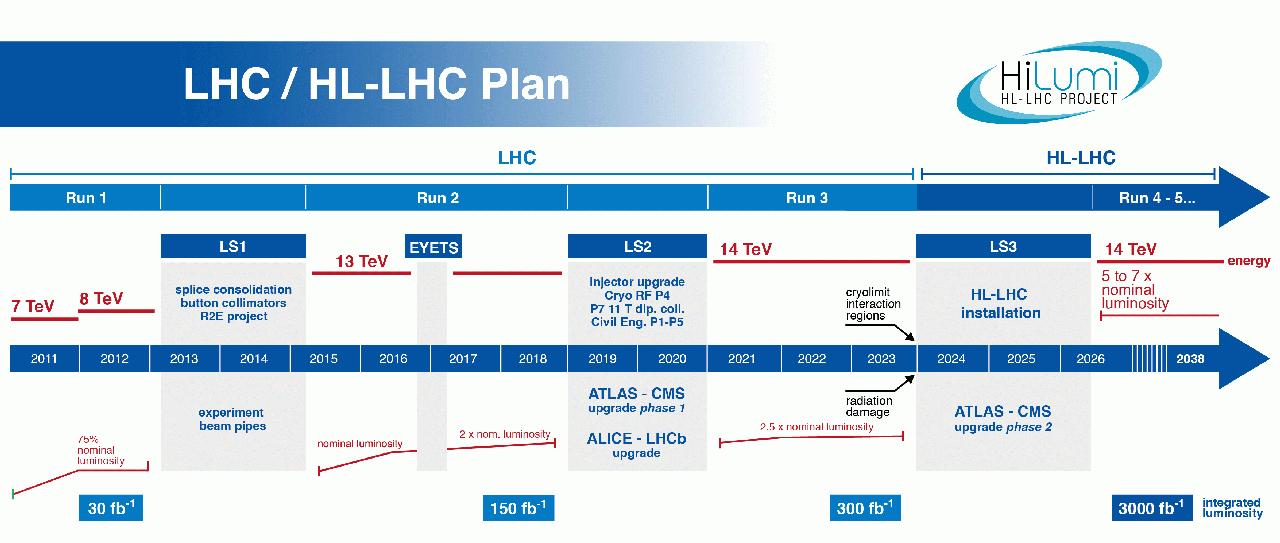}
\includegraphics[width=0.24\textwidth]{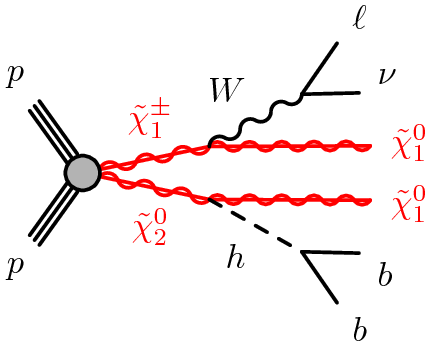}
\includegraphics[width=0.45\textwidth]{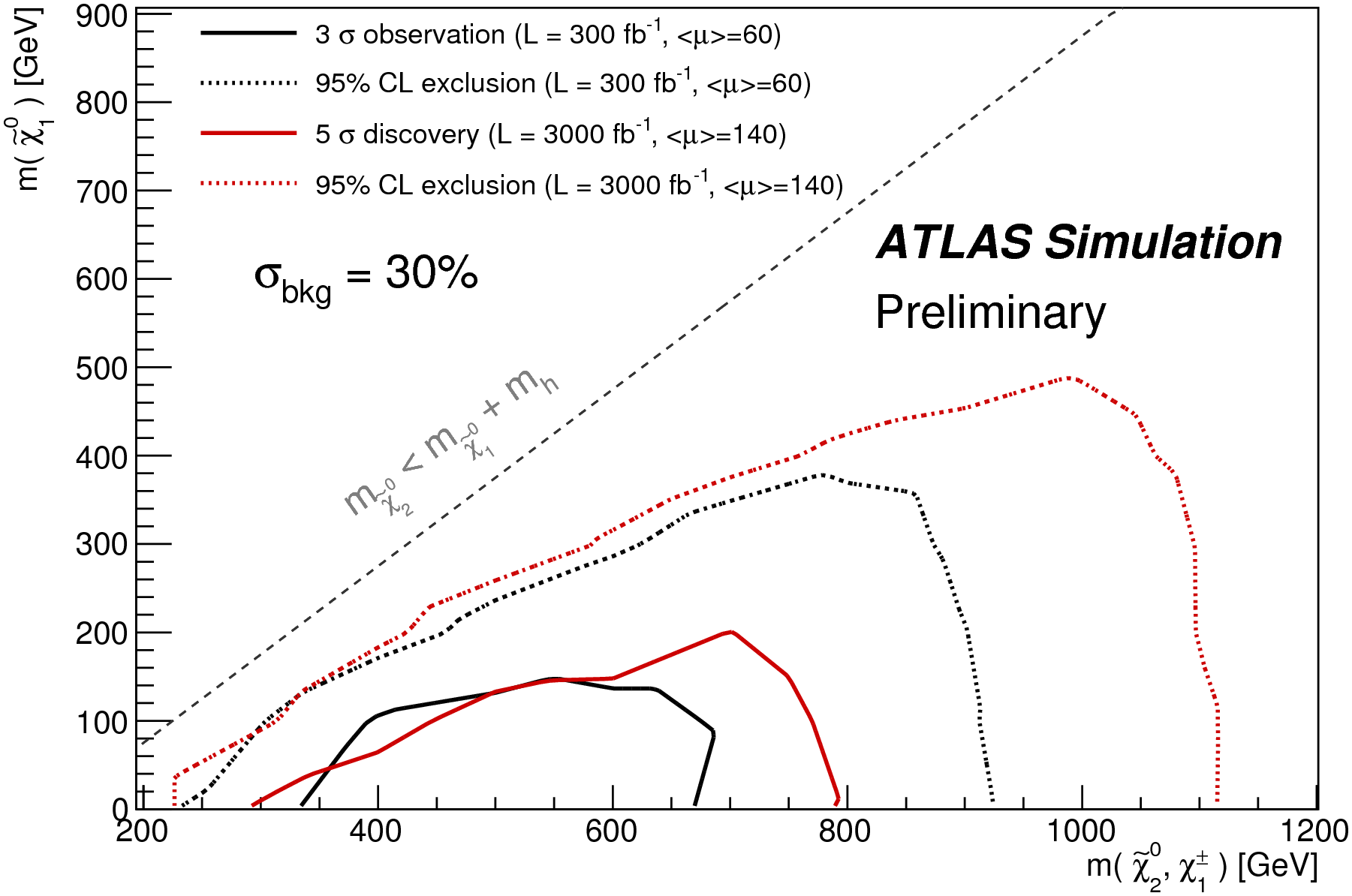}
\includegraphics[width=0.45\textwidth]{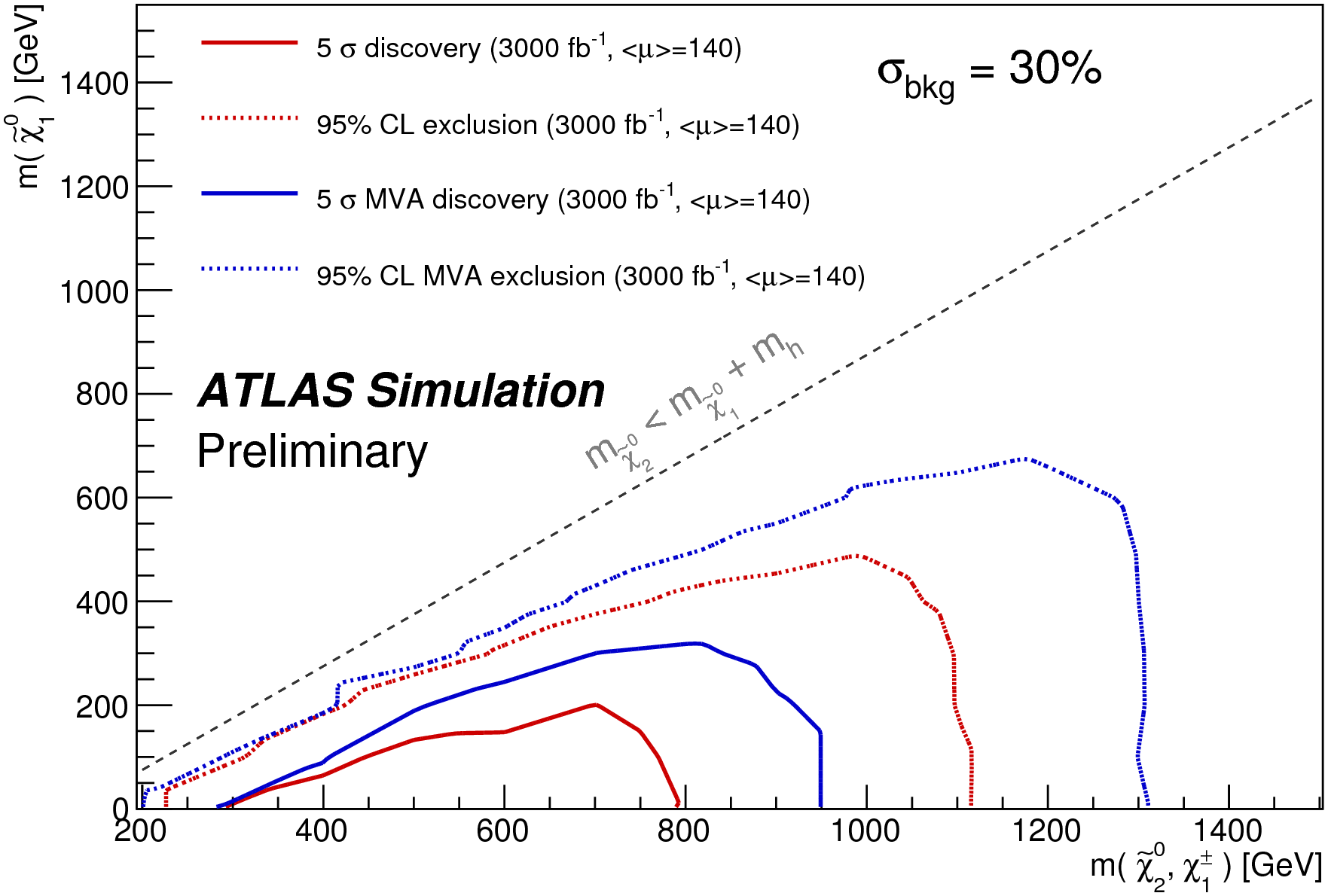}
\vspace{-2mm}
\caption{Overview of the LHC operation for the coming years.
         Prospects for a search for direct pair production
         of a chargino and a neutralino decaying via a W boson and
         the lightest Higgs boson in final states with one lepton,
         two b-jets and missing transverse momentum at the high luminosity LHC.
        }
\label{fig:prospects}
\vspace{-3mm}
\end{figure}

An overview of the LHC operation for the coming years is given in Fig.~\ref{fig:prospects}.
The prospects for SUSY searches were studied for $\sqrt{s}= 14$\,TeV for average proton-proton interactions per 
bunch crossing $\mu=200$, and a total integrated luminosity of 3000\,fb$^{-1}$. 
Smearing functions for the upgraded ATLAS detector simulation and truth level particle corrections for detector 
effects were implemented,
assuming 30\% systematic uncertainties on the background.
As an example, Fig.~\ref{fig:prospects} also shows the prospects for a search 
with the ATLAS detector for direct pair production of a chargino and a 
neutralino decaying via a W boson and the lightest Higgs boson in final states 
with one lepton, two b-jets and missing transverse momentum at the high luminosity 
LHC~\cite{ATL-PHYS-PUB-2015-032}.

\section{Conclusions and outlook}
\label{sec:conclusions}

SUSY searches in ATLAS cover a large range of scenarios.
Inclusive squark and gluino searches resulted in mass limits up to above 2 TeV.
Direct stop limits increased from about 700\,GeV\,(Run-1)\,to about 950 GeV,
and direct sbottom limits increased from about 650\,GeV (Run-1) to about 950\,GeV.
Chargino mass limits range between 600 and 1100\,GeV.
Gluino limits reach up to 2\,TeV with photon signatures.
Searches of long-lived charginos, gluinos, R-parity violation (RPV) and pMSSM 
resulted in stringent limits.
There are excellent prospects to advance SUSY searches with LHC Run-2 
(completing 2015-2018 operation),
LHC Run-3 (2021-2023) and the high-luminosity HL-LHC (2026-).

\section*{Acknowledgment}
The project is supported by the Ministry of Education, 
Youth and Sports of the Czech Republic 
under projects LG15052 and LM2015058.

\bibliography{biblio}

\end{document}